# Multilayer Babinet Metamaterial to Initiate Nonreciprocal Topological Phenomena and Generalized Faraday Rotation

Balázs Bánhelyi[1], Miklós Waldhauser[2], Virág Szünstein[2], Ákos Sebők-Pap[2], Olivér Ardelán[2], Anna Kőházi-Kis[2], Dávid Vass[2], András Szenes[2], David Keene[3], Maxim Durach[3], Mária Csete[2]

[1]Department of Computational Optimization, University of Szeged, Árpád tér 2, H-6726 Szeged, Hungary;
[2]Department of Optics and Quantum Electronics, University of Szeged, Dóm tér 9, H-6726 Szeged, Hungary;
[3]Department of Biochemistry, Chemistry and Physics, Georgia Southern University, 301 Veazy Hall, Suite 2000, Statesboro, GA 30458
Correspondence: mcsete@physx.u-szeged.hu

**Abstract** Multilayers of Babinet complementary periodic structures constructed with miniarrays of spherical plasmonic nanoresonators were optimized to ensure Generalized Faraday Rotation. Nonreciprocal rotation and asymmetric transmission were achieved in spectrally overlapping regions due to the reach physics involving (i) symmetry breaking via coupled localized modes, (ii) Brillouin zone-folding stemmed from constituent sub-lattices forming in-plane twisted coupled loops, (iii) interlayer coupling between Babinet complementary patterns. The nanophotonical phenomena include (i) quasi-BIC resonances, (ii) hierarchically coupled localized and propagating modes that results in time-periodic Floquet modulation, (iii) initialization of synthetic potentials tuneable independently via intra and inter-layer parameters. The unique bianisotropic composites result in a synthetic vector gauge and emulated magnetic field manifesting itself in tilted-precessing magnetic dipoles and the accompanying modulation being time-periodic, inherently ensures a synthetic dimension. The asymmetric transmission is enhanced in the classical sense along quantized flat bands, and in mixed and forward bases inside finite wavelength-and-tilting intervals overlapping with nonreciprocal polarization rotation. The transmitted pulse re-shaping proves beating of nearby resonant modes, the loss can be compensated with active ad-layers thereby resulting in Faraday isolator capability. The multilayers synthetize topological phenomena in high-dimensional synthetic parameter spaces.

## 1. Introduction

Topological phenomena allow tailoring of the dispersion characteristics and to achieve various counterpart nanophotonic phenomena via metamaterials [1]. In quantum information processing applications desired capabilities are the control of the polarization and intensity, with propagation direction selectivity on demand, e.g. to perform Faraday isolation.

To achieve large transmission in tailorable spectral intervals, electromagnetically induced transparency (EIT) like phenomena were primarily proposed. Coupled nanoresonator induced transparency (CRIT) can be achieved in case of strong-coupling between modes of different complex frequency [2]. A certain degree of detuning is essential to prevent degradation of transparency in presence of losses, indicating a trade-off, because of the achievable $Q$ is inversely proportional to the $\delta\omega$ [3,4]. The hybridization of plasmons excitable on coupled embedded disk and surrounding ring results in Fano lines and the interaction of narrow dark and broad bright resonances leads to plasmon induced transparency [5-7]. Hierarchical hybridization of localized modes and surface lattice resonances on plasmonic particles forming honeycomb pattern results in asymmetric near-field [8].

It is a great challenge to break reciprocity in the far-field responses, e.g. asymmetrical transmission can be achieved via bianisotropic metamaterials [9]. For nonreciprocal polarization rotation gyromagnetic or gyroelectric materials and Weyl semimetals were proposed, however such materials have limited compatibility with QIP elements and are difficult to tune their performance even in twisted multilayered configuration [10-13]. Metamolecules offer solutions, as the chiral and Tellegen metamaterial elements result in polarization dependent, while the omega and artificial moving elements allow for propagation direction dependent optical response. When these elements are combined, one has to consider that chiral and artificial moving metamaterials are capable of converting the frequency, while omega and Tellegen metamaterials modify only the phase [14].

However, asymmetric transmission with reversed polarization rotation could be realized with a very complex 3D metamaterial up until now [15]. In addition, modes supported by complex metamaterials are often non-radiative, therefore different methods are necessary to extract light in the case of bound-states-in-continuum (BICs) [16]. Coupling of quasi-BIC and scattering multipoles allows super-scattering beyond radiation limits, due to the lifted degeneracy between poles and zeros allowing for a superdipole emergence [17]. Nonreciprocal transmission was achieved via BICs on double layers with vertical mirror symmetry breaking, which enables phase inversion ($\pi/2$) in the Γ-M direction [18]. Metasurfaces that are made of ferromagnetic materials and support BIC exhibit strong nonreciprocity and result in Faraday rotation due to enhanced magneto-optic effect [19]. The BIC and guided mode resonances can be tuned into radiative modes via symmetry breaking and multiple periods induced band inversion phenomena and Brillouin-Zone folding with inherited properties revealing their predecessors [20].

Tailoring the topology of metamaterials has the potential to control the optical response, polarization (spin) and to ensure magnet-free nonreciprocity. The chirality of the modes excitable on hexagonal plasmonic particle arrays itself results in Berry phase modulation [21]. Primary studies revealed that an effective gauge can be initialized by varying the phase dynamically on a nanoresonator lattice [22]. Bianisotropic metamaterials acting as photonic topological insulators (PTI) were demonstrated that ensure magneto-electric coupling and are capable to emulate the spin-orbit coupling [23,24].

Such pseudo-spin engineered topological systems rely on anisotropic meta-atoms or co-existent shrunken and expanded sub-lattices of graphene-like honeycomb hexagonal patterns [21-26]. The shrinking and expansion results in band-folding and valleys (pseudo-spins) mixing, manifesting itself in four doublet bands [27]. The quality factor of radiative modes can be controlled with symmetry breaking phenomena realized via profile engineering, while the mode polarization can be tuned continuously by varying the spin-Hall and valley-Hall contribution [28,29]. Two-dimensional metasurfaces thereby enable light guiding or radiation with a degree controlled by the topology [30]. Topological defects create a cavity, where the vanishing phase allows for meeting the quantization condition, which results in appearance of equally spaced modes [31].

Metamaterials topology offers versatile concepts to achieve nonreciprocal phenomena [32]. Accordingly, various (vectorial, scalar, non-Abelian, real and complex, non-Hermitian) gauge fields have already been demonstrated in real and synthetic spaces and used in topological photonics [33].

The EIT-like phenomenon can be converted into nonreciprocal transmission by adding temporal modulation to the surface impedance [34]. The valleys degeneracy can be lifted by breaking the time-reversal symmetry via time-modulation. The methodology was extensively studied during the development of metamaterial counterpart of Floquet photonic topological insulators, and it was demonstrated that the counterparts of $\omega$ gaps are $k$ gaps in PTIs [35,36]. A cascaded photonic transition was achieved by modulated refractive index [37]. Floquet ladder was demonstrated via electro-optically modulated ring [38]. Complete polarization conversion was demonstrated by temporally switching the permittivity and permeability between isotropic and anisotropic values [39].

In the case of generic space-time variation giant nonreciprocity can be achieved, when the modulation speed approaches the speed of waves [40]. The phenomena accompanying the simultaneous variation of the permittivity and permeability phased in space and time manifest themselves in bianisotropic effective parameters mimicking moving uniaxial media, which exhibit the Fresnel drag effect [41,42]. The time-reflection can also be eliminated, when the impedance remains constant at the temporal interface and gain is achievable [13,42]. Specific examples of space-time modulated materials were thoroughly inspected, and several topological phenomena where adopted to metamaterials [43,44]. Traveling wave space-time modulation is a considerable option, since it results in magneto-electric coupling and a corresponding synthetic magnetic potential [45]. Rotating wave space-time modulation also can result in bianisotropic response and in the breaking of time reversal symmetry [46].

In nonreciprocal bianisotropic metamaterials broadband amplification can be achieved via adiabatic space-time modulation [47]. The expansion of the k-gaps via photonic-time crystals by using structural resonances was also demonstrated [48]. Recent results show that space-time duality manifests itself PTC like phenomena in spatially periodic systems subject of temporal modulation and temporal flips induce topological transition [49,50]. Two-coupled non-Hermitian loops allows for band braiding, when the amplitude and phase is simultaneously modulated [51]. Recent studies show that Landau levels are achievable in non-Hermitian type-II Dirac systems in case of coexistent gauge $\boldsymbol{E}$ and $\boldsymbol{B}$ fields [52].

The potential of nonreciprocal phenomena in optical isolation was already demonstrated [53-56]. To compensate losses in non-Hermitian systems and to ensure sufficiently high transmission, metamaterial elements are often combined with active media.

Topological metamaterials were also applied to tailor lasing and to generate BICs and Dirac-vortex cavities [57]. To achieve topological-cavity surface emitting laser (TCSEL) a Dirac-vortex pattern was generated via Kekulé transformation of honeycomb lattices and used also for Majorana states and on-chip vector beam tailoring [58,59]. Vectorial vortex thermal light was also generated, which originates from BICs, by introducing BZF as a part of topological design [60].

A metamaterial possessing all these capabilities is an ideal candidate for Faraday isolation. In our previous studies we have presented multilayer plasmonic metamaterials, which enable asymmetric transmission [61,62]. In present work similar multilayers of Babinet complementary periodic convex and concave structures, constructed with complex miniarrays of spherical plasmonic nanoresonators, were optimized to achieve nonreciprocal polarization rotation beside asymmetric transmission.

The key topological as well as near-field and far-field nanophotonic phenomena were identified via steady-state and time-domain computations, and the material loss was overcome by adding an active cover layer, thereby demonstrating Faraday rotation and isolation capabilities.

## 2. Optimization Method

Multilayer metamaterials were constructed by aligning Babinet complementary subwavelength periodic patterns of spherical nanoresonator miniarrays in convex-concave-convex succession. The primary spherical nanoresonator and periodic pattern parameters were in accordance with our previous studies [61,62].

The optimization was performed to achieve the generalized Faraday rotation (GFR) phenomenon by illuminating the multilayers with linearly (p-polarized) light. Accordingly, the objective function was defined to reach $\sum=(\beta_{forward}-\alpha_{forward}+\beta_{backward}-\alpha_{backward})\sim 90°$ cumulative polarization rotation, where the polarization ellipses' orientation corresponding to the incoming and outgoing beams is indicated by $\alpha_{forward\&backward}$ and $\beta_{forward\&backward}$ in case of forward and backward propagation directions, respectively (schematics in Fig. 1c and Fig. 2c).

The GFR condition can be fulfilled by meeting the following criteria: (i) the illumination of the multilayer is realized with a backward-propagating beam in azimuthal orientation corresponding to the outgoing polarization orientation achieved in case of p-polarized light illumination with forward propagating beam ($\alpha_{backward} := \beta_{forward}$); (ii) the polarization rotation occurs in the same direction in both propagation directions, which means nonreciprocal rotation; (iii) the difference of co-polarized transmitted beam intensities – hereinafter referred to as asymmetric transmission ($AT$) - is maximal in the azimuthal orientations meeting the GFR criterion ($AT_{co-polarized}^{balanced\,/\,differential} = AT_{co-polarized}^{mixed\,/\,forward\ bases} = |T_{co\text{-}polarized_forward} - T_{co\text{-}polarized_backward}| :=$ maximal); which implies that (iv) the transmitted intensity approaches unity (could be achieved only via non-lossy media) /vanishes in one /opposite propagation directions ($T_{co\text{-}polarized_forward}$ /$T_{co\text{-}polarized_backward}$ := maximal /minimal or minimal /maximal); as well as (v) the ellipticity of outgoing beams approaches zero. The applied multilayer is plasmonic, so an upper threshold exists for the achievable transmittance. In addition, ensuring completely linearly polarized beam propagation throughout such metamaterial of unique symmetry properties is challenging.

Therefore, the polarization ellipses were evaluated based on the Stokes parameters to determine the polarization rotation and ellipticity using the method described in our previous studies [61,62] (for more detail see Supporting Information).

To optimize and thereafter to evaluate the capabilities of the designed multilayers, the polarization-selective readout of the COMSOL Multiphysics RF module was used, and the co-polarized and cross-polarized transmitted signals were extracted, while the dissipative loss was determined based on resistive heating of the plasmonic components made of gold. The optimization was performed by GLOBAL algorithm implemented into COMSOL Multiphysics software [63]. In addition to the symmetrically varied distance between the composing layers, the azimuthal orientation was also varied during p-polarized light illumination in forward propagation direction ($\gamma_f$) and was inherited in backward propagation direction ($\gamma_b$) according to the polarization rotation occurring in forward direction.

Based on the literature such bianisotropic artificial metamaterials result in asymmetric optical responses, while for the achievement of nonreciprocity an important criterion is the asymmetry of the co-polarized transmission [9].

Accordingly, the co-polarized components of the transmission were determined and compared in the azimuthal orientations corresponding to $\gamma_f$ and $\gamma_b$, so as by defining the $AT_{co}^{classical}$ for specific azimuthal orientations in forward and backward propagation directions in the classical sense: $AT_{co}^{classical_\gamma_f} = T_{co}^{f_\gamma_f} - T_{co}^{b_\gamma_f}$ and $AT_{co}^{classical_\gamma_b} = T_{co}^{f_\gamma_b} - T_{co}^{b_\gamma_b}$, where indices "f" and "b" always refer to forward and backward propagation direction, respectively.

Assuming that the nonreciprocal rotation is accompanied by asymmetric transmission in both involved propagation directions, the cumulative impact of bianisotropy was considered by computing the balanced asymmetrical transmission in mixed bases: $AT_{co}^{balanced} = T_{co}^{f_\gamma_f} - T_{co}^{b_\gamma_b}$. To achieve GFR configurations, during the primary optimizations, the following complex objective function was examined, which includes terms related to the balanced asymmetrical co-polarized transmission, as well as rotation and ellipticity ($\tau$) at the transmission sides: *OPT-FOM*= $AT_{co}^{balanced}/(|90°-|(\beta_b-\alpha_b)+(\beta_f-\alpha_f)|| \times (|\tau_f|+|\tau_b|))$. Thereafter during comparison of the optical responses of distinctly different optimized multilayers (Pattern-I and Pattern-II), the *GFR-FOM*= $|AT_{co}^{balanced}| \cdot \frac{1}{|90°-\Sigma||+1}$ spectra were also determined, and the reminiscent ellipticity was evaluated separately, considering that ellipticity on the minimal transmission side does not have an impact on the GFR performance.

Though the *OPT-FOM* can yield multilayers exhibiting nonreciprocal rotation, the $AT_{co}^{balanced}$ quantity approaches zero, when the forward /backward propagation allows for small transmission and the backward /forward propagation predominantly results in a cross-polarized component. Therefore, the differential asymmetrical transmission ($AT_{co}^{differential}$) in forward bases was also monitored, which is the difference of the co-polarized component in forward propagation direction and the transmission in backward propagation direction projected onto the cross-polarized component of forward direction ($AT_{co}^{differential} = T_{co}^{f_\gamma_f} - T_{full\to cross}^{\gamma_b\to\gamma_f} \sim T_{full}^{f_\gamma_f} - T_{full\to cross}^{\gamma_b\to\gamma_f}$). The *GFI-FOM* constructed with this quantity was also evaluated (*GFI-FOM* $= \frac{1}{(T_{full}^{f_\gamma_f}+1)}\frac{1}{(1-T_{full\to cross}^{\gamma_b\to\gamma_f})+1} \cdot \frac{|T_{full}^{f_\gamma_f} - T_{full\to cross}^{\gamma_b\to\gamma_f}|}{|90°-\Sigma||+1} \cdot \frac{1}{|\tau_f|+|\tau_b|+1}$) for the optimized multilayers. About further details on corresponding cross-polarized spectra and *FOMs*, please see the Supporting Information. To catch the tiny asymmetric transmission, it is straightforward to consider the wavelength dependency of the forward outgoing and backward incoming azimuthal orientation, which is experimentally challenging, therefore the optical responses in $\gamma(\lambda)$ configuration are presented in the Supporting Information.

## 3. Results and discussion

### 3.1 Parameters of the optimized multilayers

The primary miniarray was constructed with a central nanoring of *R*=25 nm radius with embedded central nano-disk/hole of 5 nm radius in the concave/convex monolayer, which was surrounded by satellite nanocrescents of *r*=12.5 nm, aligned along a circular quadrumer of $\rho$=100 nm radius. These nano-objects have perimeters of (10; 25; 50; 200)×π.

The centered rectangular pattern with a unit cell of $p_{x/y}$=300/173 nm side-lengths could be fabricated using integrated colloid sphere and interference lithography, due to its large degrees of freedom [64-66]. However, in this case the quadrumer is slightly rotated ($\Omega$~96°) and the nanocrescents are misaligned compared to the horizontal direction ($\omega$=106°). These parameters were inherited during the optimization of the presented multilayers. Important inherited spectral and topological properties are that (i) the resonance wavelength of the nanoring and nanocrescent (nanocrescent and propagating plasmon polaritons) differs by ~10 nm in U/C and C/U resonance, developing if the ***E***-field is oriented along the axis and tip on the convex/concave layer (in case of coupled U resonance on the concave layer if the ***E***-field is oriented along the tip); (ii) there are two sub-lattices (A with and B without nanoring). The (i/ii) facilities symmetry breaking /period doubling type perturbations [20].

The optimization performed by using the *OPT-FOM* resulted in two very similar multilayers that meet the GFR criterion. Namely Pattern-I and Pattern-II, which are constructed with dielectric spacers of $h_{diel_I}$=23.92 nm and $h_{diel_II}$=23.95 nm that correspond to $h_{multilayer_I}$=164.84 nm and $h_{multilayer_II}$=164.9 nm complete sub-wavelength slab thicknesses. Although it is counterintuitive considering the minor difference between the multilayers' parameters, the optical and near-field responses differ characteristically (data are presented in Table S1 in Supporting Information). This is due to the different configurations involving the azimuthal orientation in both propagation directions, and indicates that the interplay between the (i) and (ii) type perturbations provides dual-solutions.

### 3.2 Optical responses of Pattern-I and Pattern-II under linearly polarized light illumination in backward propagation direction

In Pattern-I the $\sum$=-89.08° cumulative rotation is reached at $\lambda_{GFR}$=637.45 nm, when linearly polarized light illumination is realized both in $\gamma_{forward_GFR}$=95.39° and $\gamma_{backward_GFR}$=120.1° azimuthal orientations (Fig. 1a-c). This GFR configuration results in smaller $\beta_{forward}-\alpha_{forward}$=-24.67° rotation in forward propagation direction, while approximately two-times larger $\beta_{backward}-\alpha_{backward}$ =-64.41° rotation is achieved in backward propagation direction. In Pattern-II the cumulative rotation of $\sum$=93.66° slightly more pronouncedly mismatches the target polarization rotation value at $\lambda_{GFR}$=637.22 nm.

Counterintuitively, the slightly different spacer allows for meeting the GFR condition in very similar $\gamma_{forward_GFR}$=95.23°, but in fundamentally different $\gamma_{backward_GFR}$=65.62° azimuthal orientations in the forward and backward propagation direction (Fig. 2a-c). The same signs of the $\beta-\alpha$ orientation differences in opposite propagation directions prove that both GFR configurations result in nonreciprocal polarization rotation (Fig. 1a and Fig. 2a).

The co-polarized transmission is weakly dependent on the propagation direction in these azimuthal orientations, accordingly small $AT_{forward}^{classical_\gamma_f}$=9.5×10$^{-6}$ and $AT_{backward}^{classical_\gamma_b}$=1.8×10$^{-5}$ classical asymmetrical transmission values are reached in Pattern-I (Fig. 1b). Pattern-II shows more propagation direction dependent transmission and significantly and slightly larger classical asymmetrical transmission of $AT_{forward}^{classical_\gamma_f}$=5.5×10$^{-5}$ and $AT_{backward}^{classical_\gamma_f}$=2.5×10$^{-5}$ in forward and backward propagation direction (Fig. 2b). This is because the transmission is very low in both propagation directions revealing a non-radiative, i.e. BIC-related phenomenon [15-20].

Moreover, the classical $AT^{classicalc_\gamma_f/\gamma_b}_{forward/backward}$ exhibits a global /local minimum at the $\lambda_{GFR}$=637.45 nm in forward /backward propagation directions in Pattern-I and a local minimum at the $\lambda_{GFR}$=637.22 nm in both propagation directions in Pattern-II.

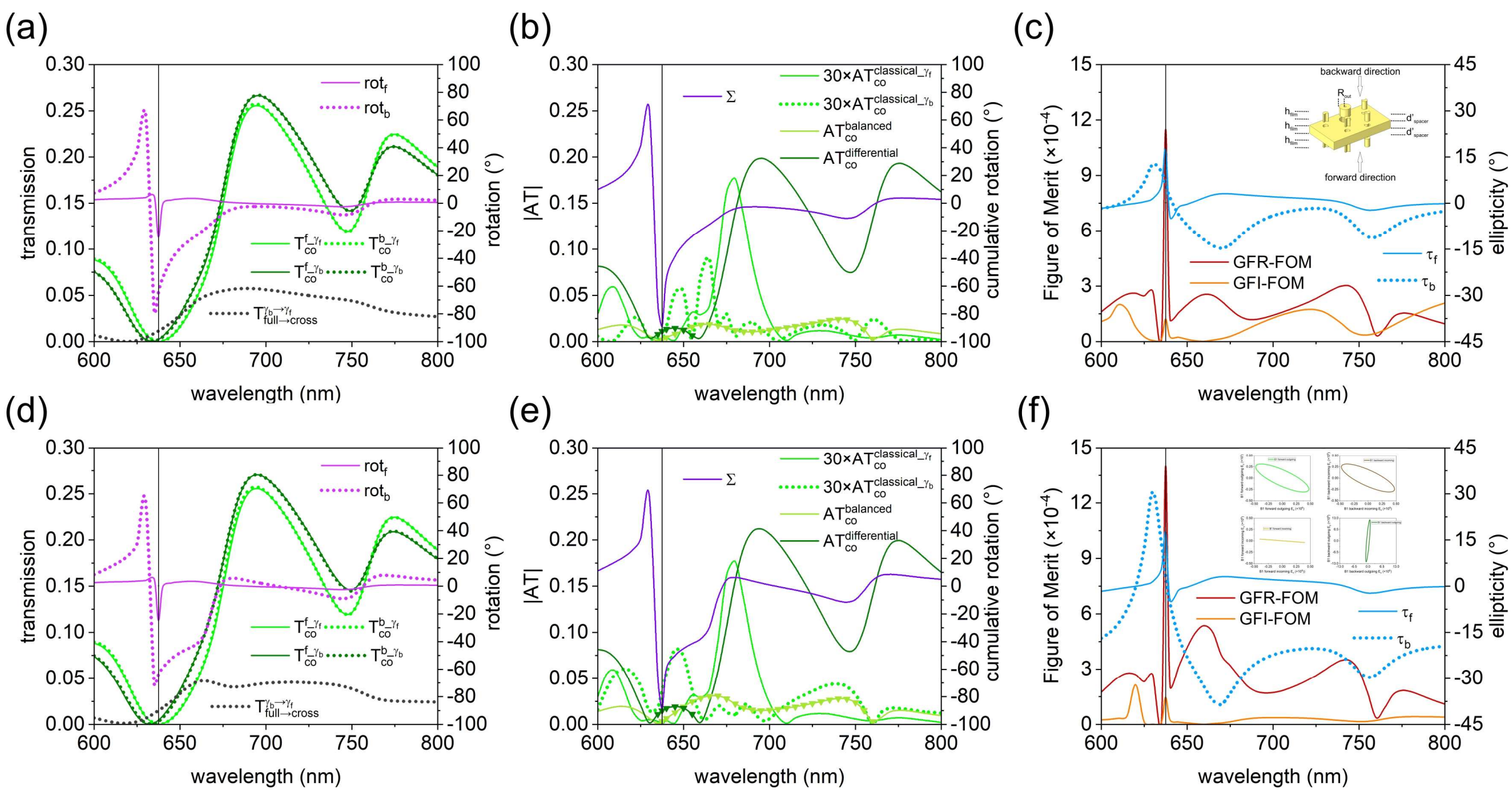

**Figure 1.** Optical responses of Pattern-I, transmission side. (a, d) Co-polarized transmission and polarization rotation in forward and backward propagation direction as well as the backward transmission projected onto the forward cross-polarized component, (b, e) cumulative rotation and the classical, balanced and differential co-polarized asymmetric transmission (triangular symbols indicate regions, where the backward transmission is larger), (c, f) ellipticity of outgoing beams, *GFR-FOM* and *GFI-FOM*. (a-c) Linearly, (d-f) elliptically polarized light for backward illumination. Insets: (c) the schematic of the studied multilayer indicating its geometrical parameters and the illumination directions, (f) the polarization ellipses corresponding to the forward incoming (bottom left) and outgoing (top left), backward incoming (top right) and outgoing (bottom right) plane waves.

The balanced asymmetrical transmission takes on more than two-orders of magnitude larger ($AT^{balanced}_{co}$=2.2×10$^{-3}$) value in the mixed bases in Pattern-I close to a local maximum, revealing capability for propagation direction dependent transmission, when the GFR condition is met (Fig. 1b). In Pattern-II a global maximum develops in the balanced asymmetrical transmission close to the GFR condition, which allows for slightly larger value of $AT^{balanced}_{co}$=2.4×10$^{-3}$, and thus predicts more well-defined nonreciprocal phenomena (Fig. 2b). Indeed, the differential transmission in the forward bases, indicating the difference between the co-polarized transmission in the forward direction and the backward propagating transmission projected onto the cross-polarized component in forward propagation direction, is one more order of magnitude larger ($AT^{differential}_{co}$= 1.14×10$^{-2}$) compared to the backward bases in Pattern-I close to a local maximum (Fig. 1b). In Pattern-II even larger ($AT^{differential}_{co}$= 1.28×10$^{-2}$) differential transmission is reached close to a larger local maximum (Fig. 2b). The outgoing beams show ellipticity of $\tau_f$=17.52° and $\tau_b$=5.92° in the forward and backward propagation direction in Pattern-I (Fig. 1c, 1f inset). In Pattern-II the ellipticity is slightly more pronounced ($\tau_f$=20.09°) in forward direction, while approximates the linearly polarized case in backward propagation direction ($\tau_b$=-1.48°, Fig. 2c, 2f inset). In both cases, the coincidence of GFR with narrow and broadband modulation in forward and backward ellipticity is observable.

### 3.3 Optical responses considering the forward outgoing ellipticity in backward propagation direction through optimized layers Pattern-I and Pattern-II

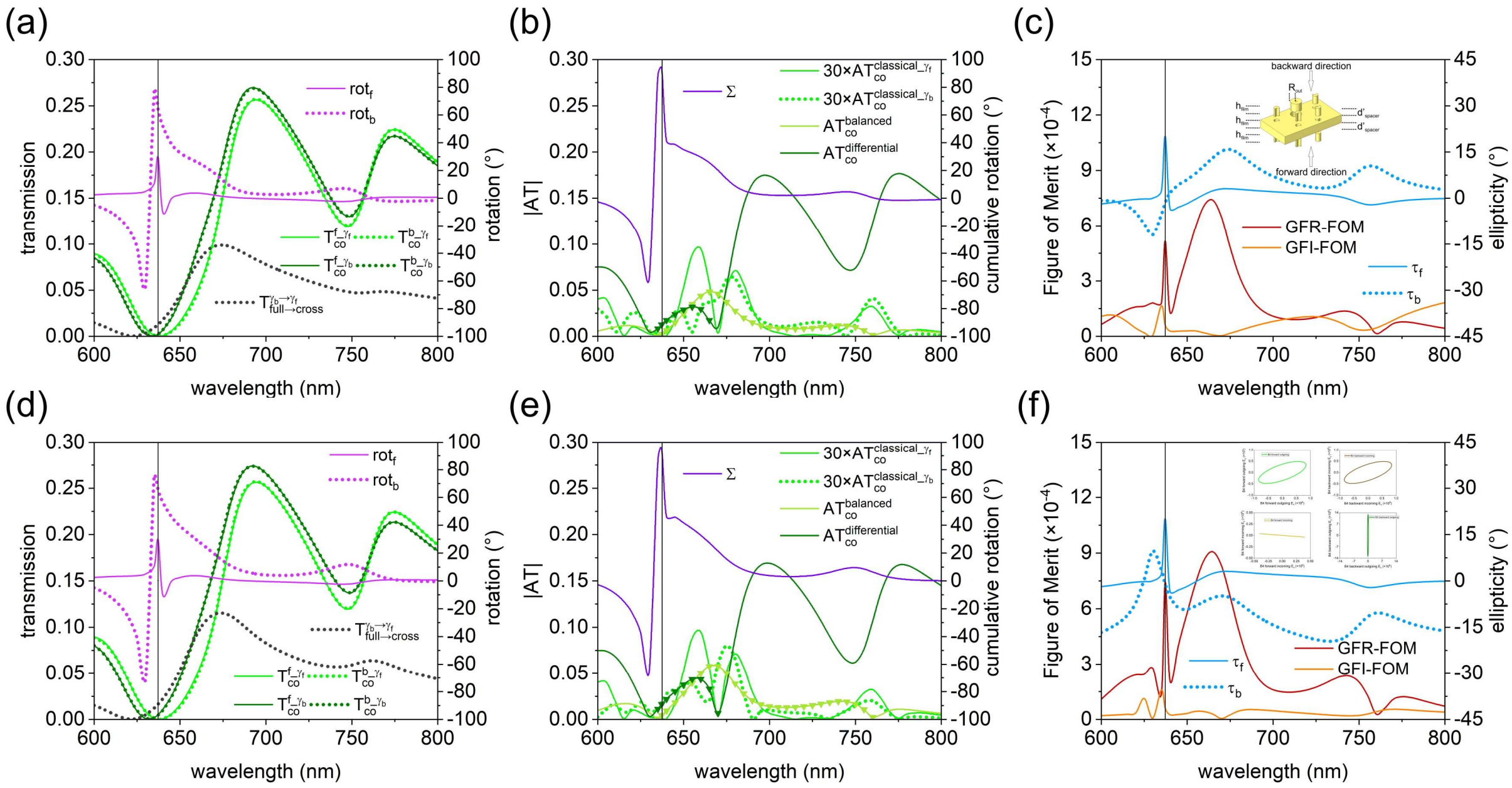

**Figure 2.** Optical responses of Pattern-II, transmission side. (a, d) Co-polarized transmission and polarization rotation in forward and backward propagation direction as well as the backward transmission projected onto the forward cross-polarized component, (b, e) cumulative rotation and the classical, balanced and differential co-polarized asymmetric transmission (triangular symbols indicate regions, where the backward transmission is larger), (c, f) ellipticity of outgoing beams, *GFR-FOM* and *GFI-FOM*. (a-c) Linearly, (d-f) elliptically polarized light for backward illumination. Insets: (c) the schematic of the studied multilayer indicating its geometrical parameters and the illumination directions, (f) the polarization ellipses corresponding to the forward incoming (bottom left) and outgoing (top left), backward incoming (top right) and outgoing (bottom right) plane waves.

When the ellipticity of the outgoing beam in forward propagation direction is considered for, namely the beam illuminating the Pattern-I multilayer in backward propagation direction is imparted by elliptical polarization of $\tau_f$=17.52°, the outgoing beam shows slightly smaller $\beta_{backward}-\alpha_{backward}$=-63.95° rotation, accordingly the cumulative rotation is also slightly smaller (∑=-88.63°), however the resulted $\tau_b$=2.99° ellipticity becomes advantageously smaller (Fig. 1d-f). Pattern-II also meets the GFR criterion less perfectly, when the illumination in backward propagation direction is realized with a beam of elliptical polarization ($\tau_f$=20.09°). This is because the outgoing beam shows slightly larger $\beta_{backward}-\alpha_{backward}$ =65.43° rotation, accordingly the cumulative rotation is also slightly larger (∑=95.04°), though the ellipticity is further reduced to the value of $\tau_b$=-1.07° (Fig. 2d-f).

Both the balanced ($AT_{co}^{balanced}$=3.32×10$^{-3}$) and differential ($AT_{co}^{differential}$= 1.49×10$^{-2}$) asymmetric transmission values become larger in Pattern-I, due to that the local maximum is transformed into a global maximum and the local maximum is enhanced (Fig. 1e). Further enhanced balanced global ($AT_{co}^{balanced}$=4.5×10$^{-3}$) and differential local ($AT_{co}^{differential}$= 1.64×10$^{-2}$) maximal asymmetric transmission values are reached in Pattern-II, when elliptically polarized light is used in backward propagation direction (Fig 2e). The counterpart *AT* values are larger in Pattern-II. Moreover, both *FOMs* qualifying the GFR and GFI capability become larger at the global (*GFR-FOM*=1.15×10$^{-3}$ < 1.40×10$^{-3}$) and local (*GFI-FOM*= 1.22×10$^{-4}$ < 1.47×10$^{-4}$) maximum in Pattern-I (Fig. 1c-to-f). Similarly, enhanced *FOMs* are achieved in Pattern-II, the only difference is that in *GFR-FOM* a smaller local maximum is enhanced (*GFR-FOM*=5.16×10$^{-4}$ < 7.44×10$^{-4}$), while a nearby smaller local maximum is transformed into a global maximum in *GFI-FOM* (*GFI-FOM*= 6.09×10$^{-5}$ < 6.16×10$^{-5}$) (Fig. 2c-to-f).

All merits are better for Pattern-II, except the meeting of the ∑=90° GFR criterion, consequently the *GFR-FOM* and *GFI-FOM* are also smaller, in addition the ellipticity reached in forward propagation direction remains larger (Fig. 1-to-2).

The optical responses determined in case of linear and elliptical illumination for the cross-polarized components both in wavelength independent (Fig. S1a-f and Fig. S2a-f) and in $\gamma(\lambda)$ configuration (Fig. S1m-r and Fig. S2m-r) as well as for the co-polarized component in $\gamma(\lambda)$ configuration (Fig. S1g-l and Fig. S2g-l) are presented in the Supporting Information (Table S1).

### 3.4 Dispersion characteristic of Pattern-I and Pattern-II

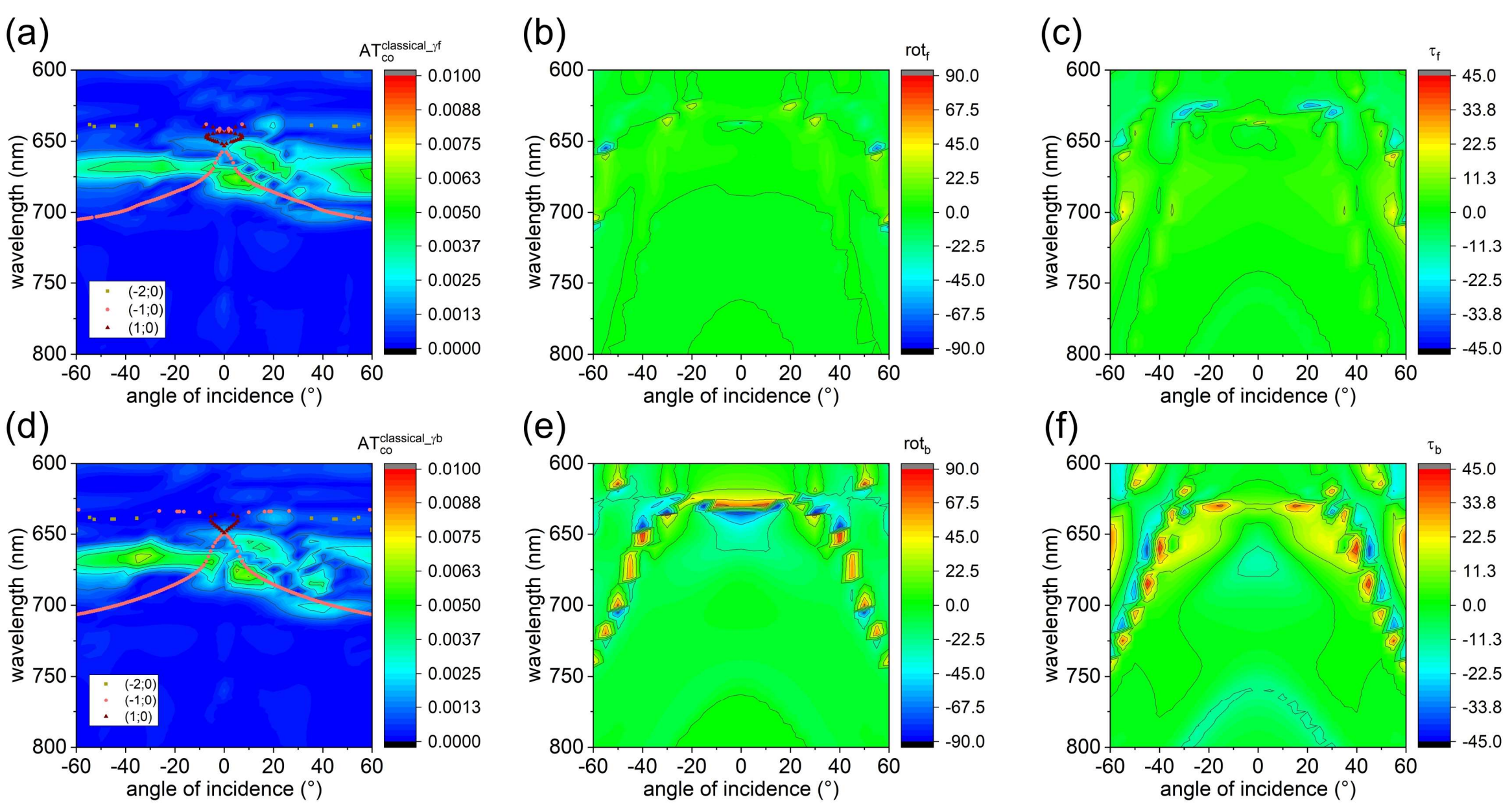

**Figure 3.** Dispersion characteristic of classical co-polarized asymmetric transmission and non-reciprocal rotations of Pattern-I. (a, d) Classical co-polarized asymmetric transmission with fitted plasmonic branches (indicated with dots), (b, e) polarization rotation (c, f) ellipticity; in (a-c) forward and (d-f) backward propagation direction.

The dispersion characteristics of the convex and concave monolayers were already presented in our previous papers [61,62]. The dispersion maps of the concave monolayer taken in γ=90° azimuthal orientation (Γ-Y direction of the centred rectangular unitcell) indicate complementary localized modes as well as scattered photonic modes (Surface Lattice Resonance) and SPPs coupled in (-1,0) order that branch extends towards the wavelength region of λ<600 nm at the Γ point. The strong-coupling with spectrally overlapping localized modes resonant on the nanoring and nanocrescents form Fano lines at perpendicular incidence. The intuitive expectation is that the impact of these phenomena is inherited by the multilayers as well, by contributing to the symmetry breaking perturbations [20].

For Pattern-I and Pattern-II, the co-polarized and cross-polarized transmission components were mapped separately (for the direct co-polarized and cross-polarized components see Supporting Information Fig. S3a,b,d,e; Fig. S4a,b,d,e). On the dispersion maps of the co-polarized and cross-polarized transmittance in the azimuthal orientations of $\gamma_{forward_GFR_I/II}$ (95.39° /95.23°) and $\gamma_{backward_GFR_I/II}$ (120.1° /65.62°), the GFR phenomenon appears in the region where predominantly global transmission minima appear inside a gap above inverted bands.

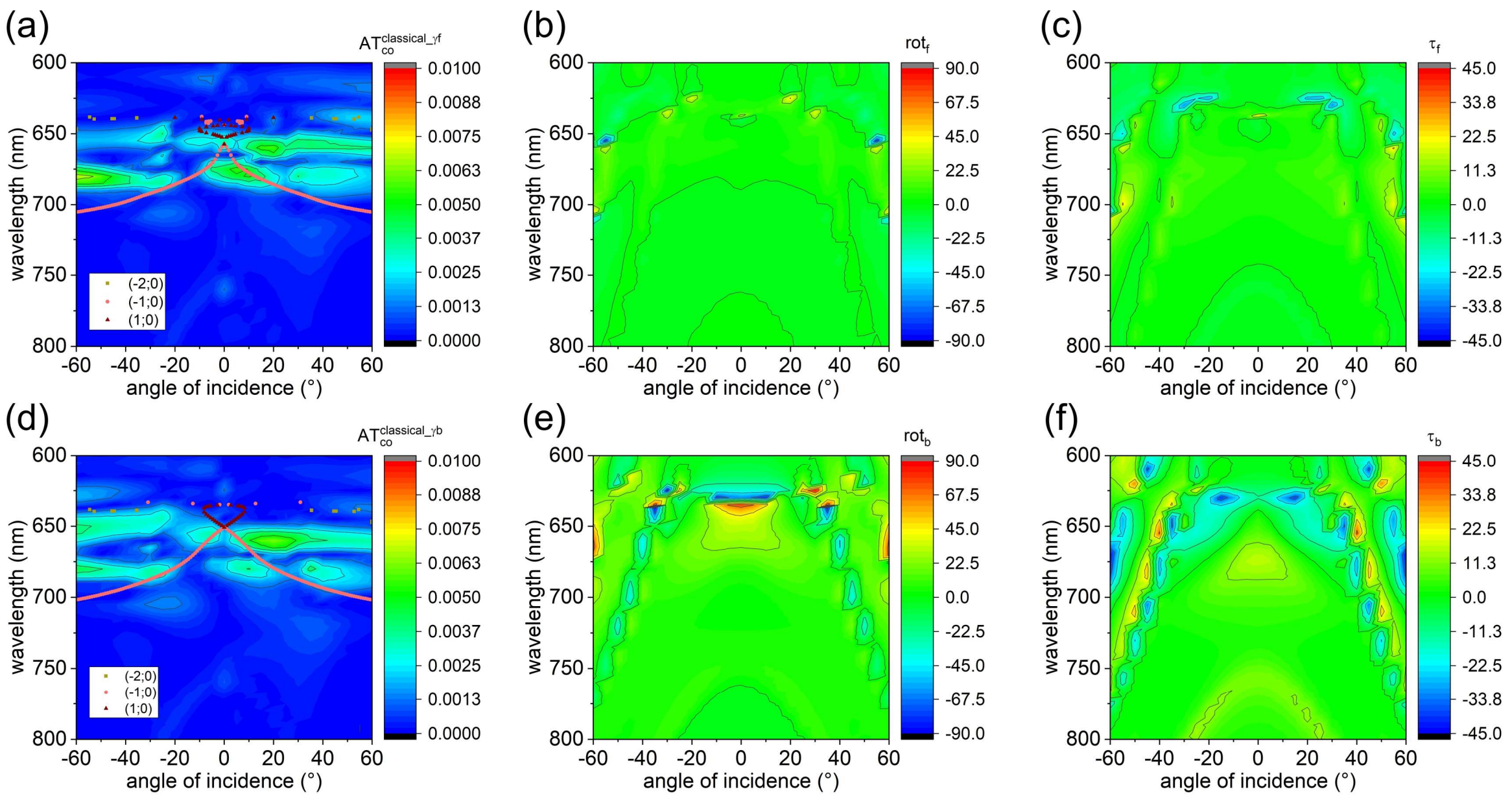

**Figure 4.** Dispersion characteristic of classical co-polarized asymmetric transmission and non-reciprocal rotations of Pattern-II. (a, d) Classical co-polarized asymmetric transmission with fitted plasmonic branches (indicated with dots), (b, e) polarization rotation (c, f) ellipticity; in (a-c) forward and (d-f) backward propagation direction.

Exceptional is the cross-polarized transmittance in the backward propagation configuration, which shows the GFR at the boundary of a pass-band in both multilayers of Pattern-I and Pattern-II (see Supporting Information Fig. S3e and Fig. S4e).

The classical asymmetric co-polarized and cross-polarized transmission show maxima on two continuous and four less well-defined broken quantized flat bands in both propagation directions, which show left-right asymmetry around perpendicular incidence, namely more pronounced intersecting bands appear with left /right-tilting along the $k$-axis in Pattern-I/II (Fig. 3a,d; Fig. 4a,d and Fig. S3c,f; Fig. S4c, f). These ladder-like quantized flat bands reveal a co-existent space-time-modulations on the hierarchical topology [31,37,38,47]. The regions where global maxima of significantly larger $AT_{co}^{classical_\gamma_f/\gamma_b}$ values are taken, appear in considerably red-shifted interval (~680 nm /650 nm for Pattern-I/II) close to perpendicular incidence /at small tilting in both propagation directions (Fig. 3a,d and Fig. 4a,d).

The band structure is governed by Brillouin zone-folding and band inversion due to the period doubling perturbation stem from co-existent A and B sub-lattices (Supporting Information Fig. S3a,b,d,e and Fig. S4a,b,d,e) [20,27,60]. The nonreciprocal phenomenon related space-time modulation justifies the use of specific retrieval methods, however these are not accurate exactly at the spectral location of the modulation [49,50,67]. Therefore, the branches determined by considering the dispersion (wavelength dependent optical properties) of the effective medium as well as the azimuthal orientations corresponding to forward and backward propagation directions of the GFR configurations, were fitted only to the $AT_{co}^{classical}$ maps (Fig. 3a,d and Fig. 4a,d, for retrieved optical parameters please see Supporting Information Fig. S9a,b and Fig. S10a,b) [66]. On the $AT_{co}^{classical}$ dispersion maps modulations appear in the region of (+/-1,0) grating coupling of plasmonic modes, on both patterns in both propagation directions (Fig. 3a,d and Fig. 4a,d). The impact of (-2, 0) SPP band is more noticeable in both propagation directions for Pattern-I (Fig. 3a). In addition, steeper branches intersect the quantized flat bands, which have pronounced impact on polarization rotation.

The GFR corresponds to local maxima in the $AT_{cross}^{classical_\gamma_f/\gamma_b}$ at perpendicular incidence in both azimuthal orientations of Pattern-I, and the regions, where significantly larger $AT_{cross}^{classical_\gamma_f/\gamma_b}$ values are taken, appear at large right /left tilting and small left /right tilting in forward and backward propagation direction in Pattern-I/II (Fig. S1b and Fig. S3c,f; Fig. S2b and Fig. S4c,f).

The difference in Pattern-II compared to Pattern-I is that the $AT_{cross}^{classical}$ value is slightly smaller /larger in forward /backward propagation direction at perpendicular incidence and the intermediately red-shifted global maximum appearing (slightly above 650 nm at small tilting) in backward propagation direction is broader in wavelength but narrower in tilting. The wavelength and tilting parameter interval, where large values are taken, is complementary in the sense that indicates reversed side-preference in Pattern-I and Pattern-II, in both propagation directions (Fig. S3c,f to Fig. S4c,f).

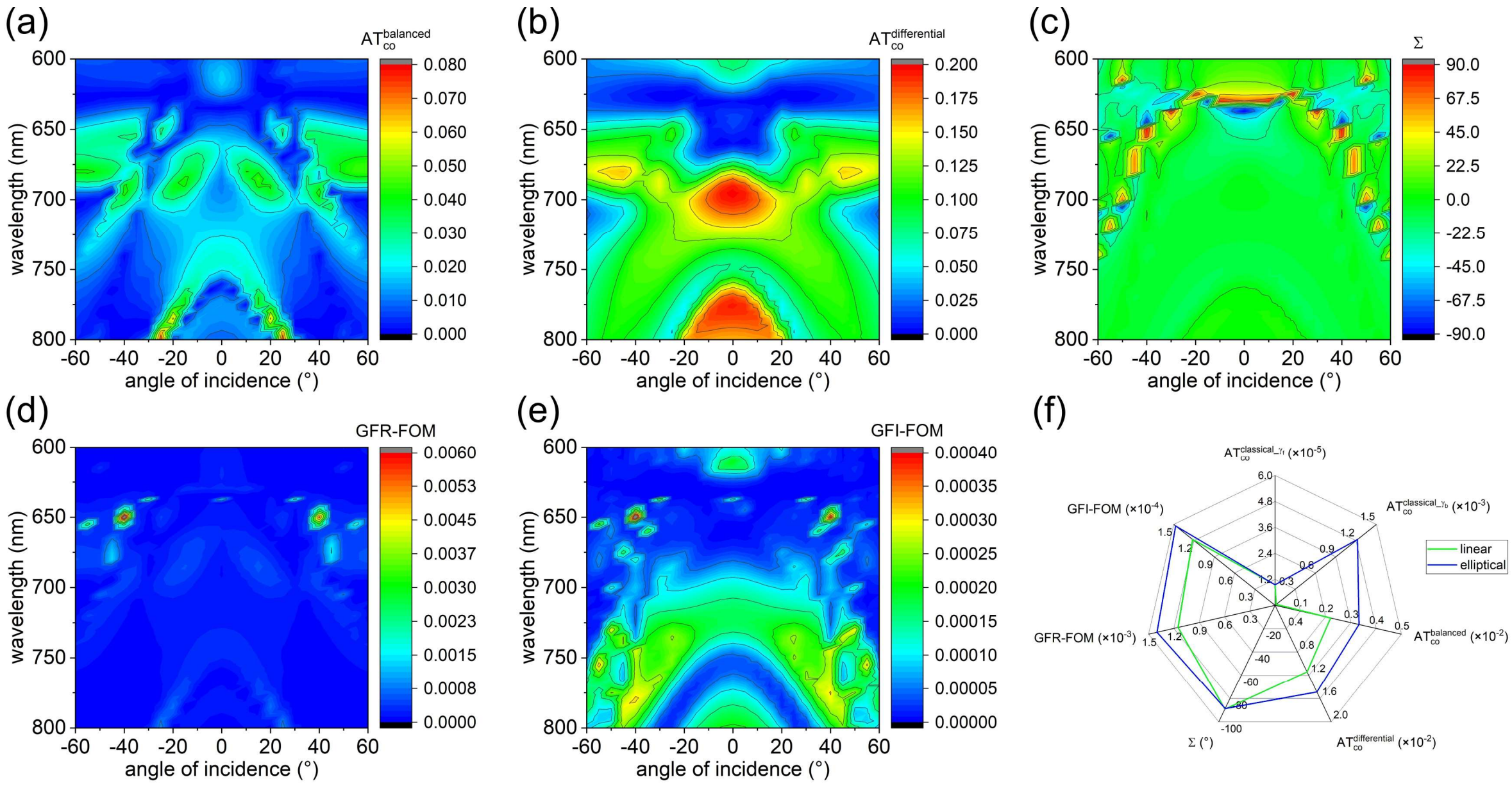

**Figure 5.** Dispersion characteristic of co-polarized asymmetric transmissions and cumulative rotation and FOMs qualifying both propagation directions in GFR configuration and proving non-reciprocity of Pattern-I. (a) Balanced and (b) differential co-polarized *AT*, (c) cumulative rotation, (d) *GFR-FOM*, (e) *GFI-FOM*, (f) evaluation-web.

The modulations in rotation and ellipticity in forward propagation direction are less well defined and appear along a flat band bounded by the branches related to the (-1,0) SLR and SPP band well-known from the studies of monolayers in Γ-Y direction, in accordance with that the azimuthal orientation of $\gamma_{forward_GFR_I/II}$ (95.39°/95.23°) is close to 90° (Fig. 1a,c and 3b,c; Fig. 2a,c and 4b,c) [61,62]. In backward propagation direction the plane of incidence along $\gamma_{backward_GFR_I/II}$=120.1°/65.62° is parallel to the downward /upward diagonal of the unit cells in real space, which allows for perturbation due to the folded bands in the rotation and ellipticity of both patterns (Fig. 3e,f and Fig. 4e,f). The branches, which originate from Brillouin zone-folding along the diagonals, intersect a flat band extending up to perpendicular incidence at the GFR wavelength. These Brillouin zone-folding related perturbations play unambiguously important role in the polarization modulation (Fig. 3e,f and Fig. 4e,f). The polarization characteristics also exhibit six-fold ladder revealing a time-periodic modulation, which is spectrally slightly shifted to those on the classical asymmetric transmission dispersion maps.

However, the GFR related miniband is coincident almost with the minimum at the boundary of a differential pass-band of $AT_{co}^{balanced}$ taken in the mixed bases, which is weaker in Pattern-I and stronger in Pattern-II (Fig. 1b and Fig. 2b, Fig. 5a, Fig. 6a). These balanced asymmetrical transmission pass-bands are inherently perturbed by strong tilted branches appearing in the mixed bases, which are also related to the Brillouin zone-folding due to the co-existent two sub-lattices [20,27,60]. The GFR appears close to the spectral region, where the $AT_{cross}^{balanced}$ shows a global maximum, at the upper boundary of a strong differential pass-band (Fig. S5a, Fig. S6a). This is evidently due to the pass-band appearing in the cross-polarized transmittance in backward propagation direction (Fig. S3e, Fig. S4e). In accordance with the larger red-shifted global maximum (appearing above 650 nm), the value taken at the GFR is also larger in Pattern-II.

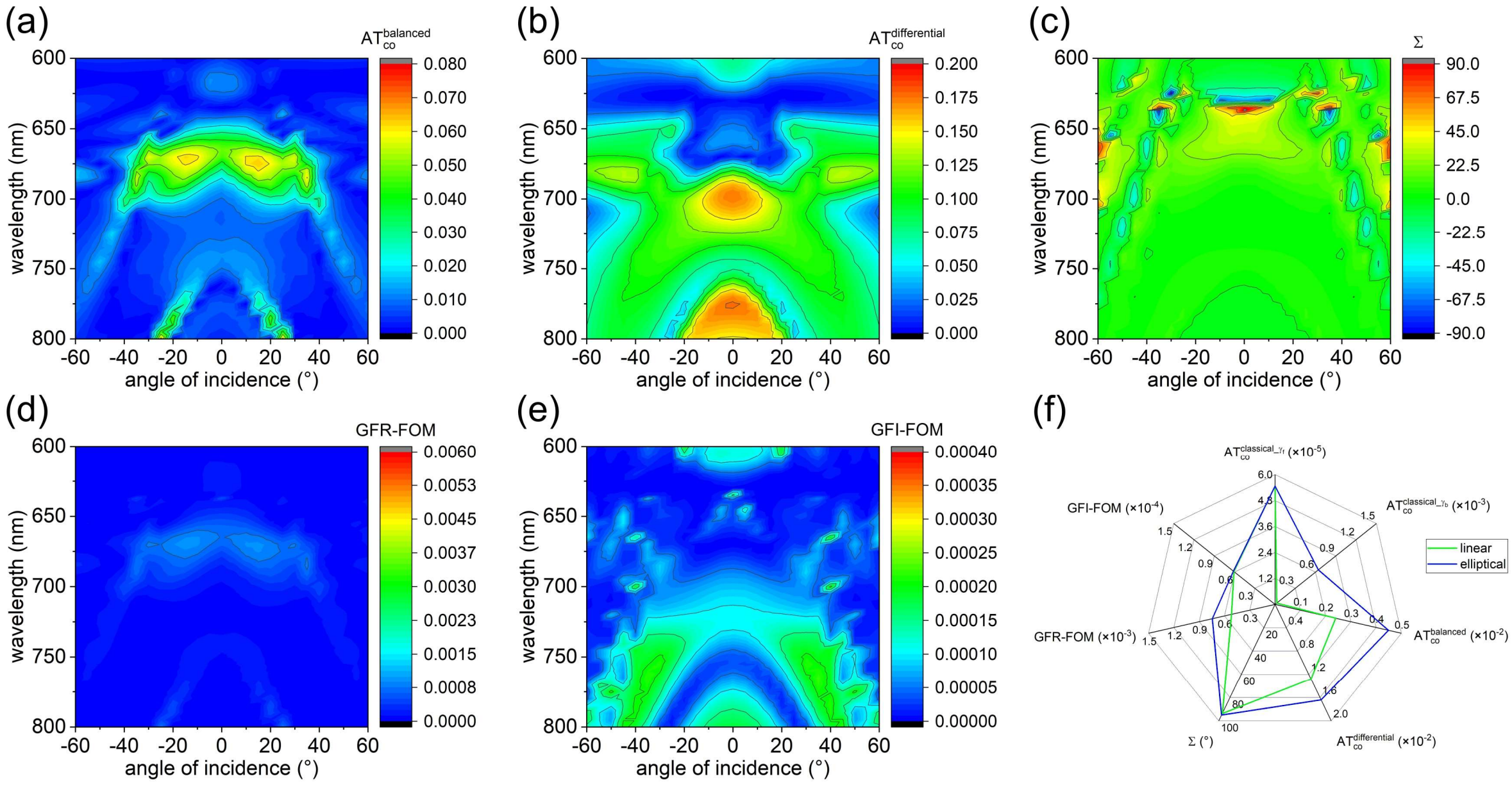

**Figure 6.** Dispersion characteristic of co-polarized asymmetric transmissions and cumulative rotation and FOMs qualifying both propagation directions in GFR configuration and proving non-reciprocity of Pattern-II. (a) Balanced and (b) differential co-polarized AT, (c) cumulative rotation, (d) *GFR-FOM*, (e) *GFI-FOM*, (f) evaluation-web.

The more pronounced difference between the dispersion maps of the co-polarized transmission in the forward direction and the backward propagating transmission projected onto the cross-polarized component in forward propagation direction results in a narrow miniband of local maxima in the differential asymmetrical transmission ($AT_{co}^{differential}$) at the GFR in both Patterns (Fig. 1b and Fig. 2b; Fig. 5b and Fig. 6b). This is limited by the extension in tilting and wavelength of the backward transmission band projected onto the forward cross-polarized component in Pattern-I and Pattern-II (Fig. 5b, Fig. S3a-to-5c; Fig. 6b, Fig. S4a-to-6c). While in Pattern-I this local $AT_{co}^{differential}$ maximum is centred nearby the GFR wavelength, in Pattern-II just the side band of a larger but pronouncedly red-shifted local maximum overlaps with the GFR condition at perpendicular incidence (Fig. 1b and Fig. 2b; Fig. 5b and Fig. 6b).

The $AT_{cross}^{differential}$ is predominantly determined by the backward transmission band projected onto the co-polarized component in Pattern-I and Pattern-II, rather than limited by the extension of cross-polarized transmittance in tilting and wavelength (Fig. S5b, Fig. S3b-to-5f; Fig. S6b, Fig. S4b-to-6f).

The GFR is coincident with the centre of the global minimum of the transmission projected onto the co-polarized component in both Patterns (Fig. S1b and Fig. S2b; Fig. S5b and Fig. S6b).

In accordance with the well-defined global maximum-minimum and minimum-maximum pairs appearing as a function of the wavelength in backward rotation, counterpart extrema develop in the ∑ of Pattern-I and Pattern-II, throughout finite, but almost tilting independent bands close to perpendicular incidence (Fig. 1b, 3e, 5c and Fig. 2b, 4e, 6c). Remarkable difference is that the order of extrema is different in the two optimized patterns and is in accordance with the usual resonance phenomena related phase shifts in Pattern-II. The results of extended parametric studies indicate that the nonreciprocal phenomena might result in 90° cumulative polarization rotation in different spectral intervals, which overlap with regions, where both the balanced and the differential asymmetric co-polarized transmission are significant.

Therefore, good differential co-polarrized *AT* is achievable, even though the azimuthal orientations corresponding to the configurations meeting the GFR criterion are not accompanied by large classical asymmetric transmission (Fig. 5a-c and Fig. 6a-c). The achieved GFR performance is governed by the degree of overlap between the *AT* bands and polarization rotation. As a result, narrow band of local maxima appears on all monitored *FOMs* close to the GFR in both Patterns (Fig. 1c,f and Fig. 5d,e; Fig. 2c,f and Fig. 6d,e, see Supporting Information Fig. S1c,f and Fig. S5d,e; Fig. S2c,f and Fig. S6d,e). Although, there are regions, where the *FOMs* take on larger values, these are not coincident with locations, where simultaneously larger AT and better meeting of the GFR condition occurs, as compared to perpendicular incidence.

In Pattern-I the GFR configuration corresponds unambiguously to the global optimum in *GFR-FOM* at perpendicular incidence, moreover it corresponds to the global maximum through wide parameter regions, larger values are taken only at large tilting (Fig. 1c,f and Fig. 5d). In comparison, in Pattern-II the appearance of a red-shifted global maximum predicts the existence of a configuration operable in different wavelength interval at intermediate tilting (Fig. 2c,f and Fig. 6d), moreover, a wide band of maxima appears through intermediate tilting at slightly larger wavelength. The *GFI-FOM* maps show local maxima at the GFR along similar discontinuous bands (Fig. 1c,f, 5e and Fig. 2c,f, 6e). The overall performance is very similar in Pattern-I and Pattern-II (Fig. 5f and Fig. 6f; Fig. S5f and Fig. S6f).

### 3.5 Impact of tilting

Considering that the asymmetric transmission is achieved in well-defined flat bands through intermediate tilting, the sensitivity of the GFR configuration to the angle of incidence was inspected. Small tilting was found to be advantageous in both patterns. Moreover, better meeting of the GFR condition (∑=89.92°) was possible at 2.5° in Pattern-I, where the asymmetric transmission both in the mixed and forward bases ($AT_{co}^{balanced}$ and $AT_{co}^{differential}$) was larger, the ellipticity was smaller in both propagation directions, which allowed for reaching larger *FOMs* (Fig. S7a-c).

In contrast, in the case of Pattern-II tilting caused weaker matching of the GFR condition, only the balanced and differential *AT* and the ellipticity in forward direction could be improved, proving the ultimate advantage of perpendicular incidence for either tilting sides (Fig. S8a-c). All these features reveal that quasi-BICs govern the achievable *AT* in both Patterns.

The side-dependent impact of tilting was also inspected, and well defined difference was found in case of Pattern-I, which exhibits different enhanced *FOM* maxima at similar tilting, while similar sole global maximum appears at perpendicular incidence in Pattern-II independent of tilting-side (see Supporting Information Fig. S7 and Fig. S8 for more detail).

### 3.6 Propagation direction dependent effective optical properties of the multilayer

Primarily a standard metamaterial retrieval method was used to determine the effective optical properties of the complete multilayer in case of linearly polarized illumination [68]. Considering that there are discontinuities in the effective parameters of time-modulated and equivalent bianisotropic materials, we used these retrieval results to demonstrate strong dispersion in bands surrounding the GFR spectral region [67].

The effective parameters of Pattern-I and Pattern-II show overlapping electric and magnetic resonances in the vicinity of GFR phenomenon. The narrow bandwidth of the imaginary part reveals a BIC-related resonance, which is accompanied by a steep modification of the real part in the permittivity and permeability, as well as in the index of refraction (Fig. S9a,b; Fig. S10a, b). The modulation is slightly stronger in the forward, compared to the backward propagation direction; and in case of Pattern-I compared to Pattern-II. The index of refraction is smaller /larger than unity and the imaginary part indicates similarly strong absorption in the forward propagation direction, while the index of refraction is intermediate in backward propagation direction and the imaginary part proves moderate loss in Pattern-I/II. The imaginary parts correlate with the smaller absorption, which allows for higher transmittance in backward direction.

Both the Tellegen and chirality parameters exhibit well-defined extrema at the spectral location of polarization rotation and the absolute values of both are larger in Pattern-I, except the imaginary part of the Tellegen [69]. Namely, the absolute values of Tellegen and chirality parameters are slightly different, except the real part of chirality, which is an order of magnitude larger in Pattern-I both in forward and backward propagation directions (Fig. S9c, Fig. S10c). Considering that for linearly polarized illumination all asymmetrical transmissions as well as the rotation (except for linearly polarized illumination in backward direction) are larger in Pattern-II, correlation holds only for the rotation and the imaginary part of the Tellegen factor in forward direction, when Pattern-I and Pattern-II is compared. This indicates that the multilayers are endowed with all possible characteristic of bianisotropic materials (Fig. S9c; Fig. S10c).

Finally, the refractive index matrix method was used to qualify the effective properties, which proved that all of the Tellegen, artificial moving, chirality and omega parameters are uniquely modulated at the GFR, based on these the multilayer is an unconventional bianisotropic material (Fig. S9f, Fig. S10f) [68,69]. The nonreciprocal responses are governed by extended topological phenomena rather than simply by metamolecule related bianisotropic composition. In addition, the bounding convex layers are impacted by the bulk dielectric environment, which justifies consideration of SLR phenomena as well.

### 3.7 Near-field phenomena

*Charge distribution:* When the multilayer is illuminated in $\gamma$=0° /90° azimuthal orientation the cv (cx) /cx (cv) composing monolayers resonate in C (U) type modes.

These principal orientations were inspected in more detail in our previous studies [61,62,64-66], for periodic patterns constructed with individual objects in monolayers. The counterintuitive near-field phenomena, especially the reversal rotation on the nano-objects composing the miniarray, is observable already on the monolayers weakly (pronouncedly) in $\gamma_{forward\ (backward)}$ azimuthal orientation of GFR configuration.

Namely, in forward configuration the rotation is counter-clockwise on all constituents, except on the lower (#3&4, see insets in Fig. 7m and Fig. 8m) nanocrescents in the miniarray. In backward configuration, the rotation remains counter-clockwise /clockwise and clockwise /counter-clockwise rotation on the nanoring and nanocrescent in Pattern-I/II, when the miniarray is constructed (please see Fig. S11a-f and Video S1 and Video S2; Fig. S12a-f, and Video S7 and Video S8 in the Supporting Information).

The near-field studies revealed that different modes play an important role in the GFR phenomenon at the spectrally overlapping nonreciprocal rotation and asymmetrical transmission on the dispersion characteristics (Fig. 7a,b,d,e & Video 1-4, and Fig. 8a,b,d,e & Video 5-8). In GFR configurations the reversal charge rotation on the nanoring and nanocrescents is preserved, however, the dominance of this depends on the propagation direction and on the optimized multilayer composition. The characteristic in forward propagation direction is very similar for the two patterns, while in backward propagation direction the rotation direction on the nanorings is opposite. In forward propagation direction on the Pattern-I (Pattern-II) the charge rotation on the nanorings is counter-clockwise, except the incoming convex layer, where flipping occurs; while on the nanocrescents the rotation is synchronously counter-clockwise, except the entering concave surfaces, where clockwise charge rotation is noticeable.

In backward propagation direction on the Pattern-I (Pattern-II) the charge rotation on the nanorings is counter-clockwise (clockwise), except the incoming convex layer, where clockwise (counter-clockwise) rotation occurs; while on the nanocrescents the rotation is synchronous, except the #1&2 and #4 (#2 and #3&4) crescents on the entering and leaving concave surfaces (Fig. 7a,d & Video 1, 2 and Fig. 8a,d & Video 5, 6).

Both patterns exhibit more pronounced reversal rotation on the nanoring and nanocrescents in backward propagation direction, where larger polarization rotation occurs and larger classical asymmetric transmission is achieved (except the $AT_{co}^{classical}$ in backward propagation direction in case of linearly polarized illumination of Pattern-II). The reversal rotation on half of the nanocrescents on the entering concave surface correlates with moderate nonreciprocity in Pattern-I, while the appearance of reversal rotation on half of the nanocrescents on the leaving concave surface indicates slightly stronger capability for nonreciprocal phenomena in Pattern-II. This is due to the better phase coverage of reversal rotations, considering the continuous charge rotation on the nanoring on the leaving surface.

In addition, reversal rotation is observable on the concave layer between the A and B sub-lattices, except the entering surfaces in backward propagation direction in Pattern-I/-II, where a well-defined phase-delay develops between counter-clockwise / clockwise rotations, and on the leaving surface in Pattern-II, where the clockwise rotations are almost synchronous in the sub-lattices (Fig. 7a,d & Video 1, 2 and Fig. 8a,d & Video 5, 6; see Supporting Information Fig. S11 g-j & Video S3-S6 and Fig. S12 & Video S9-S12).

Remarkable difference is that Pattern-I/II operates with reversal /phase-shifted rotations in the A and B-sub-lattices acting as coupled loops.

The rotation on distinctly different A and B sub-lattices makes these structures suitable to achieve time-periodic Floquet modulation that serves as a synthetic dimension along the frequency axis, which is the precondition of to generate quantized photonic Landau levels and to realize photonic Landau-Zener tunnelling, as well as to achieve band braiding [33,38,51,52].

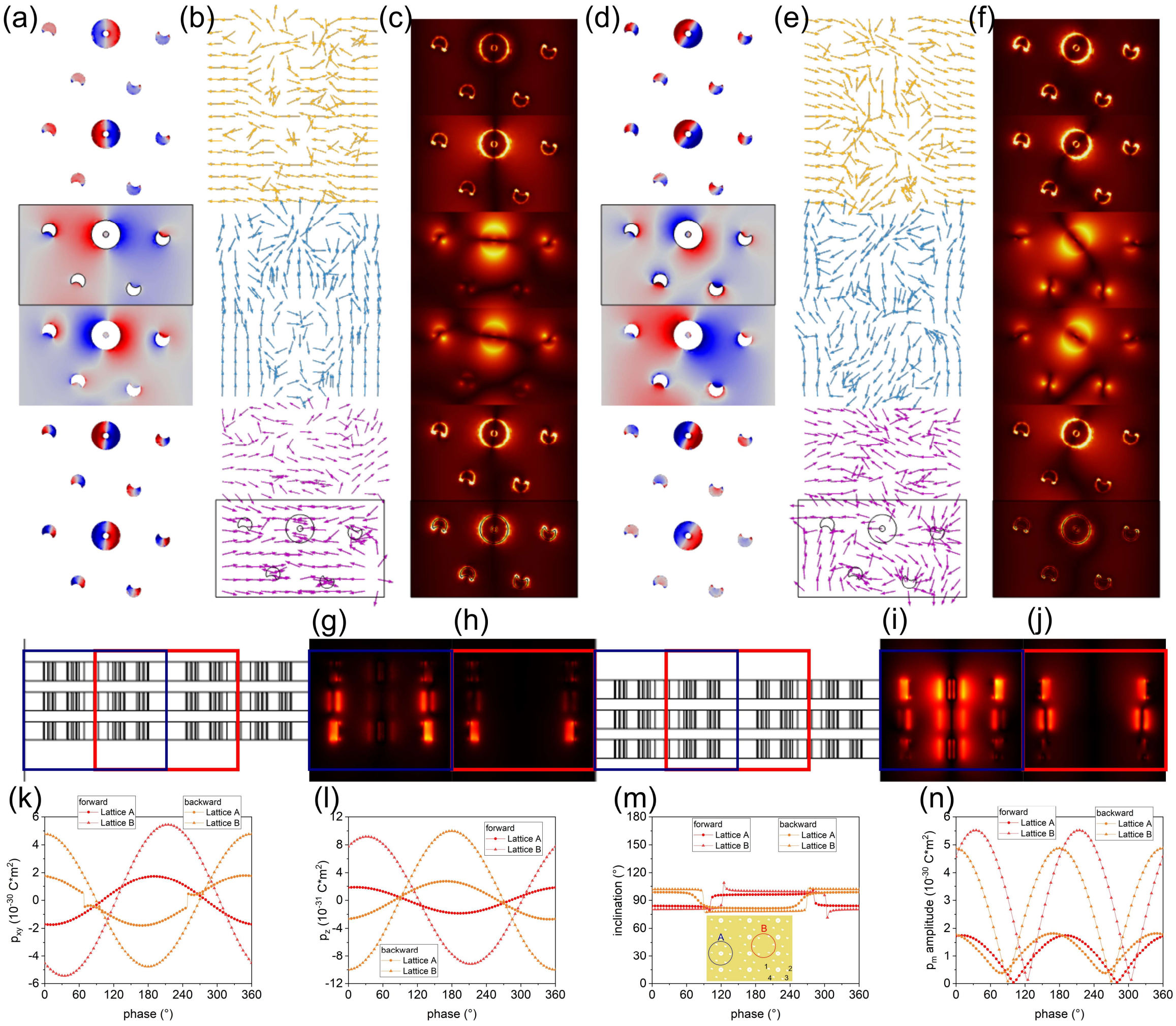

**Figure 7**. Near-field phenomena of Pattern-I. (a, d) Charge distribution on consecutive layer surfaces (Video 1, 2), (b, e) magnetic field (yellow), displacement current (blue) and vorticity distribution on entering and leaving concave surfaces (Video 3, 4), (c, f) $|E_z|$ on convex and $|B_z|$ on concave entering and leaving surfaces, (g, h, i, j) $|B_z|$ on vertical cross-sections through A-B sub-lattices, with schematics on the left side; a, b, c, g, h/d, e, f, i, j) forward /backward propagation direction. Emulated magnetic field related $\boldsymbol{p}_m$ dynamics on Pattern-I. (k) $p_{xy}$, (l) $p_z$, (m) inclination, (n) amplitude (please note the existence of offset). Inset on (m) segment: multi-unitcell schematic figure in the xy-plane, indicating the A and B sub-lattices.

The orientation of the charge distribution is predominantly along $\gamma_{backward_I}$ =120.1° ($\gamma_{backward_II}$ 65.62°) on the entering concave surface of the Pattern-I (Pattern-II), when they are illuminated in backward direction by linearly /elliptically polarized beams with $\boldsymbol{E}$-field oscillation /main axis corresponding to the outgoing beams in forward propagation direction ($\alpha_{backward_I}$ := $\beta_{forward_I}$=-30.06° and $\alpha_{backward_II}$ := $\beta_{forward_II}$=24.38°). The distinctly different near-field dynamics is governed predominantly by the modes excited directly via light illuminating the entering sides with $\boldsymbol{E}$-field oscillation along the downward /upward diagonal of the unit cell and perturbed mediately by the propagating modes coupled both on the entering and leaving sides of the concave layer.

*Predominant modes:* The modes on the complementary convex and concave layers can be identified on the $|E_z|$ and $|B_z|$ maps, respectively. The forward illumination configuration is close to the C (U) orientation on the convex (concave) layer, and very similar modes are excited on the corresponding layers of Pattern-I ($\gamma_{forward_I}$=95.39°) and Pattern-II ($\gamma_{forward_II}$=95.23°). Namely, horizontal (vertical) dipoles develop on the convex (concave) nanoring and horizontal dipoles (quadrupoles) are observable on the convex (concave) nanocrescents (Fig. 7c and Fig. 8c).

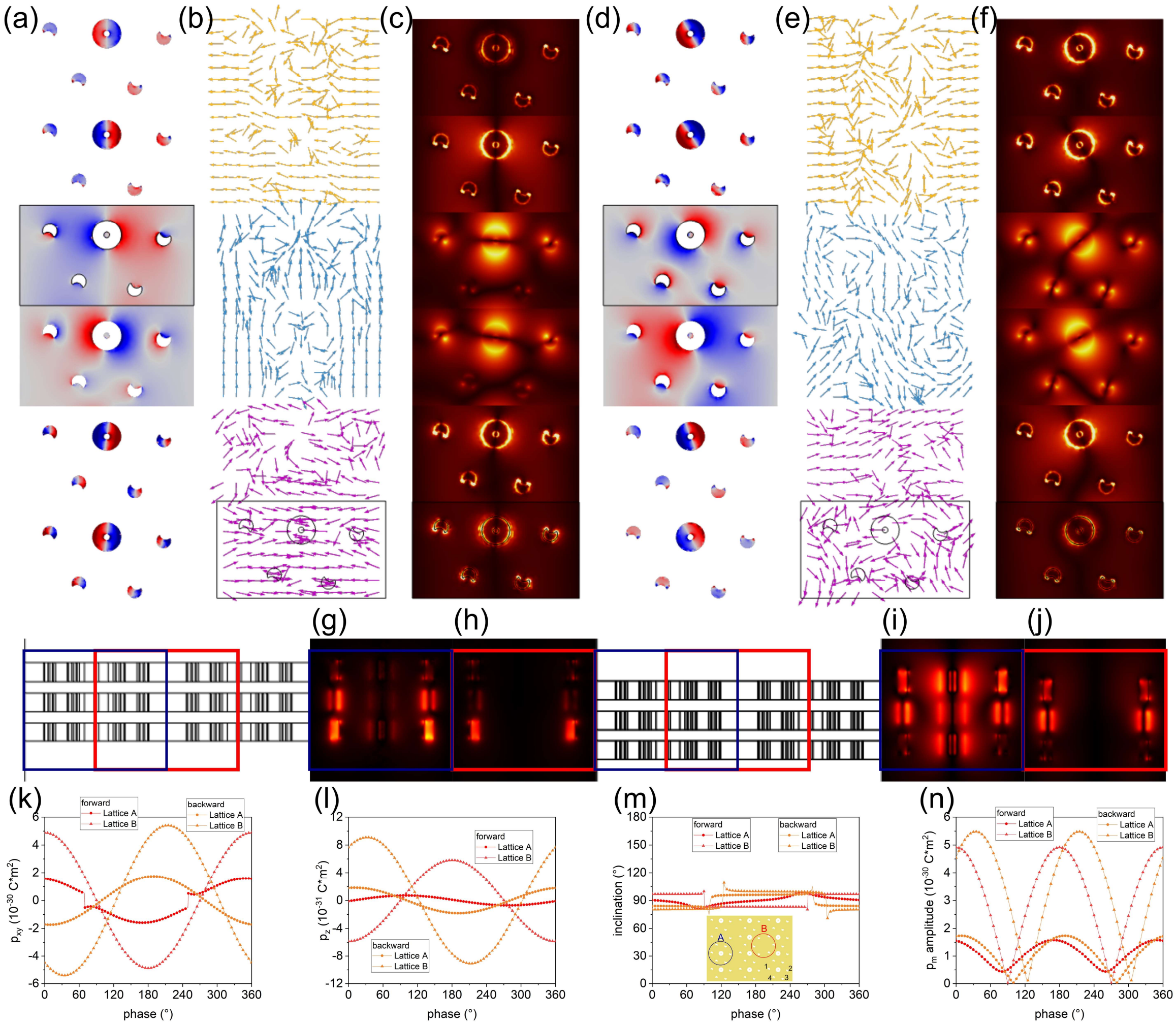

**Figure 8.** Near-field phenomena of Pattern-II. (a, d) Charge distribution on consecutive layer surfaces (Video 5, 6), (b, e) magnetic field (yellow), displacement current (blue) and vorticity distribution on entering and leaving concave surfaces (Video 7, 8), (c, f) $|E_z|$ on convex and $|B_z|$ on concave entering and leaving surfaces, (g, h, i, j) $|B_z|$ on vertical cross-sections through A-B sub-lattices, with schematics on the left side; a, b, c, g, h/d, e, f, i, j) forward /backward propagation direction. Emulated magnetic field related $\boldsymbol{p}_m$ dynamics on Pattern-II. (k) $p_{xy}$, (l) $p_z$, (m) inclination, (n) amplitude (please note the existence of offset). Inset on (m) segment: multi-unitcell schematic figure in the xy-plane, indicating the A and B sub-lattices.

In contrast, in backward propagation direction the $|E_z|$ and $|B_z|$ distribution on the nanoring significantly differ, since the dipoles are oriented along $\gamma_{backward_I}$=120.1° and ~30° on the convex and concave nanoring in case of Pattern-I, while the $|E_z|$ and $|B_z|$ lobes appear along $\gamma_{backward_II}$=65.62° and ~155° in Pattern-II (Fig. 7f and Fig. 8f). (Slightly different orientation is noticeable on the entering concave surfaces, where the dipoles are oriented at ~45° /135° in case of Pattern-I/II). As a result, the convex and concave mode distributions are perpendicular almost perfectly.

This is in accordance with the EM duality in case of orthogonal entering polarizations. The nanocrescents exhibit mixed higher order modes, which are coupled with the propagating SPPs on the concave layers.

Inspection of the $|E_z|$ and $|B_z|$ on the vertical cross-section of the multilayers indicates that there is no difference in layer's dominance in forward direction (Fig. 7g,h and Fig. 8g,h), while in backward direction the entering convex layer /middle concave layer is relatively more dominant on both of the A and B sub-lattices of Pattern-I/II (Fig. 7i,j and Fig. 8i,j). This correlates with the larger transmission through Pattern-II.

*Field enhancements:* The near-field enhancement (*NFE*), as well as the volume (*V*) corresponding to a decrease by a factor of 1/e and their product ($NFE \times V$) was also determined and compared, both for the electric and magnetic field, and the *NFE* achieved in Pattern-I and Pattern-II was also compared (see Supporting Information, Table S1 and Table S2)

### 3.8 Emulated magnetic field, related vorticity and helicity

In order to recognize displacement current (***D***) loops and related $\boldsymbol{p}_m$ magnetic dipoles, the ***D*** and ***H*** field distributions were inspected on the surfaces of the concave layer. The $H_z$ component (parallel with the propagation) was negligible in both patterns, therefore the $H_{xy}$ components were inspected in more detail (Fig. 7b,e and Fig. 8b,e).

The fingerprint of ***D*** loops is recognizable also on the xy plane cross-section, pronouncedly around the nanorings and noticeably around the nanocrescents as well (Fig.. 7b,e and Fig. 8b,e, blue). The $H_{xy}$ component shows a magnetic dipole along $\gamma\sim 5°$ at the nanoring in case of forward propagation direction on both multilayers, while the magnetic dipoles are oriented at $\gamma = 30°$ and $\gamma = 155°$ on the concave layer of Pattern-I and Pattern-II, respectively (Fig. 7b, e and Fig. 8b, e, yellow). The orientation of the polarization ellipses determined based on the Stokes parameters (Fig. 1f and Fig. 2f), the ***D***-field and the accompanying $H_{xy}$ components are in accordance with the small and large rotation in the forward and backward propagation direction, considering either linearly or elliptically polarized light illumination in the backward propagation direction. The consecutive polarization orientations for the forward and backward incoming and outgoing beams are $\beta_{forward_I} = -5.39°$; $\alpha_{backward_I} := \beta_{forward_I} = -30.06°$; $\beta_{backward_I} = -94.47°$ for the Pattern-I, while $\beta_{forward_II} = -5.23°$; $\alpha_{backward_II} := \beta_{forward_II} = 24.38°$; $\beta_{backward_II} = 88.43°$ were registered in case of Pattern-II (Fig. 1f and Fig. 2f). These components also prove the most important difference, which is that the polarization rotation occurs in different directions in the two patterns (Fig. 7b, e, Video 3, 4 –to– Fig. 8b, e, Video 7, 8). The vorticity ($\mathrm{Re}(\boldsymbol{E} \times \boldsymbol{H}^*)$) and helicity ($\mathrm{Re}(\boldsymbol{E} \cdot \boldsymbol{H}^*)$) were also inspected on these surfaces (Fig. 7b,e and Fig. 8b,e; see Supporting Information Fig. S13 and Fig. S14, all in purple). The distribution of these quantities is more structured in case of Pattern-II.

### 3.9 Induced magnetic dipole

The amplitude and $p_{xy}$ components as well as the inclinations of the $\boldsymbol{p}_m$ corresponding to the A and B sub-lattices in forward /backward propagation direction in Pattern-I are very similar to those in backward /forward propagation direction in Pattern-II. Only exception is the inclination of the $\boldsymbol{p}_m$ in A sub-lattice, which phase evolution differs in backward propagation direction in Pattern-I and in **f**orward propagation direction in Pattern-II.

The $p_z$ component of the $\boldsymbol{p}_m$ corresponding to the A and B sub-lattices in forward propagation direction in Pattern-I is also very similar to that in backward propagation direction in Pattern-II, while the $p_z$ component in backward propagation direction in Pattern-I is almost two-times larger compared to that in forward propagation direction in Pattern-II, and the A sub-lattices show 90° phase difference (Fig. 7k-m, Fig. 8k-m). The shift between the phase evolution in A and B sub-lattices in backward propagation direction indicates the impact of co-existent space-time modulation (Fig. 7k and Fig. 8k).

### 3.10 Correlations between optical responses and the $\boldsymbol{p}_m$ magnetic dipole offset, in- and out-of-plane as well as orthogonal components in sub-lattices

The average of the $\boldsymbol{p}_m$ shows an offset, which increases in (sub-lattice – forward /backward propagation direction nomination) A-f<B-b<B-f<A-b succession in Pattern-I, and B-f<A-b<B-b<A-f succession in Pattern-II. (Fig. 7n, Fig. 8n). In case of Pattern-I the simultaneously larger asymmetric transmission and rotation in backward propagation direction correlate with the larger average and maximal $p_z$ and co-polarized components, as well as with the larger average inclination and cross-polarized component, in both sub-lattices.

The offset, maximal and average amplitude and $p_{xy}$ projection, and maximal cross-polarized component correlate /anticorrelate for the A/B sub-lattice with $AT$ and $\sum$, while the maximal inclination shows anti-correlation for both of the A and B sub-lattices (Fig. 7k-n, see Supporting Information Fig. S9d,e).

In comparison, the asymmetric transmission /nonreciprocal rotation is larger in forward /backward propagation direction in Pattern-II, which correlates with the offset, co-polarized component in A sub-lattice /B sub-lattice, while the values for the amplitudes, in- and out-of-plane and cross-polarized components of $\boldsymbol{p}_m$ that are larger in backward propagation both in A and B sub-lattices, correlate with the polarization rotation (except the average cross-polarized in the B sub-lattice) (Fig. 8k-n, see Supporting Information Fig. S10d, e).

These results indicate that the synthetic gauge resulting in $AT$ and nonreciprocal polarization rotation is more closely related to the A sub-lattice in Pattern-I**,** while the larger $AT$ in forward direction /nonreciprocal polarization rotation in backward direction is promoted by $\boldsymbol{p}_m$ the A/B sub-lattice in Pattern-II. The relatively enhanced role of the B lattice in Pattern-II is confirmed also by the more structured vorticity and helicity characteristics (see Supporting Information Fig. S13 – to – Fig. S14, c segments especially).

For both inspected multilayers the average inclination as well as the average and maximal $p_z$ component correlate with the polarization rotation, when the A and B sub-lattices are considered separately. This also indicates that the origin of the nonreciprocity is a vector gauge, which is revealed by a tilted-precessing magnetic dipole (Fig. 7n and Fig. 8n, Table S3). Even though the multilayer thickness is just slightly different for the Pattern-I and Pattern-II, the effective material parameters and the induced gauge are also different. Considering that the nonreciprocal polarization rotation, accompanying the GFR phenomenon, requires a vector gauge and a DC $\boldsymbol{B}$-field component along the dominant modes' propagation direction, however the results reveal that the subject of rotation is partially photonic and plasmonic mode.

The correlations with the average and maximal amplitude and in-plane component for the A sub-lattice in Pattern-I, and for both sub-lattices in Pattern-II reveal that the plasmonic modes have an important role.

The different $\boldsymbol{p}_m$ characteristic shows enhanced role of the plasmons in A sub-lattice in Pattern-I, and balanced role of photonic and plasmonic modes in A and B sub-lattices in Pattern-II [72].

### 3.11 Correlation between displacement and optical response asymmetry on the multilayers

Considering that nonreciprocal phenomena are accompanied by modification of the line of forces, and our former studies on similar patterns have shown correlation between optical responses and normal component displacement asymmetry, correlation study was realized [42,61,62].

Pattern-I and Patter-II were inspected for one and two-sided linearly polarized illumination, and beside the nonreciprocal rotation the $AT_{co}$ and outflow asymmetry were compared in the former and latter case with the asymmetry of the normal component of the displacement vector. In case of Pattern-I the $\delta D_{z_convex\&concave}$ on both of the convex and concave layers exhibits correlation with the asymmetry of the reversal |outflow|; while $\delta D_{z_concave}$ correlates with asymmetry of the $AT_{co}^{classical}$ in the forward configuration for the concave layer, but only when the maximum values are compared. In case of Pattern-II the $\delta D_z$ on the convex /concave layer exhibits correlation with the rotation and asymmetry of the cross-polarized classical $AT$ and reversal of |outflow| / asymmetry of the co-polarized classical $AT$ in backward / forward configuration. (Table S4).

### 3.12 Time-domain studies

The multilayer was illuminated by 12 fs laser pulses of central frequency corresponding to the GFR condition of Pattern-I and Pattern-II calculated with steady-state studies. The structure was illuminated with pulses incident perpendicularly and thereafter at a 45° polar angle to inspect time modulation (Fig. 9a-f and Fig. 10a-f with respective insets). All transmitted pulses' complete envelope is longer than the 12 fs pulse according to the resonant behaviour, though the constituent pulses are shorter than the incoming one in case of splitting. The co-polarized component is significantly stronger at the transmission sides for both propagation directions in all bases, to varying degrees in the two different multilayers.

In forward configuration, both the co-polarized and cross-polarized transmitted pulses split into two consecutive pulses. The individual pulses envelope's amplitude decreases – except the cross-polarized component in Pattern-II, in which the second pulse is unambiguously enhanced (Fig. 9a and Fig. 10a). At a 45° incidence angle, no splitting is observed in the cross-polarized component (insets of Fig. 9a and Fig. 10a). In backward configuration in the backward bases, the co-polarized component is split into two consecutive transmitted pulses, and their envelope's amplitude decreases, while the cross-polarized pulse is significantly prolonged and exhibits a delayed maximum at the gap of the co-polarized pulses independent from the angle of incidence (Fig. 9c and Fig. 10c and insets). In backward configuration in the forward bases both the co-polarized and cross-polarized components are split into two consecutive transmitted pulses, their envelope's amplitude decreases, and the components are more commensurate in Pattern-II (Fig. 9b and Fig. 10b).

When illuminated with a pulse incident at a 45°, the second peak in the cross-polarized components will be larger than the first, for both Patterns (Fig. 9b and Fig. 10b inset). Compared to the response in backward bases, the cross-polarized component is more modulated in forward base. In either bases, there is no remarkable difference between the time evolutions except that the gap is more well-defined in Pattern-II. In the case of 45° of illumination, the most significant difference is that the aforementioned increase in the cross-polarized component in the forward base less well defined in Pattern-II.

The Fourier transform of the time signals were also inspected to analyse the modification of the spectral distribution (Fig. 9d-f and Fig. 10d-f). The spectral amplitude of transmitted light is much smaller than the excitation. The spectrum of co-polarized light is split in all cases regardless of illumination direction, tilting angle or bases. The peaks are close to each other (~50 nm), which is the origin of the beating-like time modulation observed as multiple transmitted pulses.

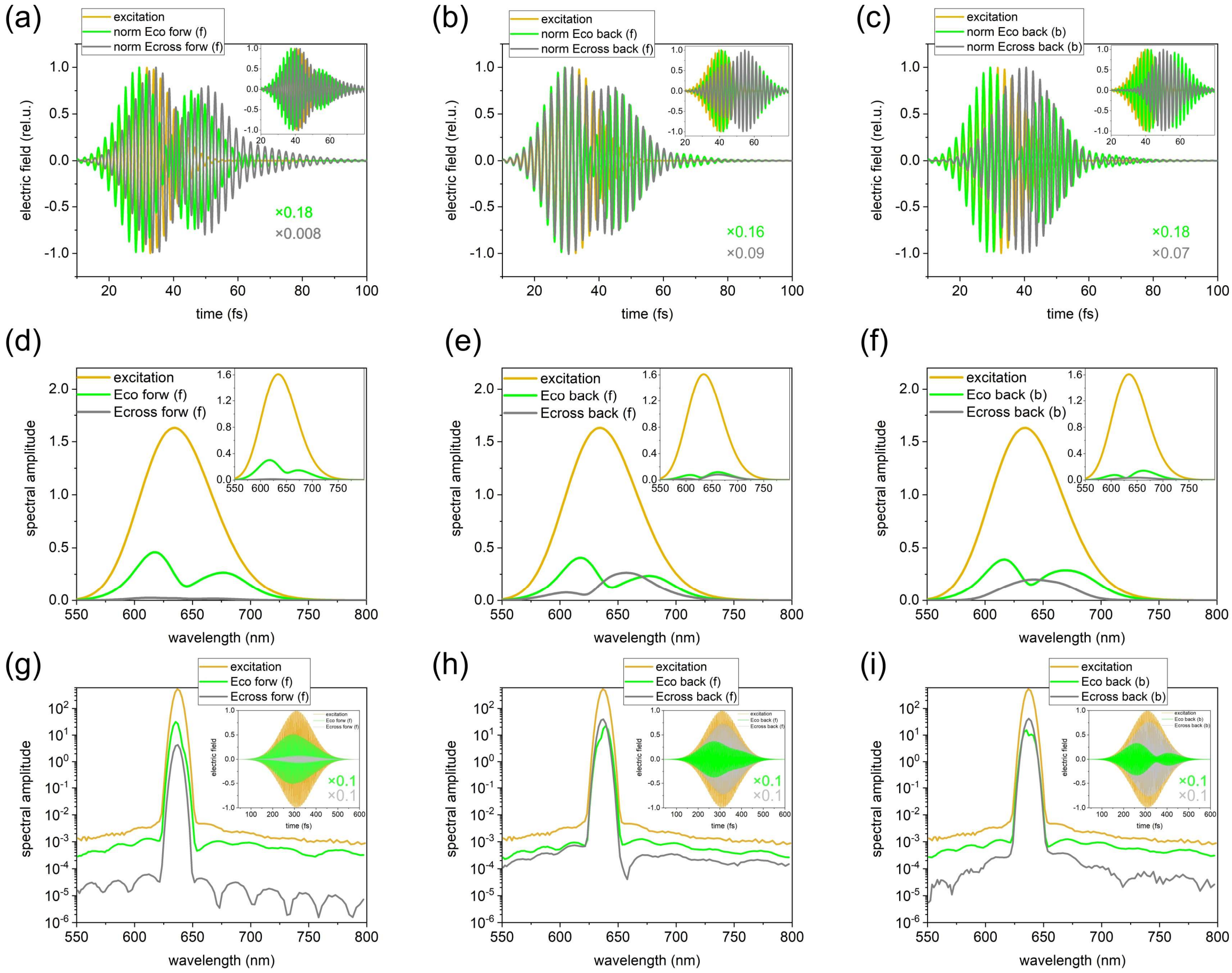

**Figure 9.** Time-dependent optical response of Pattern-I. Co-polarized and cross-polarized components of (a-c) transmitted pulses, (d-f) spectral distribution for perpendicular incidence and 45° tilting in insets; (g-i) spectral distribution and transmitted pulses in insets; in (a, d, g) forward, and (b, c, e, f, h, i) backward configuration; (c, f, i) in backward, (a, d, g and b, e, h) in forward bases; using (a-f) 12 fs and (g-i) 120 fs short pulse illumination.

The center of splitting is nearby the GFR wavelength. In the case of 45° tilting a significant difference between forward and backward directed illumination emerges, which indicates that tilting enhances the asymmetry (Fig. 9d-f and Fig. 10d-f insets).

Namely in the backward direction the gap shifts to shorter wavelengths. The only well-defined difference between Pattern-I and Pattern-II is that the second spectral peak in the former is slightly broader. The transmission of cross-polarized light in the forward direction is much smaller compared to the co-polarized component (in agreement with the smaller rotation calculated with frequency-domain simulations).

There is a split nearby the GFR wavelength, which results a beating. In backward propagation and backward base, the splitting vanishes, resulting in a nearly Gaussian, slightly prolonged response due to the absence of beating. In backward propagation and forward base the first peak is significantly less intense, which is the origin of the moderate gap in the time-domain signal in cross-polarized components and the unambiguously increasing constituting pulse amplitude at 45° tilting.

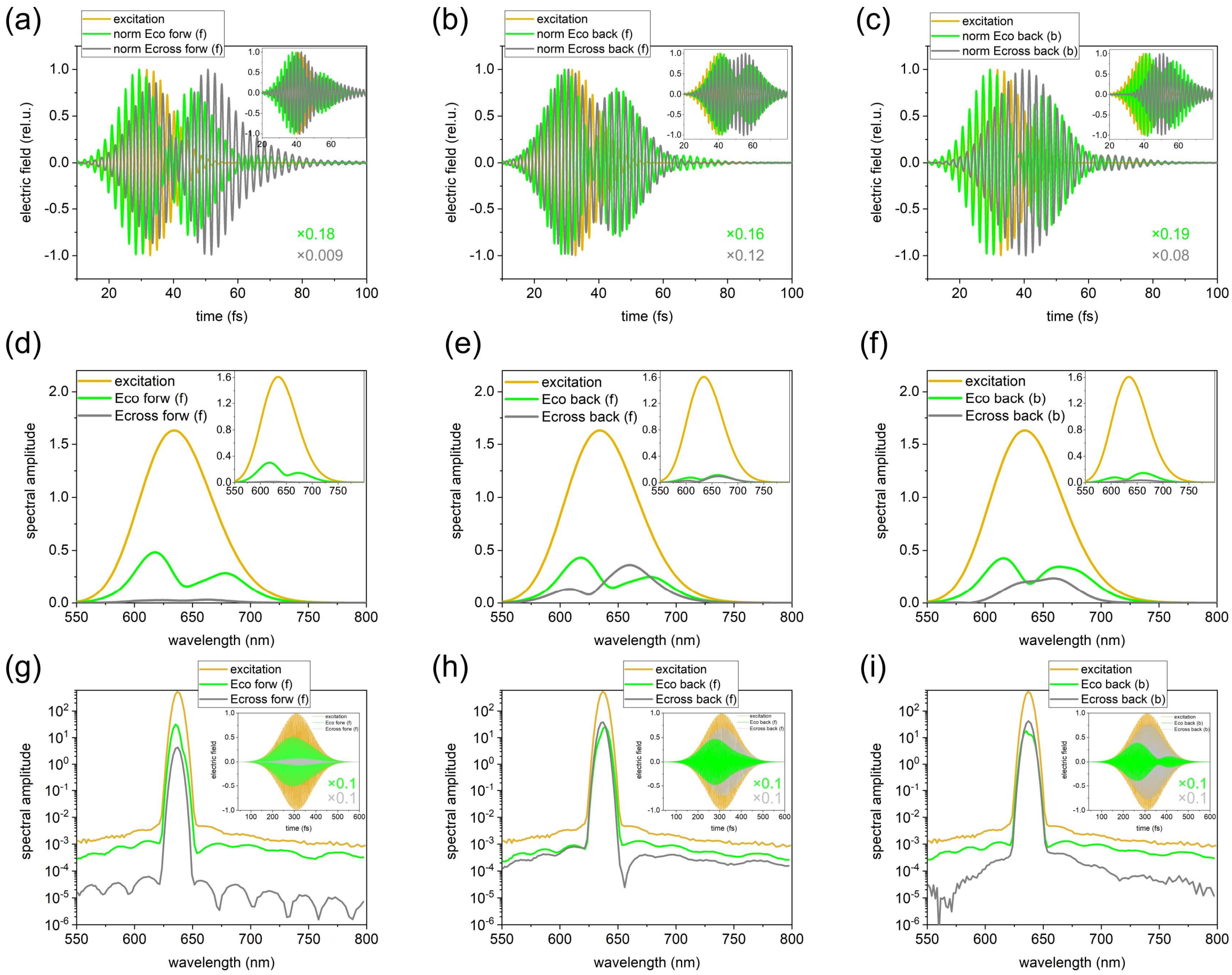

**Figure 10.** Time-dependent optical response of Pattern-II. Co-polarized and cross-polarized components of (a-c) transmitted pulses, (d-f) spectral distribution for perpendicular incidence and 45° tilting in insets; (g-i) spectral distribution and transmitted pulses in insets; in (a, d, g) forward, and (b, c, e, f, h, i) backward configuration; (c, f, i) in backward, (a, d, g and b, e, h) in forward bases; using (a-f) 12 fs and (g-i) 120 fs short pulse illumination.

The multilayer was illuminated by 120 fs pulses as well, since the narrow bandwidth allows for distinguishing signals stemming from time modulation and originating from the broad bandwidth (Fig. 9g-i and Fig. 10g-i). The time evolution of the electric field shows less modulated transmitted pulses compared to the case of 12 fs excitation. The co-polarized pulse in the forward direction has a near-Gaussian shape but prolongs. In contrast, the cross-polarized light is delayed in the multilayer but preserves its Gaussian shape. In the backward direction and in backward base pulse splitting is observed, as with 12 fs pulse. The envelope of the consecutive pulses decreases in amplitude. The cross-polarized component preserves the Gaussian shape with a delay of 10 fs.

In the forward base the splitting in co-polarized component is less pronounced, while the cross-polarized component preserves the Gaussian shape. There is no noticeable difference between the two patterns due to the very similar geometry.

The spectral distribution of the forward transmitted signal indicates single strongly /slightly blue-shifted co- /cross-polarized peak, and several side-bands revealing a weak but well -defined time modulation corresponding to characteristic frequency $\omega$=7.2×10$^{13}$ rad/s (geometric resonance of ~25 nm). In comparison, the spectral distribution of the backward transmitted signal indicates split in co-polarized peak, explaining the significant prolongation and splitting of the time evolution.

The cross-polarized signal appears at the center of this split, with no noticeable splitting, in accordance with the Gaussian envelope of the time evolution. The difference between backward and forward bases is that the ratio of the side-peaks is interchanged. The peaks in backward base are more commensurable in amplitude, resulting in a more pronounced modulation in the time-domain signal. The spectral splits indicate that photonic transitions occur on different frequency scales. When time-periodic Floquet modulation is imparted, beat-notes can be registered at harmonics of modulation frequencies in accordance with the literature [38].

### 3.14 Active layer compensation

To compensate the losses in the multilayer and at the same time to preserve both the nonreciprocal rotation and the good transmission contrast, dielectric cover layers seeded with dye molecules were added, while asymmetry was allowed for their thickness – and the optimization of the layer thickness, dye concentration ($N$) and pump $\boldsymbol{E}$-field strength ($E_{pump}$) was performed, inheriting the azimuthal orientation of Pattern-II in forward direction. The active multilayers were inspected both in wavelength independent configurations, which is similar to the previous interpretations, and in wavelength dependent configurations (Fig. 11a-c and d-f, Table S5). An active multilayer with side-layer thicknesses scaled to 0.63× and 0.71× of the decay length (242 nm), $N$=3.07×10$^{26}$ m$^{-3}$ and $E_{pump}$=2.33×10$^{6}$ V/m ensured that the GFR condition is still met, namely the ∑ = -84.57° cumulative rotation was achieved; with enhanced balanced (1.35) and differential co-polarized asymmetric transmission (3.13), that was enabled by the strongly enhanced total backward transmission (3.24) as well as its enhanced projection onto the cross-polarized component of the forward direction (3.18), while the forward transmission (7.4×10$^{-2}$) as well as its co-polarized components (4.7×10$^{-2}$) remained small. These together ensure 67.47 transmission contrast in the forward bases (Fig. 11a-c). The sign of the rotations in both propagation directions is switched compared to the passive Pattern-II composition, while the relation between the forward and backward rotations remains the same, namely larger rotation occurs in backward propagation direction. The inspected *GFR-FOM* (0.21) and *GFI-FOM* (5.97×10$^{-3}$) are also significantly enhanced. Although the resonant modes at the GFR are quasi-BICs, moderate $Q$-factor (116/281 in wavelength independent /$\gamma(\lambda)$ configuration) corresponds to the narrow transmission peaks for backward propagation direction is due to the relatively broad emission spectra of the used dye.

Thereafter a Kekulé transformation type dimerization was also applied, namely the radius of the upper convex quadrumer was decreased to 57.74 nm on the top layer and was rotated by -30° (Pattern-II-dimerized), and the dye molecule concentration, $E_{pump}$ and the side-layer thicknesses remained the same.

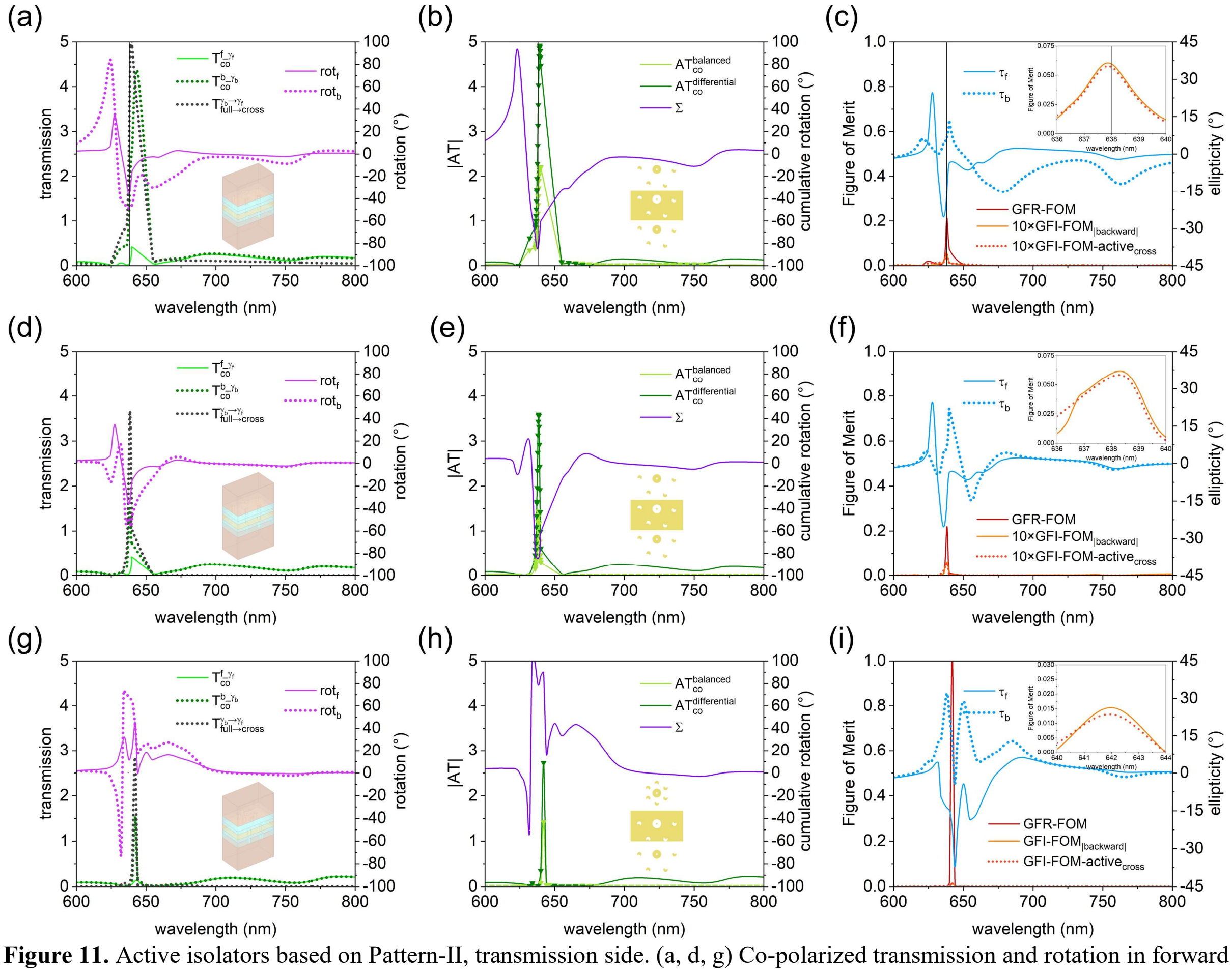

**Figure 11.** Active isolators based on Pattern-II, transmission side. (a, d, g) Co-polarized transmission and rotation in forward and backward propagation direction, as well as the backward transmission projected onto the forward cross-polarized component, (b, e, h) cumulative rotation, the balanced and differential co-polarized asymmetric transmission (triangular symbols indicate regions, where the backward transmission is larger), (c, f, i) ellipticity of outgoing beams, *GFR-FOM*, *GFI-FOM*$_{|backward|}$ and *GFI-FOM-active*$_{cross}$. (a-c) Wavelength independent, (d-f) wavelength dependent active Pattern-II and (g-i) wavelength dependent active dimerized Pattern-II configuration.

This cover layer enabled that the GFR condition is better met, namely the $\sum = 89.65°$ cumulative rotation was achieved; with more enhanced balanced (1.42) and less enhanced differential asymmetric co-polarized transmission (2.73), that was enabled by moderately enhanced total backward transmission (2.99) as well as its commensurate enhanced projection onto the cross-polarized component of the forward direction (2.85), while the forward transmission ($1.3\times10^{-1}$) as well as its co-polarized component ($1.2\times10^{-1}$) slightly increased, but remained an order of magnitude smaller compared to the backward transmission. These together ensure 23.44 transmission contrast in the forward bases (Fig. 11g-i). The sign of the rotations in both propagation direction remains the same compared to the passive Pattern-II compositions, while the relation between the forward and backward rotations becomes balanced, it is almost 45° in both propagation directions. The inspected *GFR-FOM* (1.05), *GFI-FOM* ($1.3\times10^{-2}$) are an order of magnitude enhanced compared to the active Pattern-II configuration. The *Q*-factor ($1.5\times10^{4}$ in $\gamma(\lambda)$ configuration) of the narrow transmission peaks for backward propagation direction is significantly enhanced.The cross-polarized components and the determined FOMs for wavelength independent (Fig. S15a-c) and dependent (Fig. S15d-f) backward illumination for active Pattern-II configuration and for wavelength dependent active dimerized configuration (Fig. S15 g-i) are presented in the Supporting Information.

These sub-wavelength active multilayers are proposed as an extremely thin GFR elements, moreover can serve as nonreciprocal isolators.

## 4. Conclusions

An efficient optimization methodology was developed to achieve generalized Faraday rotation, which involves both nonreciprocal rotations and considerable asymmetrical transmission. The subject of optimization was a multilayer of Babinet complementary periodic structures constructed with miniarrays of spherical nanoresonators, while the tool was an optimization algorithm that allows for setting criteria, and definition of complex objective functions constructed with all metamaterial configuration parameters that can be monitored simultaneously. The key idea is to allow arbitrary azimuthal orientations, that differ for the two propagation directions, thereby facilitating the determination of those orthogonal directions in bianisotropic multilayers, which ensure the spectral overlap of the nonreciprocal polarization rotation and asymmetric transmission phenomena.

The topological effects involved in the Brillouin zone folding and mode coupling phenomena are (i) the coupling of the nanodisk/nanohole, nanoring and nanocrescents in convex/concave miniarrays resulting in symmetry breaking perturbation and quasi-BICs (ii) co-existent two sub-lattices – A with, while B without nanoring – providing period-doubling perturbation, in-plane twisting and supporting charge rotation with phase difference, (iii) the unique arrangement of the satellite nanocrescents, which ensures that each spin-distinguishing sub-lattices are centrally symmetric but possess distinctly different deformations facilitating hierarchical time-periodic Floquet modulation and (iv) coupling of localized and propagating modes and strong-coupling in the near-field between Babinet complementary consecutive layers that are arranged in convex-concave-convex sequence, (v) bianisotropy stemmed from the unique pseudo non-Hermitian composition supporting unconventional dispersion characteristic and space-time modulation. Even though the spacers between basic constituent layers are symmetric, the polarized light illumination results in distinctly different, propagation direction and tilting-side dependent, optical responses. Moreover, the hierarchical parameter-sensitivity of the coupling phenomena on miniarrays, sub-lattices and constituent layers manifests itself in two optimized multilayers possessing very similar geometries, meaning few tenth of nanometers thickness difference, yet these metamaterials exhibit fundamentally different nanophotonic phenomena, including the time-periodic modulation characteristic and the related synthetic gauge.

Nonreciprocal rotation and the asymmetry of co-polarized transmission are demonstrated in the optimized GFR configuration, novelty of which is that it involves two different azimuthal orientations that deviates from conventional reciprocal-lattice directions. Moreover, the mapping of the dispersion in these azimuthal orientations proves that the asymmetry in transmission considered in classical sense manifests itself in quantized flat bands, which are intersected by the grating-coupling related branches governed by bianisotropic effective parameters in the specific configuration. These features are propagation-direction as well as tilting-side dependent, revealing a completely novel metamaterial. The trade-off between asymmetrical co-polarized transmission and nonreciprocal rotation is caused by that these optical responses are related to the quasi-BICs of the localized modes on constituent nanoresonators and scattered photonic and propagating plasmonic modes corresponding to different bands folded into the first Brillouin zone.

The multilayer is plasmonic, accordingly can be considered as pseudo non-Hermitian, and the GFR appears on the band originating from folding and flattening, which sensitively depend on parameters.

The near-field phenomena, including displacement current loops and vortical magnetic fields, prove the existence of a synthetic gauge and an emulated magnetic field, which manifests itself in tilted-precessing magnetic dipoles, yet producing an offset, proving a potential to act as a horizontal and vertical torque, which facilitates nonreciprocal polarization rotation. The nonreciprocity proves the existence of synthetic vector-gauge, while the time-periodic Floquet modulations adds a synthetic parameter along the frequency.

The time-evolution of the charge distribution on the sub-lattices shows reversal rotation on the nanoring and nanocrescents, thereby proving that the miniarrays act as a unique Haldane model. Moreover, reversal rotation occurs on the coupled loops of A and B sub-lattices. The spectral distribution and the temporal response is modulated due to the beating related to overlapping resonances of the nanoring – nanocrescent, A and B sub-lattices and different zone folded SPPs. In addition, remarkable configuration and propagation direction dependent vorticity features are identifiable. The coupling of localized and propagating modes causes sudden circulating charge-distribution flips in each optical cycle, ensuring that the response of the pseudo non-Hermitian metamaterial is modulated both in space and time. The time-periodic Floquet modulation reveals co-existence of scalar and vector gauge, accordingly ***E*** and ***B***-field are also at play, which indicates capability to achieve quantized Landau levels and Landau Zener Tunneling.

The multilayer composition manifests itself in a unique bianisotropic medium, where all – Tellegen, artificial moving, omega and chirality– parameters are modulated. Accordingly, the temporal responses show polarization specific pulse reshaping via beating, proving that the orthogonally-polarized components can be transiently enhanced, and complete polarization specific pulse-splitting can occur, offering the possibility of temporal cloaking. Both polarization components indicate asymmetric coupling, which indicates non-Abelian synthetic gauge characteristic.

The trade-off between the cumulative rotation, which approaches 90° within a spectral interval where the asymmetrical transmission is limited by the pseudo non-Hermiticity of the multilayer metamaterial, and the bianisotropy, which causes that the metamaterial supports elliptical eigenmodes, justifies that optimization is necessary. Despite these limitations, it is possible to re-enhance the transmission, while preserving the cumulative rotation by adding an active cover layer. Moreover, the GFR criterion can be satisfied more accurately, and the reminiscent ellipticity can be reduced, when topological tools, e.g. judicious dimerization is involved into the design.

In summary, our optimization method resulted in a completely novel bianisotropic multilayer metamaterial, which shows several topological properties that ensure unique dispersion and time-periodic Floquet modulation characteristic and high-level bianisotropy enabling Generalized Faraday Rotation via sub-wavelength thickness all-optical elements. Further applications include design of metamaterial platform supporting photonic Landau levels, and QIP elements relaying on band braiding phenomena due to the enormously high degrees of freedom achievable in synthetic parameter space.

**Acknowledgements**

This work was supported by the National Research, Development and Innovation Office (NKFIH) of Hungary, projects: "Nanoplasmonic Laser Inertial Fusion Research Laboratory" (NKFIH-2022-2.1.1-NL-2022-00002) and "National Laboratory for Cooperative Technologies" (NKFIH-2022-2.1.1-NL-2022-00012) in the framework of the Hungarian National Laboratory program, as well as the Hu-rizont "Metamaterials for Optimized Quantum Information Processing" (METAOPTQIP, 2025-1.2.1-HU-RIZONT-2025-00107).

# Supporting Information

## Multilayer Babinet Metamaterial to Initiate Nonreciprocal Topological Phenomena and Generalized Faraday Rotation

Balázs Bánhelyi[1], Miklós Waldhauser[2], Virág Szünstein[2], Ákos Sebők-Pap[2], Olivér Ardelán[2], Anna Kőházi-Kis[2], Dávid Vass[2], András Szenes[2], David Keene[3], Maxim Durach[3], Mária Csete[2]

[1]Department of Computational Optimization, University of Szeged, Árpád tér 2, H-6726 Szeged, Hungary;
[2]Department of Optics and Quantum Electronics, University of Szeged, Dóm tér 9, H-6726 Szeged, Hungary;
[3]Department of Biochemistry, Chemistry and Physics, Georgia Southern University, 301 Veazy Hall, Suite 2000, Statesboro, GA 30458
Correspondence: mcsete@physx.u-szeged.hu

## Inspected Figure of Merits to quantify Generalized Faraday Rotation (GFR) and Generalized Faraday Isolator (GFI) capabilities of the optimized multilayers.

For the qualification of the proposed passive targets two primary *FOMs* were evaluated, namely the *GFR-FOM* (Eq. S1), which quantifies asymmetrical transmission and cumulative rotation properties of the multilayer, and the *GFI-FOM* (Eq. S2), where the isolator properties and the ellipticity were also inspected.

$$GFR-FOM = \left|AT_{co}^{balanced}\right| \cdot \frac{1}{\left|90° - |\Sigma|\right| + 1} \qquad \textbf{(S1)}$$

$$GFI-FOM = \frac{1}{T_{full}^{f_\gamma_f} + 1} \cdot \frac{1}{\left(1 - T_{full\rightarrow cross}^{\gamma_b\rightarrow\gamma_f}\right) + 1} \cdot \left|T_{full}^{f_\gamma_f} - T_{full\rightarrow cross}^{\gamma_b\rightarrow\gamma_f}\right| \cdot \frac{1}{\left|90° - |\Sigma|\right| + 1} \cdot \frac{1}{|\tau_f| + |\tau_b| + 1} \qquad \textbf{(S2)}$$

Two additional *FOMs* were also evaluated, namely the *GFI-FOM*$_\tau$ (Eq. S3), which also quantifies the ellipticity of the outgoing beams in addition to the asymmetrical transmission and cumulative rotation properties, and the *GFI-FOM*$_{AT\text{-}\tau\text{-}\Sigma}{}^{-1}$ (Eq. S4), where similar weighting was used for the asymmetrical transmission, cumulative rotation and ellipticity.

$$GFI-FOM_\tau = \left|T_{full}^{f_\gamma_f} - T_{full\rightarrow cross}^{\gamma_b\rightarrow\gamma_f}\right| \cdot \frac{1}{\left|90° - |\Sigma|\right| + 1} \cdot \frac{1}{|\tau_f| + |\tau_b| + 1} \qquad \textbf{(S3)}$$

$$GFI-FOM_{AT-\tau-\Sigma}{}^{-1} = \left(\frac{90 - \left|90° - |\Sigma|\right| + 90 \cdot \left|T_{full}^{f_{\gamma_f}} - T_{full\rightarrow cross}^{\gamma_b\rightarrow\gamma_f}\right| + 1}{163 + 18 \cdot \left(\left||\tau_f| - |\tau_b|\right| + 1\right)}\right)^{\ln(|\tau_f|+|\tau_b|+1)+1} \qquad \textbf{(S4)}$$

In case of the active targets the *GFR-FOM* was used as a primary *FOM*. However, the *GFI-FOM* was slightly modified considering that when the transmission in backward direction exceeds unity, the *FOM* becomes negative, so for the second term of the *GFI-FOM*, absolute value was used (*GFI-FOM*$_{|backward|}$ (Eq. S5)).

$$GFI-FOM_{|backward|} = \frac{1}{T_{full}^{f_\gamma_f} + 1} \cdot \frac{1}{\left|1 - T_{full\rightarrow cross}^{\gamma_b\rightarrow\gamma_f}\right| + 1} \cdot \left|T_{full}^{f_\gamma_f} - T_{full\rightarrow cross}^{\gamma_b\rightarrow\gamma_f}\right| \cdot \frac{1}{\left|90° - |\Sigma|\right| + 1} \cdot \frac{1}{|\tau_f| + |\tau_b| + 1} \qquad \textbf{(S5)}$$

The novel *GFI-FOM-active*$_{cross}$ was also introduced, where the achieved transmission values were normalized (Eq. S6).

$$GFI-FOM-active_{cross} = \frac{1}{T_{full}^{f_\gamma_f} + 1} \cdot \frac{T_{full\rightarrow cross}^{\gamma_b\rightarrow\gamma_f}}{T_{full}^{f_\gamma_f} + T_{full\rightarrow cross}^{\gamma_b\rightarrow\gamma_f}} \cdot \frac{\left|T_{full}^{f_\gamma_f} - T_{full\rightarrow cross}^{\gamma_b\rightarrow\gamma_f}\right|}{T_{full}^{f_\gamma_f} + T_{full\rightarrow cross}^{\gamma_b\rightarrow\gamma_f}} \cdot \frac{1}{\left|90° - |\Sigma|\right| + 1} \cdot \frac{1}{|\tau_f| + |\tau_b| + 1} \qquad \textbf{(S6)}$$

Three further *FOMs* were evaluated, where different components of the transmission were included in order to quantify the GFR and GFI properties (Eq. S7-S9).

$$GFI-FOM-active_{mixed}=\frac{1}{T_{full}^{f_\gamma_f}+1}\cdot\frac{T_{full}^{b_\gamma_b}}{T_{full}^{f_\gamma_f}+T_{full\to cross}^{\gamma_b\to\gamma_f}}\cdot\frac{\left|T_{full}^{f_\gamma_f}-T_{full\to cross}^{\gamma_b\to\gamma_f}\right|}{T_{full}^{f_\gamma_f}+T_{full\to cross}^{\gamma_b\to\gamma_f}}\cdot\frac{1}{\left|90^\circ-\left|\Sigma\right|\right|+1}\cdot\frac{1}{|\tau_f|+|\tau_b|+1} \quad \textbf{(S7)}$$

$$GFI-FOM-active_{balanced}=\frac{1}{T_{full}^{f_\gamma_f}+1}\cdot\frac{T_{full}^{b_\gamma_b}}{T_{full}^{f_\gamma_f}+T_{full}^{b_\gamma_b}+1}\cdot\frac{\left|T_{full}^{f_\gamma_f}-T_{full\to cross}^{\gamma_b\to\gamma_f}\right|}{T_{full}^{f_\gamma_f}+T_{full\to cross}^{\gamma_b\to\gamma_f}+1}\cdot\frac{1}{\left|90^\circ-\left|\Sigma\right|\right|+1}\cdot\frac{1}{|\tau_f|+|\tau_b|+1} \quad \textbf{(S8)}$$

$$GFI-FOM-active_{inverse}=\frac{1}{T_{full}^{f_\gamma_f}+1}\cdot\left(\frac{1}{-T_{full}^{b_\gamma_b}\cdot\left|T_{full}^{f_\gamma_f}-T_{full}^{b_\gamma_b}\right|}+1\right)\cdot\frac{1}{\left|T_{full}^{b_\gamma_b}-T_{full\to cross}^{\gamma_b\to\gamma_f}\right|+1}\cdot\frac{1}{\left|90^\circ-\left|\Sigma\right|\right|+1}\cdot\frac{1}{|\tau_f|+|\tau_b|+1} \quad \textbf{(S9)}$$

**Determining the polarization rotation and ellipticity of the outgoing plane waves.**

The polarization rotation and the ellipticity of the outgoing plane wave were determined via the Stokes-parameters [1], which can be determined by the electric-field on the transmission side (Eq. S10).

$$S_0=E_x^2+E_y^2;\ S_1=E_x^2-E_y^2;\ S_2=2E_xE_y cos(\Delta\Phi), S_3=2E_xE_y sin(\Delta\Phi), \quad \textbf{(S10)}$$

where $E_x$ and $E_y$ are the x- and y-component of the transmitted ***E***-field, and $\Delta\Phi=\Phi_y-\Phi_x$ is the phase difference between them.

The orientation ($\beta$) and the ellipticity ($\tau$) of the outgoing plane wave can be given by (Eq. S11):

$$\beta=\frac{1}{2}arctan\left(\frac{S_2}{S_1}\right);\ \tau=\frac{1}{2}arcsin\left(\frac{S_3}{S_0}\right) \quad \textbf{(S11)}$$

From the incoming ($\alpha$) and outgoing ($\beta$) orientation the polarization rotation can be calculated as $rot=\beta-\alpha$.

| | Pattern-I linear | Pattern-II linear | Pattern-I elliptical | Pattern-II elliptical |
|---|---|---|---|---|
| $T_{co}^{f_\gamma f}$ ($\times 10^{-5}$) | 2.6 | 6.9 | 2.6 | 6.9 |
| $T_{cross}^{f_\gamma f}$ ($\times 10^{-5}$) | 0.78 | 3.0 | 0.78 | 3.0 |
| $T_{full}^{f_\gamma f}$ ($\times 10^{-5}$) | 3.38 | 9.9 | 3.38 | 9.9 |
| $T_{co}^{f_\gamma b}$ ($\times 10^{-3}$) | 2.3 | 2.5 | 4.6 | 5.2 |
| $T_{cross}^{f_\gamma b}$ ($\times 10^{-2}$) | 0.92 | 1.1 | 1.2 | 1.2 |
| $T_{full}^{f_\gamma b}$ ($\times 10^{-2}$) | 1.15 | 1.35 | 1.66 | 1.72 |
| $T_{co}^{b_\gamma f}$ ($\times 10^{-4}$) | 0.16 | 1.2 | 0.16 | 1.2 |
| $T_{cross}^{b_\gamma f}$ ($\times 10^{-5}$) | 6.9 | 1.8 | 6.9 | 1.8 |
| $T_{full}^{b_\gamma f}$ ($\times 10^{-4}$) | 0.85 | 1.38 | 0.85 | 1.38 |
| $T_{co}^{b_\gamma b}$ ($\times 10^{-3}$) | 2.2 | 2.5 | 3.4 | 4.6 |
| $T_{cross}^{b_\gamma b}$ ($\times 10^{-2}$) | 0.93 | 1 | 1.2 | 1.3 |
| $T_{full}^{b_\gamma b}$ ($\times 10^{-2}$) | 1.15 | 1.25 | 1.54 | 1.76 |
| $T_{full\rightarrow co}^{\gamma b\rightarrow \gamma f}$ ($\times 10^{-4}$) | 1.3 | 0.61 | 3 | 9.3 |
| $T_{full\rightarrow cross}^{\gamma b\rightarrow \gamma f}$ ($\times 10^{-2}$) | 1.2 | 1.3 | 1.5 | 1.6 |
| $AT_{co}^{classical_\gamma f}$ ($\times 10^{-5}$) | 0.948 | 5.47 | 0.948 | 5.47 |
| $AT_{cross}^{classical_\gamma f}$ ($\times 10^{-5}$) | 6.11 | 1.22 | 6.11 | 1.22 |
| $AT_{co}^{classical_\gamma b}$ ($\times 10^{-4}$) | 0.181 | 0.249 | 12.2 | 6.41 |
| $AT_{cross}^{classical_\gamma b}$ ($\times 10^{-4}$) | 1.16 | 5.29 | 4.37 | 4.02 |
| $AT_{co}^{balanced}$ ($\times 10^{-3}$) | 2.20 | 2.40 | 3.32 | 4.49 |
| $AT_{cross}^{balanced}$ ($\times 10^{-2}$) | 0.929 | 1.04 | 1.19 | 1.28 |
| $AT_{co}^{differential}$ ($\times 10^{-2}$) | 1.14 | 1.28 | 1.49 | 1.64 |
| $AT_{cross}^{differential}$ ($\times 10^{-4}$) | 1.18 | 2.95 | 0.308 | 9.04 |
| $AT_{full}^{differential}$ ($\times 10^{-2}$) | 1.14 | 1.27 | 1.49 | 1.63 |
| $rot_f$ (°) | -24.67 | 29.60 | -24.67 | 29.60 |
| $rot_b$ (°) | -64.41 | 64.05 | -63.95 | 65.43 |
| Σ (°) | -89.08 | 93.66 | -88.63 | 95.04 |
| $\tau_f$ (°) | 17.52 | 20.09 | 17.52 | 20.09 |
| $\tau_b$ (°) | 5.92 | -1.48 | 2.99 | -1.07 |
| GFR-FOM ($\times 10^{-3}$) | 1.15 | 0.516 | 1.40 | 74.4 |
| GFI-FOM ($\times 10^{-4}$) | 1.22 | 0.609 | 1.47 | 0.616 |
| GFI-$FOM_{\tau}$ ($\times 10^{-4}$) | 2.43 | 1.21 | 2.92 | 1.22 |
| GFI-$FOM_{AT\text{-}\tau\text{-}\Sigma^{-1}}$ | 0.446 | 1.420 | 0.624 | 1.530 |
| Q-factor ($T_{co}^{b_\gamma b}$) | 320.87 | 317.29 | 319.86 | 317.82 |
| Q-factor ($T_{full}^{b_\gamma b}$) | 319.76 | 318.24 | 319.68 | 317.85 |
| Q-factor ($T_{full\rightarrow cross}^{\gamma b\rightarrow \gamma f}$) | 319.61 | 317.44 | 319.61 | 317.20 |

**Table S1.** Passive multilayer optical response parameters: transmission, polarization rotation and FOMs at extrema. The co-polarized ($T_{co}$), cross-polarized ($T_{cross}$) and total ($T_{full}$) transmissions, where the superscript indicates forward (f) and backward (b) propagation direction for azimuthal orientation corresponding to forward ($\gamma_f$) and backward ($\gamma_b$) configurations. The transmission in backward propagation direction projected onto the co-polarized ($T_{full\rightarrow co}^{\gamma b\rightarrow \gamma f}$) and cross-polarized ($T_{full\rightarrow cross}^{\gamma b\rightarrow \gamma f}$) component of forward direction. The classical co-polarized ($AT_{co}^{classical_\gamma f/\gamma b} = T_{co}^{f_\gamma f/\gamma b} - T_{co}^{b_\gamma f/\gamma b}$) and cross-polarized ($AT_{cross}^{classical_\gamma f/\gamma b} = T_{cross}^{f_\gamma f/\gamma b} - T_{cross}^{b_\gamma f/\gamma b}$), the balanced co-polarized ($AT_{co}^{balanced} = T_{co}^{f_\gamma_f} - T_{co}^{b_\gamma_b}$) and cross-polarized ($AT_{cross}^{balanced} = T_{cross}^{f_\gamma_f} - T_{cross}^{b_\gamma_b}$), the differential co-polarized ($AT_{co}^{differential} = T_{co}^{f_\gamma_f} - T_{full\rightarrow cross}^{\gamma_b\rightarrow \gamma_f}$) and cross-polarized ($AT_{cross}^{differential} = T_{cross}^{f_\gamma_f} - T_{full\rightarrow co}^{\gamma_b\rightarrow \gamma_f}$), and total ($AT_{full}^{differential} = T_{full}^{f_\gamma_f} - T_{full\rightarrow cross}^{\gamma_b\rightarrow \gamma_f}$) asymmetrical transmissions. The polarization rotations and ellipticity in forward ($rot_f$ and $\tau_f$) and backward ($rot_b$ and $\tau_b$) propagation directions, the cumulative rotation ($\Sigma$). The *FOMs* qualifying the Generalized Faraday Rotation (GFR) and Generalized Faraday Isolator (GFI) properties of the passive multilayers and the *Q*-factor of the backward transmission peak in case of wavelength-dependent calculations for the quantities inside the brackets.

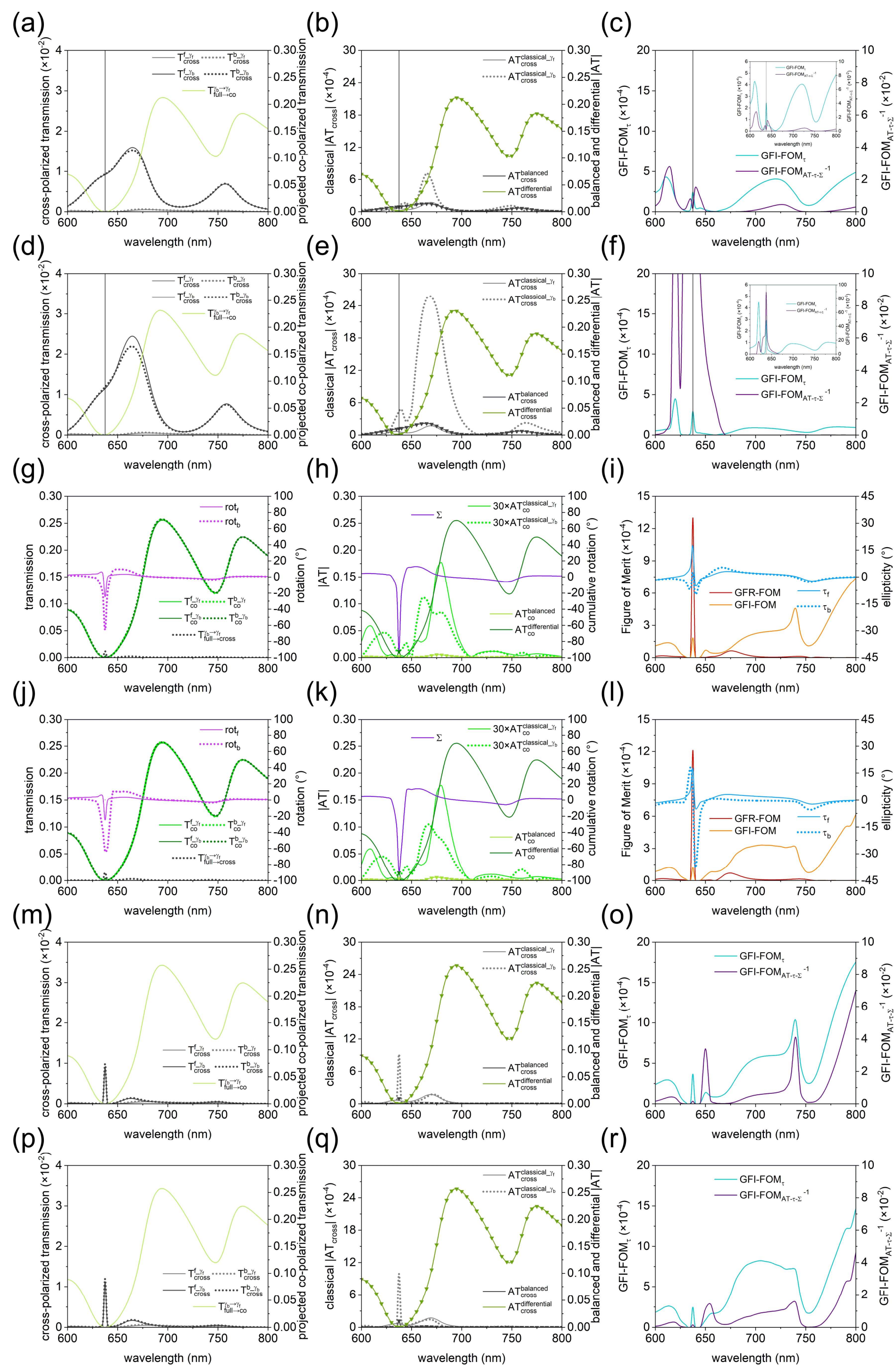

**Figure S1.** Optical responses of Pattern-I, transmission side. (a, d, m, p) Cross-polarized and (g, j) co-polarized transmission in forward and backward propagation direction, as well as the backward transmission projected onto the forward (a, d, m, p) co-polarized and (g, j) cross-polarized component, (b, e, n ,q) the classical, balanced and differential cross-polarized and (h, k) co-polarized asymmetric transmission (triangular symbols indicate regions, where the backward transmission is larger), (c, f, o, r) *GFI-FOM*$_\tau$ and *GFI-FOM*$_{AT\text{-}\tau\text{-}\Sigma}^{-1}$ and (i, l) *GFR-FOM* and *GFI-FOM* as well as the ellipticity of outgoing beams. (a-c, g-i, m-o) Linearly, (d-f, j-l, p-r) elliptically polarized light for backward illumination, (a-f) wavelength independent and (g-r) wavelength dependent configuration.

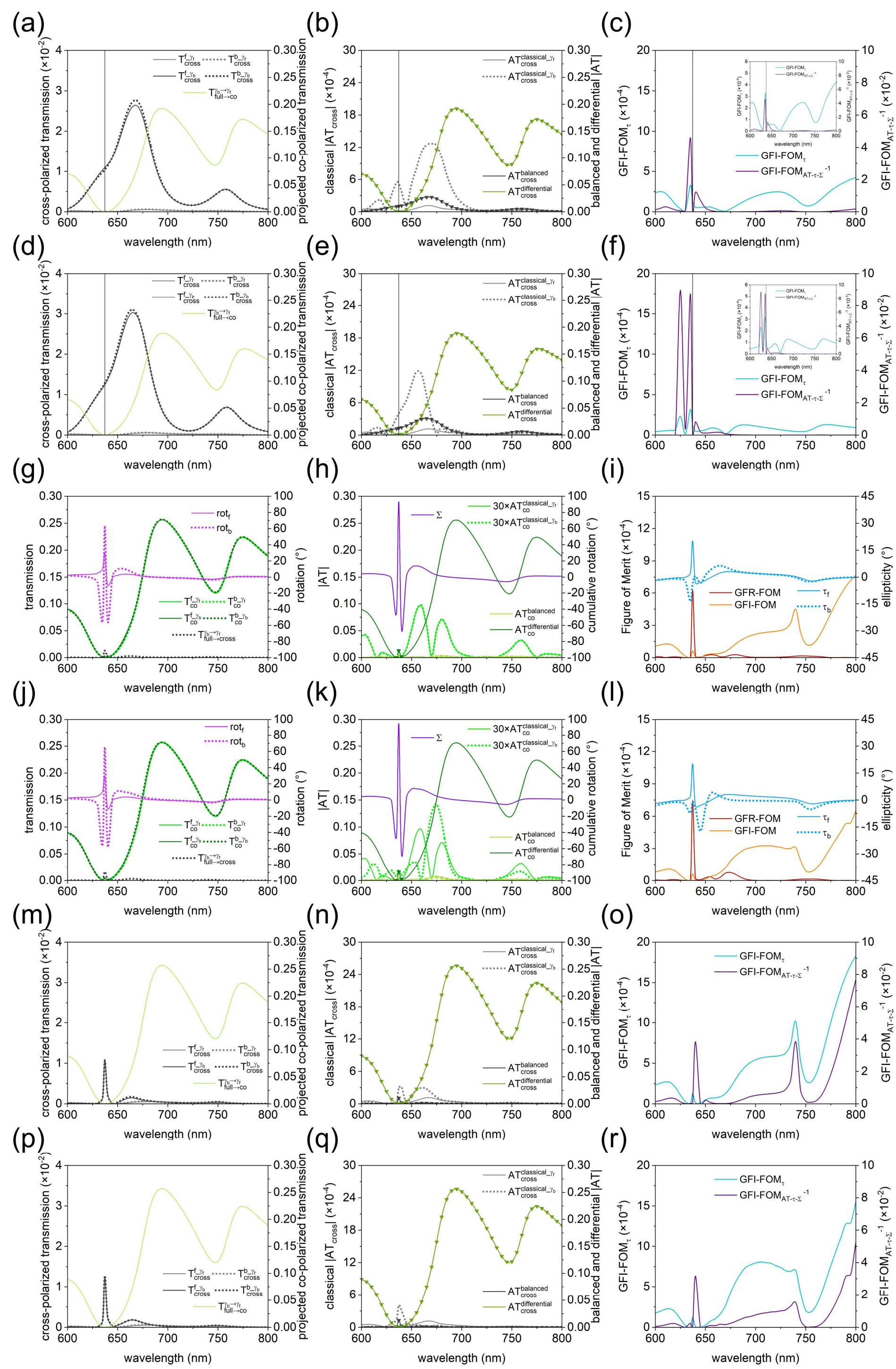

**Figure S2.** Optical responses of Pattern-II, transmission side. (a, d, m, p) Cross-polarized and (g, j) co-polarized transmission in forward and backward and propagation direction, as well as the backward transmission projected onto the forward (a, d, m, p) co-polarized and (g, j) cross-polarized component, (b, e, n, q) the classical, balanced and differential cross-polarized and (h, k) co-polarized asymmetric transmission (triangular symbols indicate regions, where backward transmission is larger), (c, f, o, r) $GFI\text{-}FOM_{\tau}$ and $GFI\text{-}FOM_{AT\text{-}\tau\text{-}\Sigma}^{-1}$ and (i, l) *GFR-FOM* and *GFI-FOM* as well as the ellipticity of outgoing beams. (a-c, g-i, m-o) Linearly, (d-f, j-l, p-r) elliptically polarized light for backward illumination, (a-f) wavelength independent and (g-r) wavelength dependent configuration.

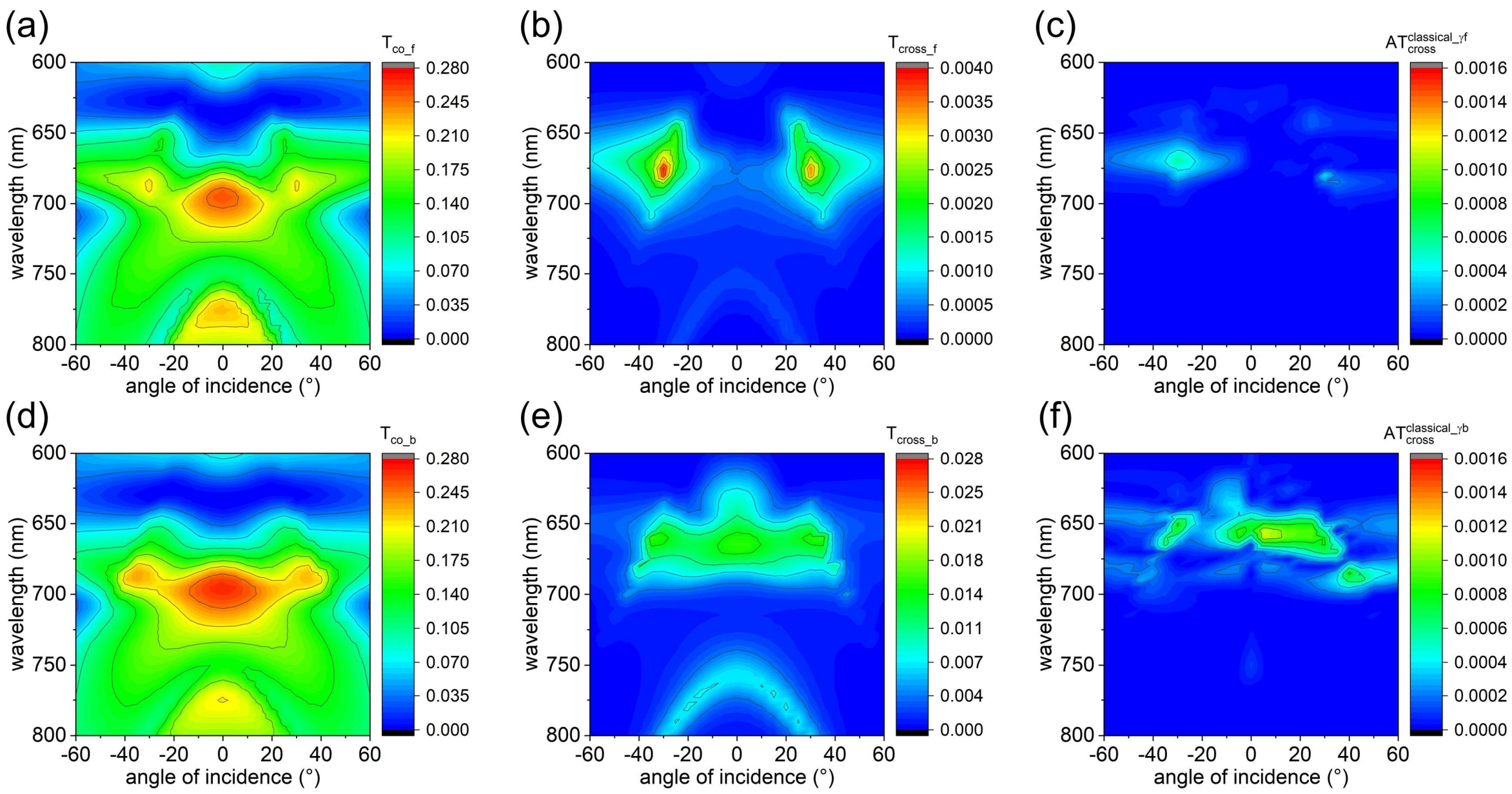

**Figure S3.** Dispersion characteristic of transmission polarization components and cross-polarized asymmetric transmission of Pattern-I. (a, d) Co-polarized transmission, (b, e) Cross-polarized transmission, (c, f) Classical cross-polarized asymmetric transmission; (a-c) forward, (d-f) backward propagation direction.

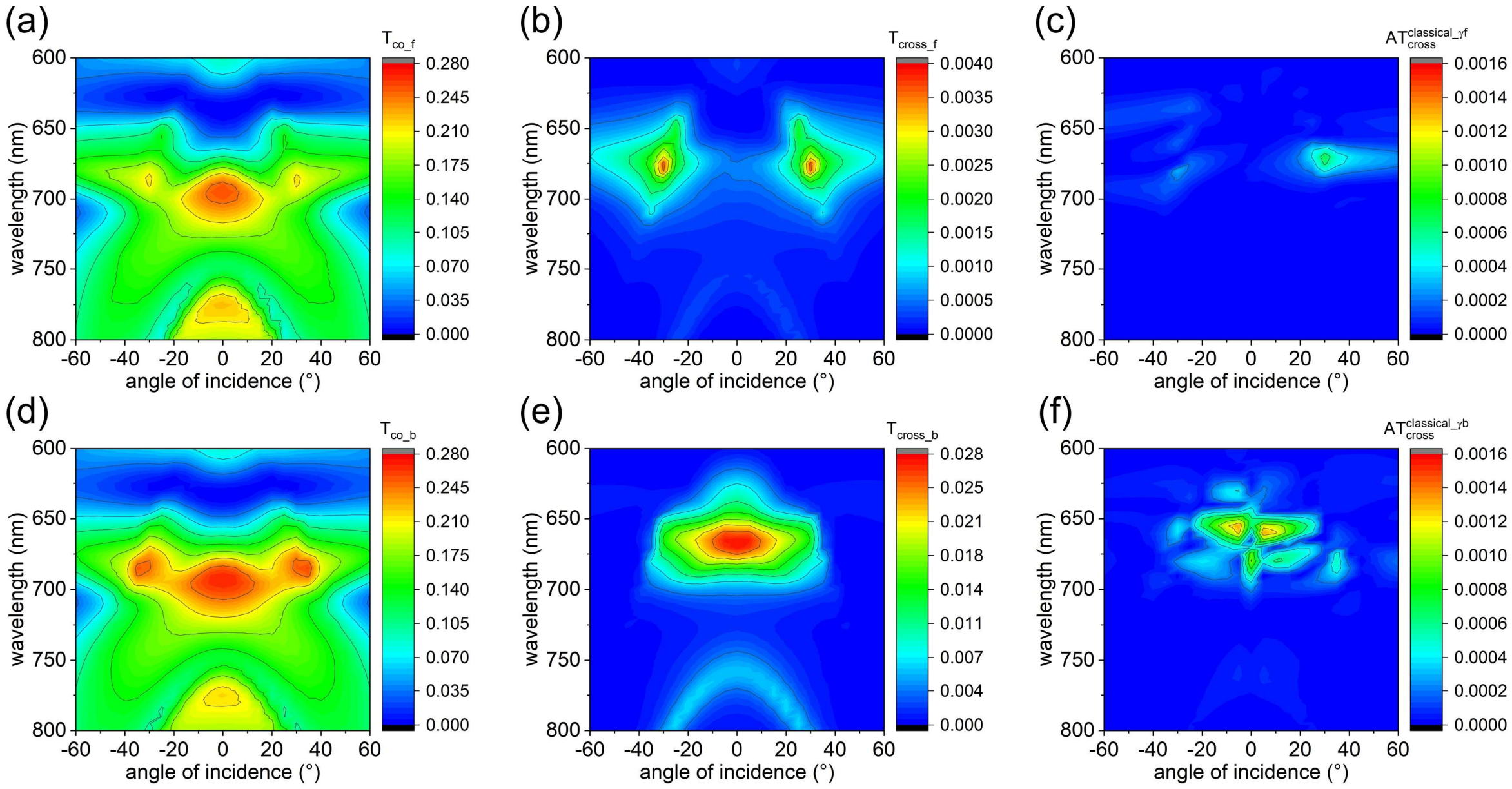

**Figure S4.** Dispersion characteristic of transmission polarization components and cross-polarized asymmetric transmission of Pattern-II. (a, d) Co-polarized transmission, (b, e) Cross-polarized transmission, (c, f) Classical cross-polarized asymmetric transmission; (a-c) forward, (d-f) backward propagation direction.

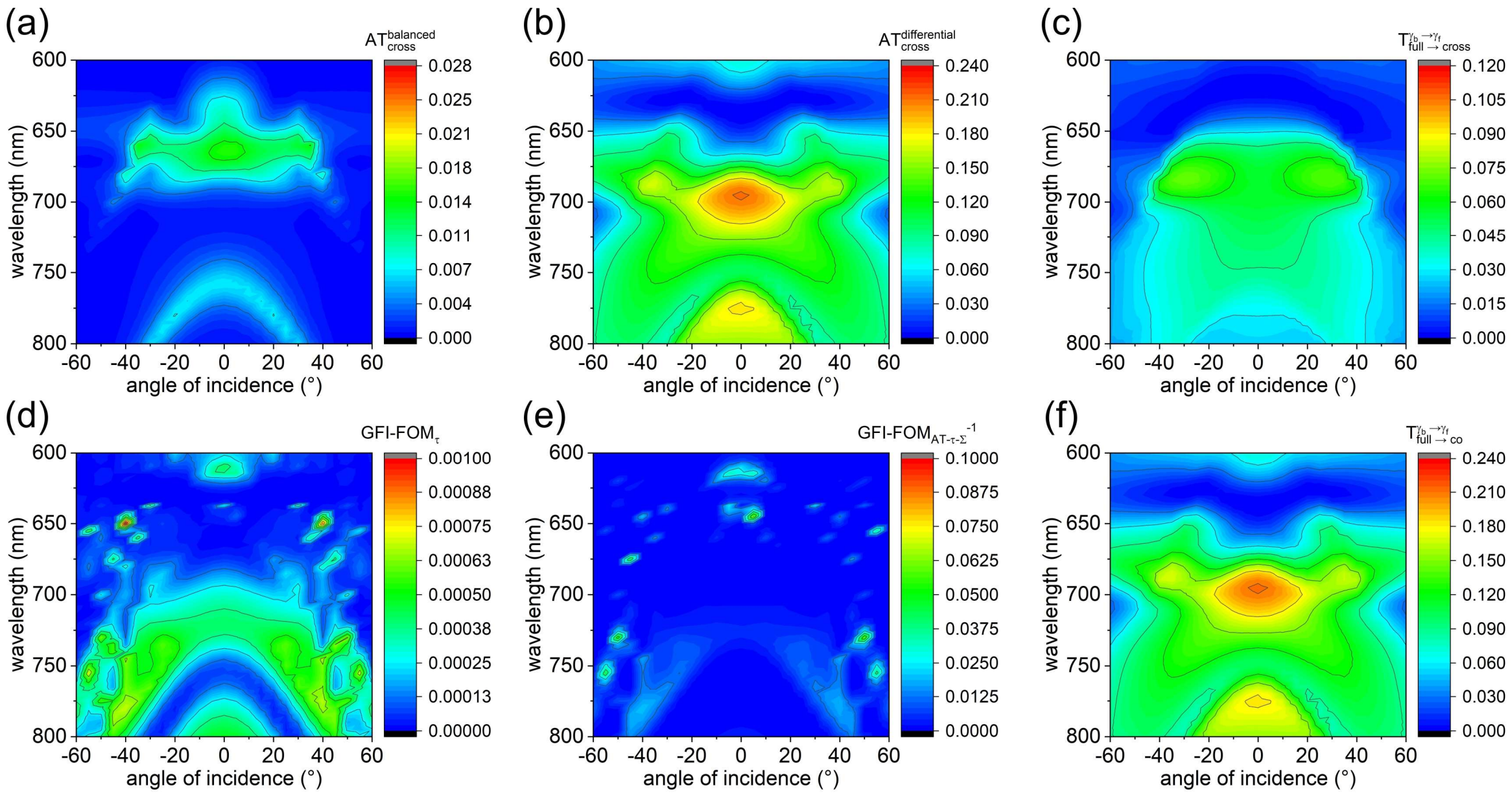

**Figure S5.** Dispersion characteristic of cross-polarized asymmetric transmissions, backward transmission projected onto forward bases and additional *FOMs* qualifying both propagation directions in GFR configuration of Pattern-I. (a) Balanced and (b) differential cross-polarized *AT*, (c/f) backward transmission projected onto cross/co-polarized component in forward bases, (d) $GFI\text{-}FOM_{\tau}$, (e) $GFI\text{-}FOM_{AT\text{-}\tau\text{-}\Sigma^{-1}}$.

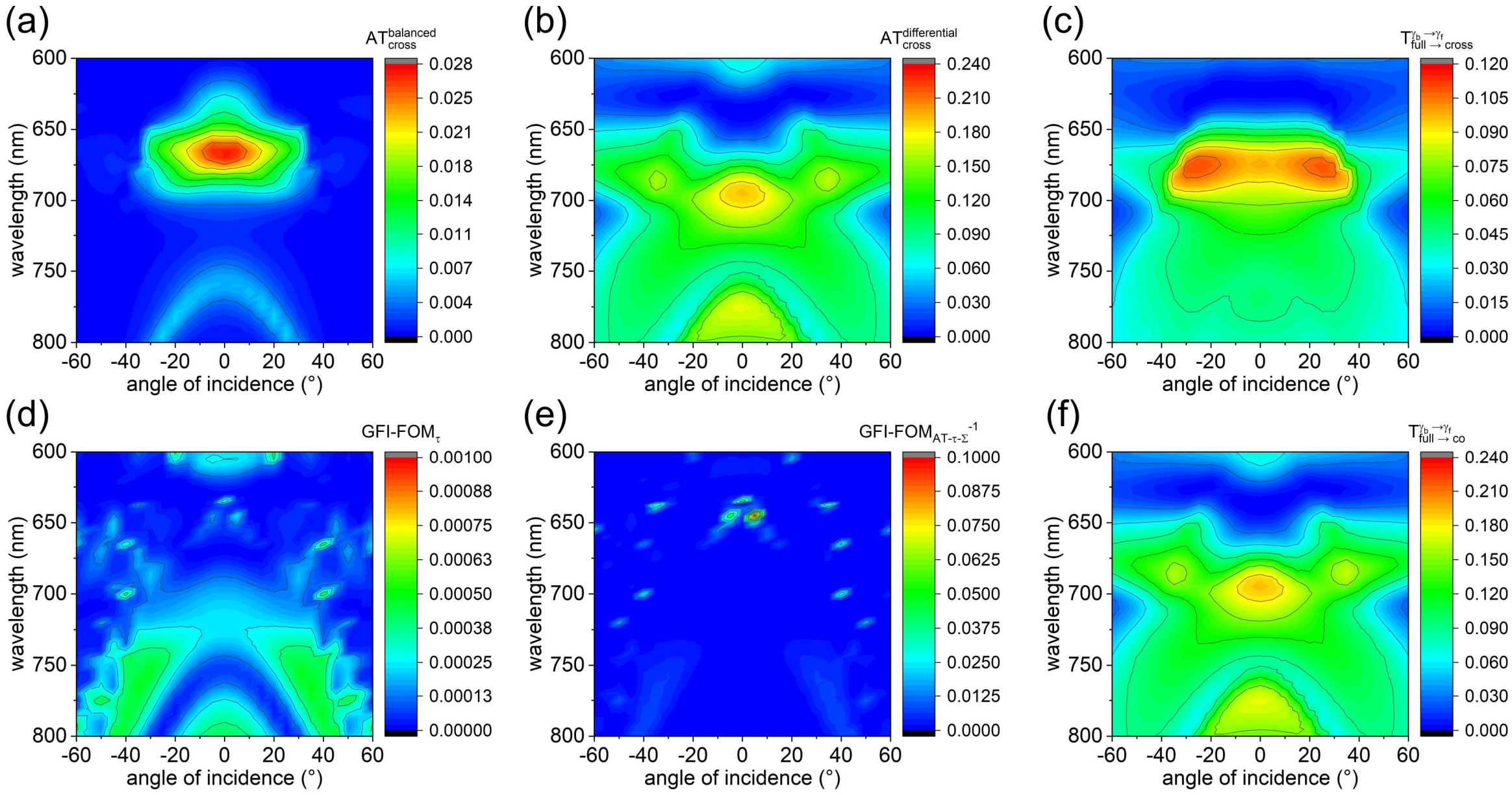

**Figure S6.** Dispersion characteristic of cross-polarized asymmetric transmissions, backward transmission projected onto forward bases and additional FOMs qualifying both propagation directions in GFR configuration of Pattern-II. (a) Balanced and (b) differential cross-polarized *AT*, (c/f) backward transmission projected onto cross/co-polarized component in forward bases, (d) $\text{GFI-FOM}_{\tau}$, (e) $\text{GFI-FOM}_{\text{AT-}\tau\text{-}\Sigma^{-1}}$.

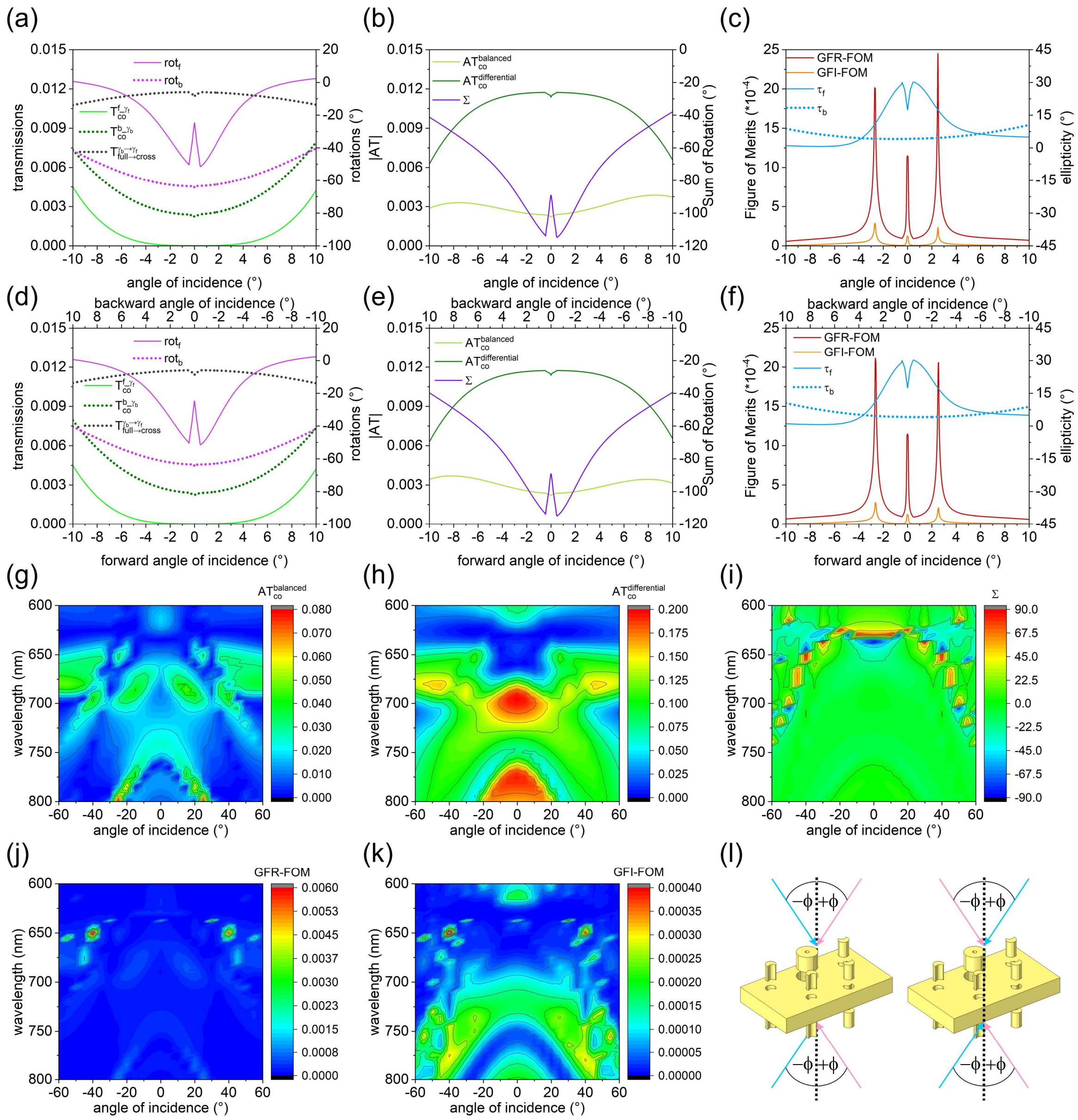

**Figure S7.** Side-dependent impact of tilting on optical responses of Pattern-I, transmission side. (a, d) Co-polarized transmission and polarization rotation in forward and backward propagation direction as well as the backward transmission projected onto the forward cross-polarized component, (b, e) cumulative rotation and the balanced and differential co-polarized asymmetric transmission, (c, f) ellipticity of outgoing beams, *GFR-FOM* and *GFI-FOM*. The dispersion characteristics of the (g) balanced and (h) differential asymmetric co-polarized transmission, the (i) cumulative rotation, the (j) *GFR-FOM* and (k) *GFI-FOM* in case of (a-c) corresponding-side and (g-j) opposite side tilting indicated on the (l) illumination schematics.

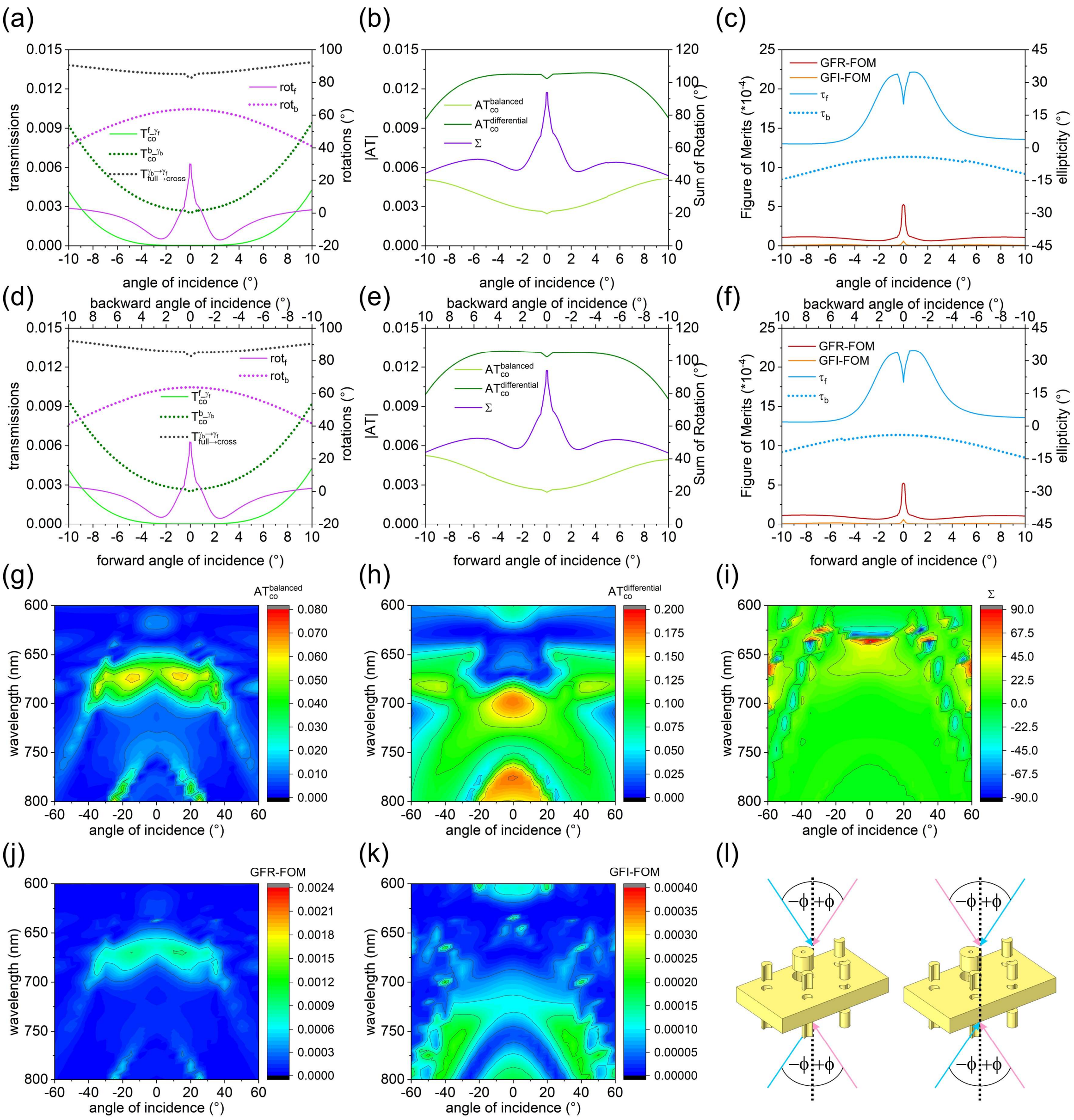

**Figure S8.** Side-dependent impact of tilting on optical responses of Pattern-II, transmission side. (a, d) Co-polarized transmission and polarization rotation in forward and backward propagation direction as well as the backward transmission projected onto the forward cross-polarized component, (b, e) cumulative rotation and the balanced and differential co-polarized asymmetric transmission, (c, f) ellipticity of outgoing beams, *GFR-FOM* and *GFI-FOM*. The dispersion characteristics of the (g) balanced and (h) differential asymmetric co-polarized transmission, the (i) cumulative rotation, the (j) *GFR-FOM* and (k) *GFI-FOM* in case of (a-c) corresponding-side and (g-j) opposite side tilting indicated on the (l) illumination schematics.

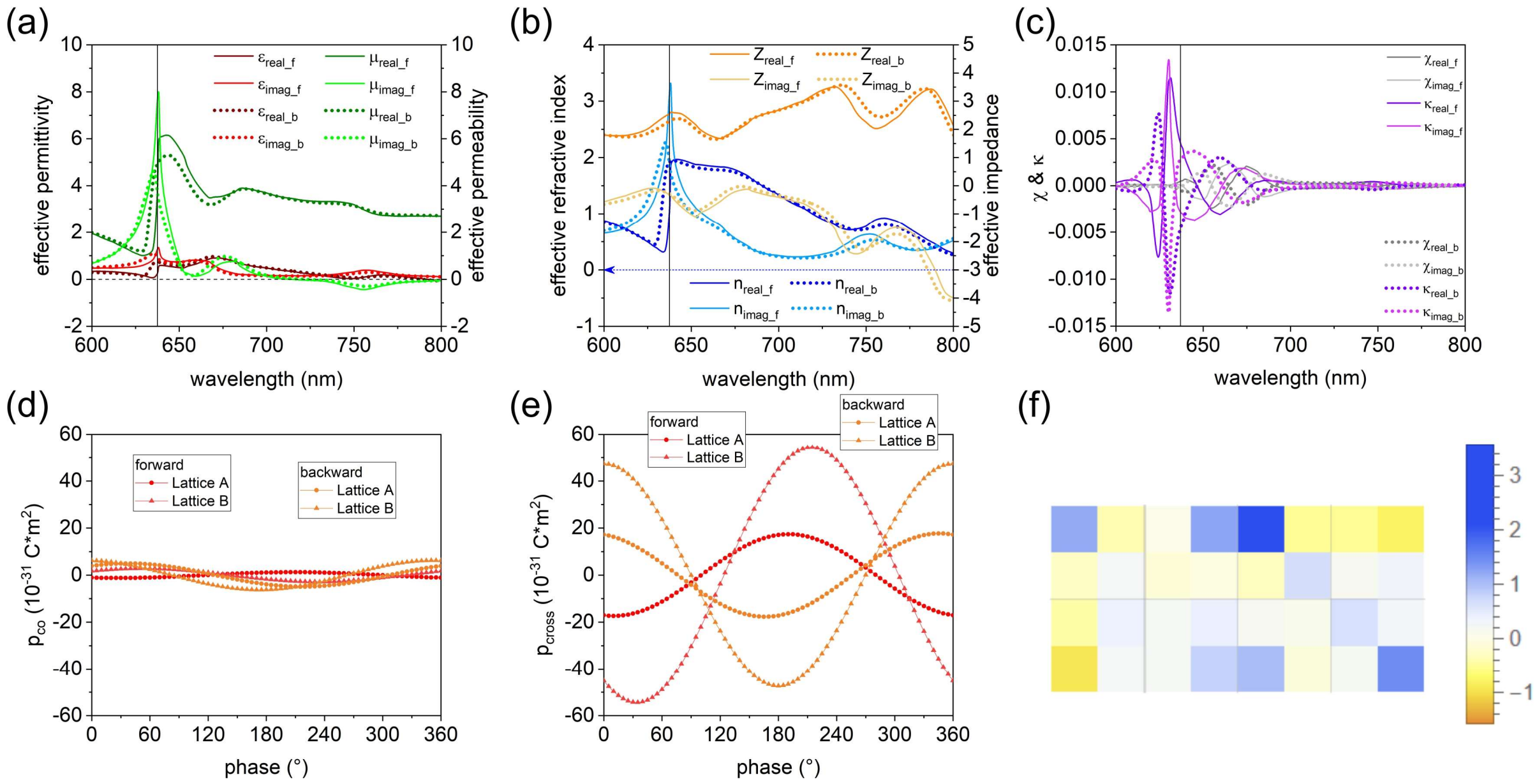

**Figure S9.** Dielectric properties of Pattern-I. (a) Dielectric permittivity and permeability; (b) index of refraction and impedance; (c) Tellegen and chirality parameter; (f) elements of the index of refraction matrix at the GFR wavelength. Emulated magnetic field related $\boldsymbol{p}_m$ dynamics on Pattern-I (d) co-polarized, (e) cross-polarized component, projected onto the GFR azimuthal orientations.

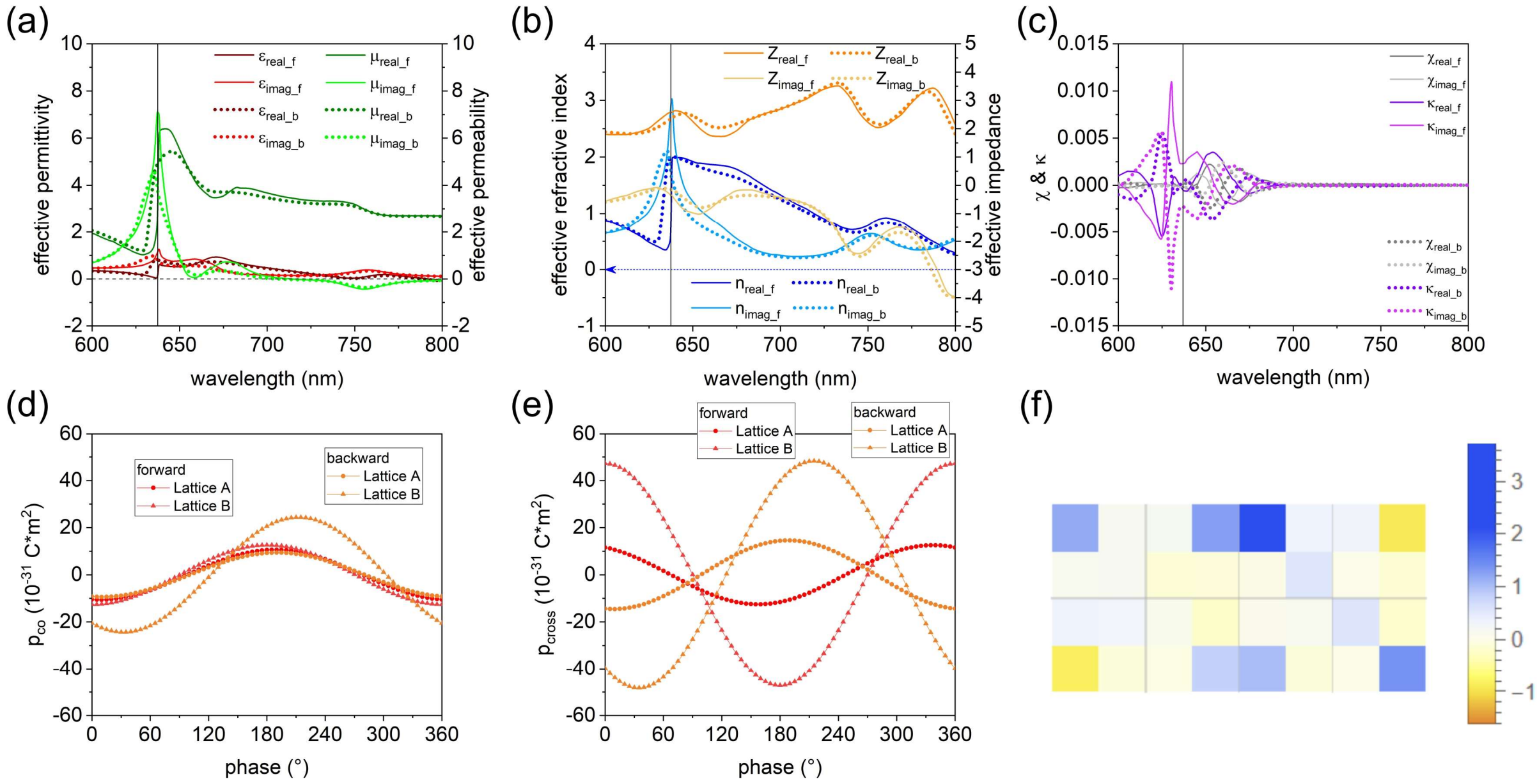

**Figure S10.** Dielectric properties of Pattern-II. (a) Dielectric permittivity and permeability; (b) index of refraction and impedance; (c) Tellegen and chirality parameter; (f) elements of the index of refraction matrix at the GFR wavelength. Emulated magnetic field related $\boldsymbol{p}_m$ dynamics on Pattern-II (d) co-polarized, (e) cross-polarized component, projected onto the GFR azimuthal orientations.

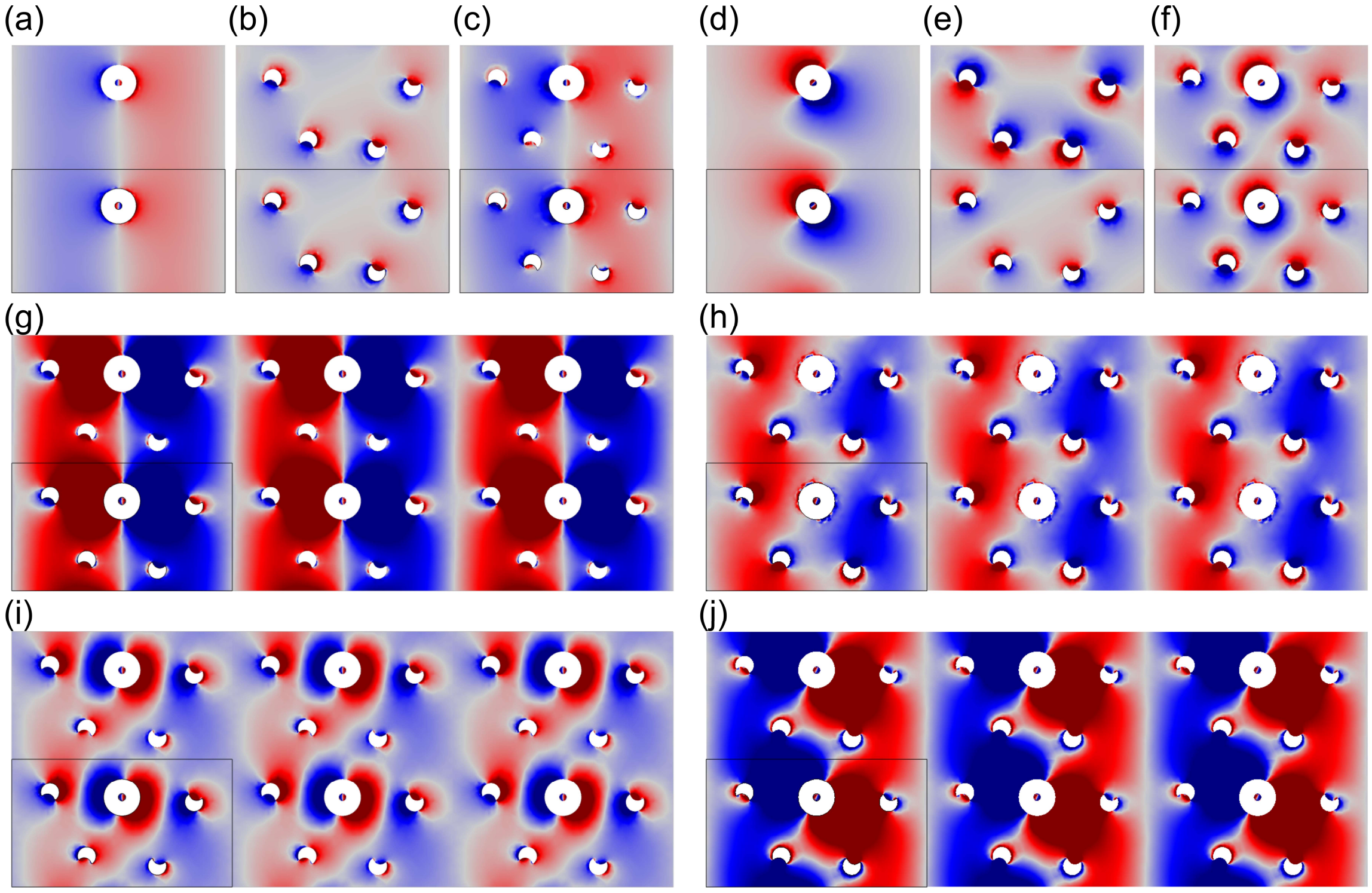

**Figure S11.** Time-evolution of the charge distribution on the periodic patterns of individual concave nanoresonators embedded into a gold monolayer. Concave (a, d) nanoring, (b, e) quadrumer, (c, f) miniarray embedded into gold monolayer (39 nm) in azimuthal orientation corresponding to (a-c) forward and (d-f) backward GFR configuration of Pattern-I (Video S1, S2). The charge distribution on the (g, h) top and (I, j) bottom surface of the concave miniarray pattern embedded into multilayer Pattern-I in (g, i) forward and (h, j) backward configuration (Video S3-S6).

### Field enhancements and phase distributions

The near-field enhancement (*NFE*), as well as the volume (*V*) corresponding to a decrease by a factor of 1/e and their product ($NFE\times V$) was also determined and compared for both the electric and magnetic fields (Table S2).

The comparison of *NFE* in Pattern-I and Pattern-II for forward and backward propagation directions was performed. The coupled modes on the multilayers result in: (i) a slightly larger $V_{NFE}$ in the forward propagation direction (except for the ***B***-field in Pattern-I, where it is the same), (ii) slightly larger average ***E*** and ***B*** fields and $NFE_{E\&B}$ values in the backward propagation direction for both patterns, which correlate with the larger polarization rotation and transmission, as well as with the larger classical co-polarized *AT* in Pattern-II, and (iii) a slightly larger $NFE\times V$ for ***E***-field in the forward /backward propagation directions for Pattern-I/II, while the same $NFE\times V$ is reached for ***B***-field in both patterns, regardless of propagation direction.

### Field enhancements

Comparison of Pattern-I and Pattern-II shows that the *V* is larger /the same for the ***E* /*B*-field** in Pattern-II compared to Pattern-I, in both propagation directions, except for the ***B***-field in the backward propagation direction, which is smaller in Pattern-II. The *NFE* is almost the same, it is slightly larger for both ***E*** and ***B***-fields in Pattern-I in the forward propagation direction, while larger *NFE* is reached in Pattern-II both for ***E*** and ***B***-fields in the backward propagation direction.

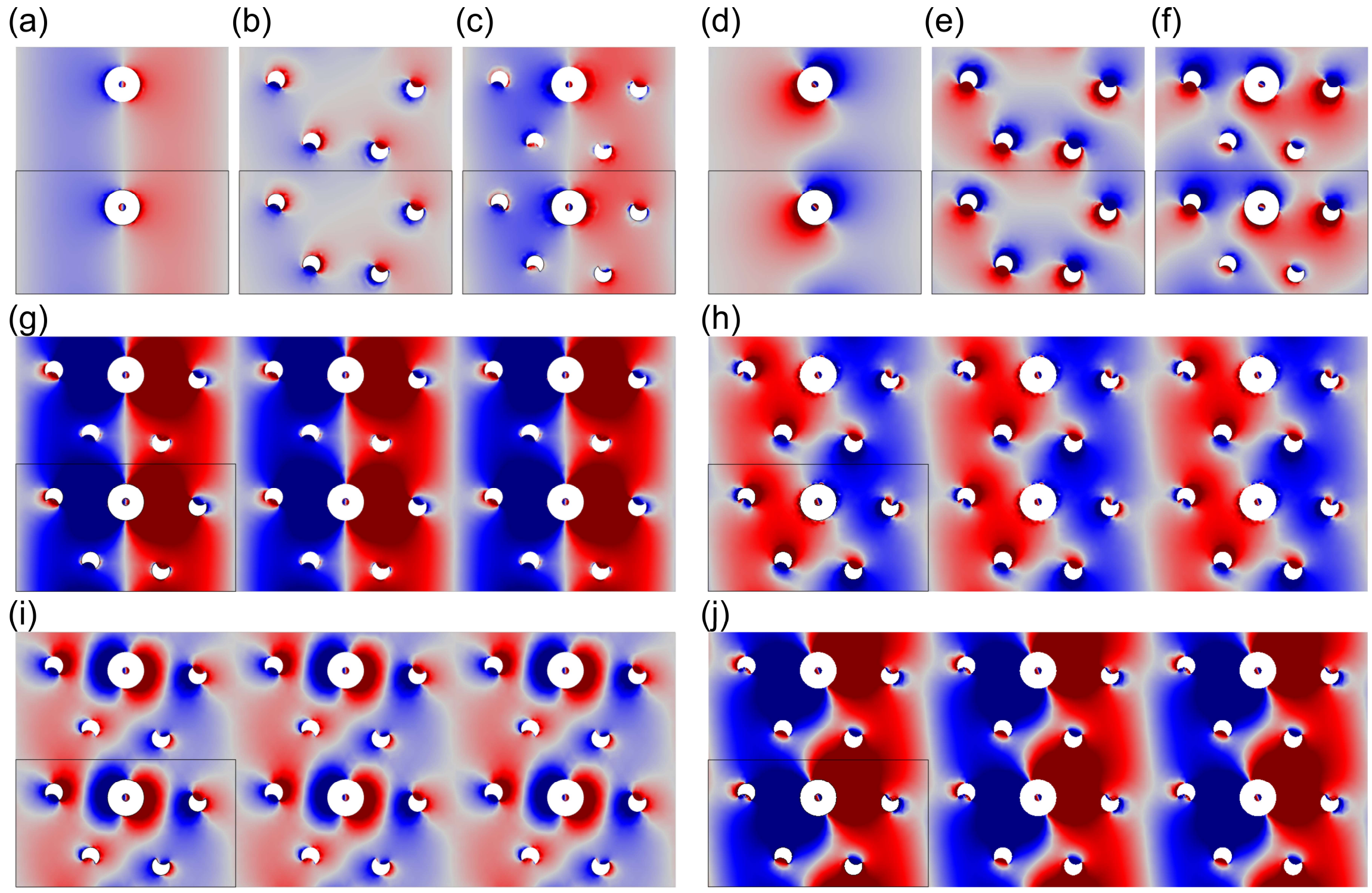

**Figure S12.** Time-evolution of the charge distribution on the periodic patterns of individual concave nanoresonators embedded into a gold monolayer. Concave (a, d) nanoring, (b, e) quadrumer, (c, f) miniarray embedded into gold monolayers (39 nm) in azimuthal orientation corresponding to (a-c) forward and (d-f) backward GFR configuration of Pattern-II (Video S7, S8). The charge distribution on the (g, h) top and (I, j) bottom surface of the concave miniarray pattern embedded into multilayer Pattern-II in (g, i) forward and (h, j) backward configuration (Video S9-S12).

The latter correlates with the larger transmission, as well as with the larger classical *AT* in Pattern-II. The *NFE* and $NFE \times V$ are slightly /more pronouncedly propagation direction dependent for both ***E*** and ***B*** fields in Pattern-I/II, which correlates with the larger classical *AT* (except for elliptically polarized illumination in forward propagation direction), as well as with the larger rotation (except for linearly polarized illumination in forward propagation direction) in Pattern-II (Table S1 and S2).

| | Pattern-I | | Pattern-II | |
|---|---|---|---|---|
| | forward | backward | forward | backward |
| $\mathbf{E}_{avg}$ ($\times 10^{8}$ V m$^{-1}$) | 1.36 | 1.37 | 1.36 | 1.39 |
| $\mathbf{B}_{avg}$ (T) | 0.41 | 0.42 | 0.41 | 0.42 |
| $NFE_{E_avg}$ | 1.39 | 1.40 | 1.39 | 1.43 |
| $NFE_{B_avg}$ | 1.65 | 1.69 | 1.64 | 1.70 |
| $V_{E_enh}$ ($\times 10^{-21}$ m$^{-3}$) | 3.11 | 3.00 | 3.12 | 3.03 |
| $V_{B_enh}$ ($\times 10^{-21}$ m$^{-3}$) | 3.89 | 3.89 | 3.89 | 3.86 |
| $NFE \times V_{E_enh}$ ($\times 10^{-21}$ m$^{-3}$) | 7.55 | 7.48 | 7.51 | 7.68 |
| $NFE \times V_{B_enh}$ ($\times 10^{-21}$ m$^{-3}$) | 10.1 | 10.1 | 10.1 | 10.1 |
| $NFE_{E_enh}$ | 2.43 | 2.50 | 2.41 | 2.54 |
| $NFE_{B_enh}$ | 2.59 | 2.60 | 2.59 | 2.62 |

**Table S2.** Field enhancements, *NFE* volumes, $NFE \times V$. The volume average of the electric ($\boldsymbol{E}_{avg}$) and magnetic ($\boldsymbol{B}_{avg}$) field and the average electric ($NFE_{E_avg}$) and magnetic ($NFE_{B_avg}$) near-field enhancement around and inside the multilayer. The volume, quantifying the region where the electric and magnetic field decreases by a factor of $e^{-1}$. The product of the near-field enhancement and the enhanced volume ($NFE \times V$) and the near-field enhancement inside the enhanced volume ($NFE_{E_enh}$ and $NFE_{B_enh}$).

## Emulated magnetic field, related vorticity and helicity

The existence of power-flow vortices indicates broken time reversal symmetry [2], while the helicity qualifies the difference between the numbers of right- and left-handed photons and represents the magneto-electric energy density [3-5].

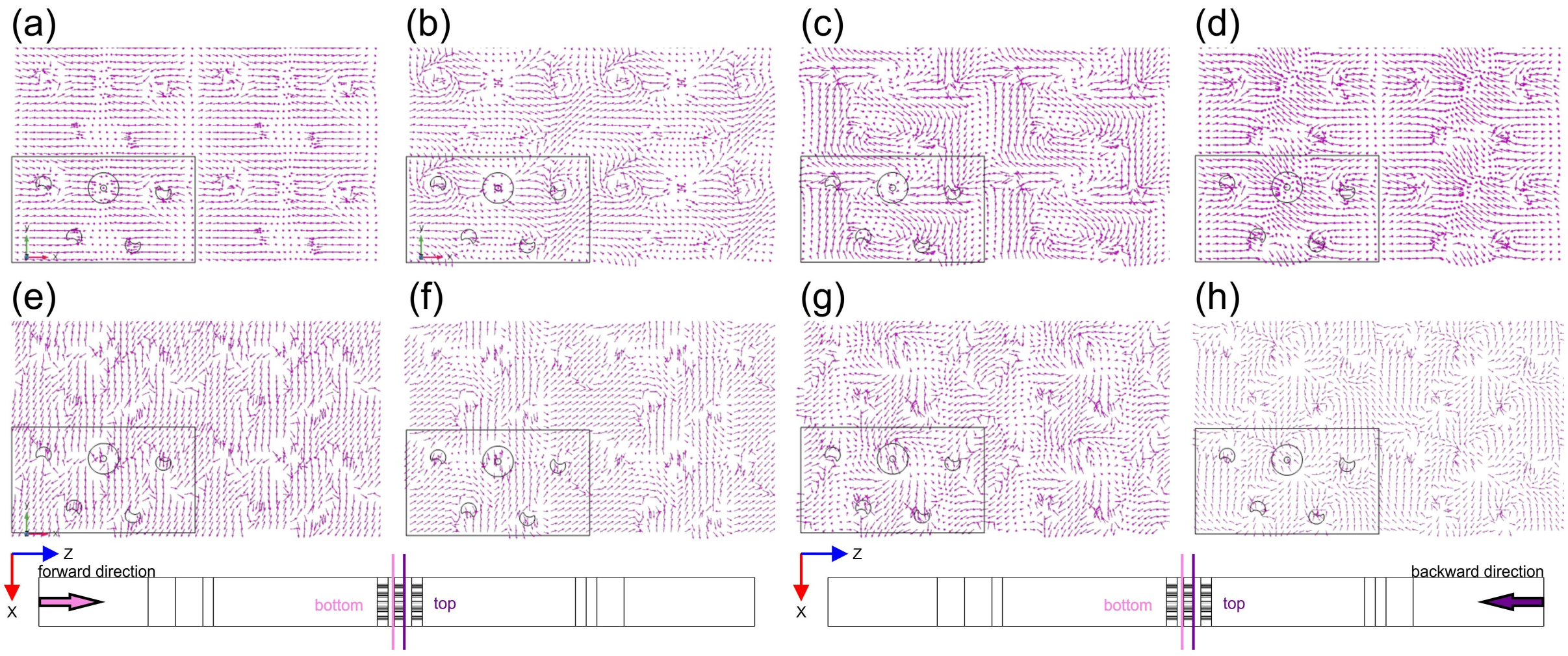

**Figure S13.** Vorticity and helicity on the concave layer surfaces embedded into Pattern-I. (a-d) Vorticity and (e-h) helicity; (a, b and e, f) forward, (c, d and g, h) backward propagation direction; on (a, c and e, g) bottom and (b, d and f, h) top surfaces. Inset about the forward and backward illumination configurations.

Intermediate and weaker /stronger vortices are recognizable around the nanocrescents in forward propagation direction, when Pattern-I and Pattern-II are compared. These vortices are stronger and more localized on the leaving (top) concave layer surfaces, especially at the nanocrescent tips on Pattern-II (Fig. S13a,b and Fig. S14a,b).

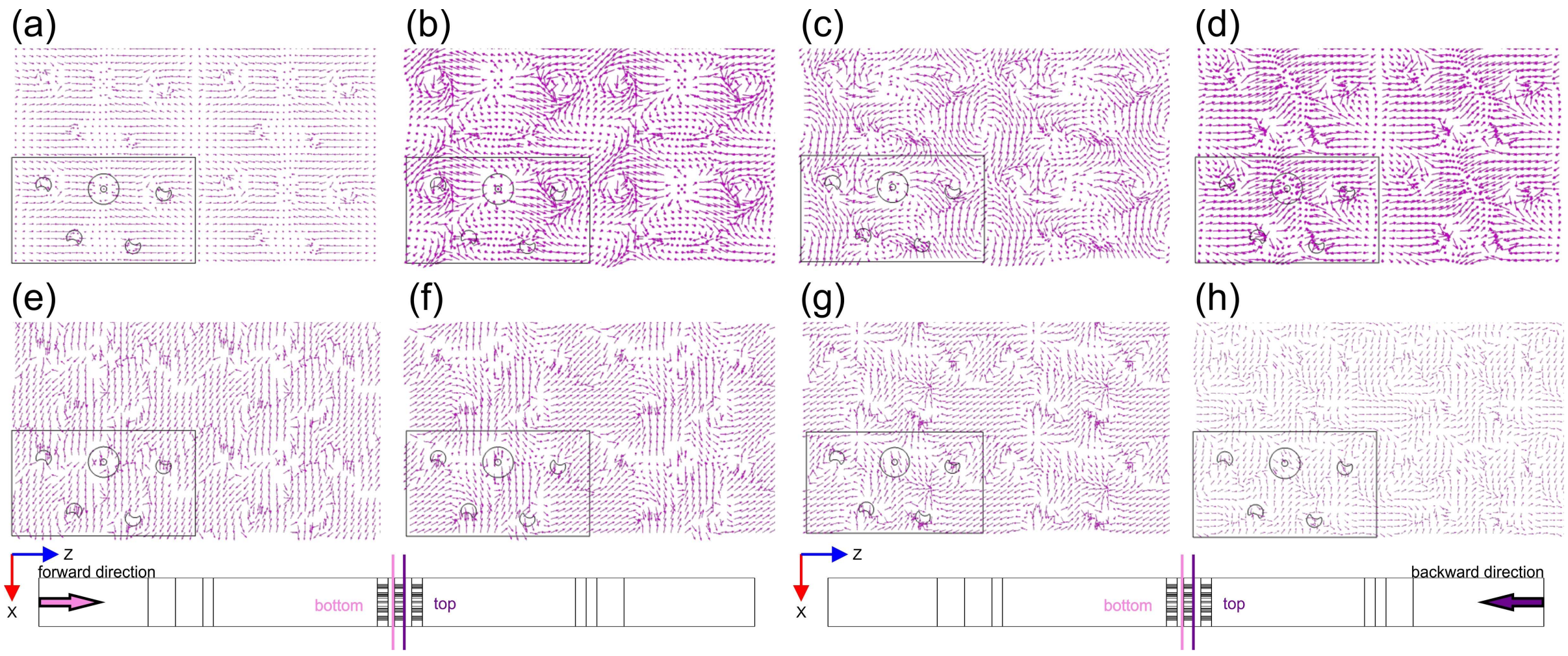

**Figure S14.** Vorticity and helicity on the concave layer surfaces embedded into of Pattern-II. (a-d) Vorticity and (e-h) helicity; (a, b and e, f) forward, (c, d and g, h) backward propagation direction; on (a, c and e, g) bottom and (b, d and f, h) top surfaces. Inset about the forward and backward illumination configurations.

Drain-source-like patterns are recognizable on the A-B sub-lattices on the entering (top) surface and strong vortices appear below the rings on the leaving (bottom) surface in backward direction. Remarkable difference is that additional vortices develop in B sub-lattice on the leaving (bottom) concave layer surface of Pattern-II (Fig. S13c,d and Fig. S14c,d).

The helicity is also well structured, four radial segments are distinguishable around the A sub-lattice, while in the B sub-lattice well-defined separated areas of tilted helicity lines appears in forward propagation direction (Fig. S13e,f and Fig. S14e,f). In comparison, in Pattern-I and Pattern-II six radial segments are distinguishable around the A and B sub-lattice in backward propagation direction. An area of tilted helicity lines appears inside the B-lattice, which is of continuously varying orientation as in the A sub-lattice /contains drain-source-like patterns on the entering (top) concave layer surface in Pattern-I/II, and a well-defined separated areas of tilted helicity lines appears on the leaving (bottom) concave layer surface, which is more extended around the neighbouring nanocrescents in Pattern-II (Fig. S13g,h and Fig. S14g,h).

**Induced magnetic dipole –Pattern-I and Pattern-II: comparison of A and B sub-lattices**

| | Pattern-I | | | | Pattern-II | | | |
|---|---|---|---|---|---|---|---|---|
| | lattice-A | | lattice-B | | lattice-A | | lattice-B | |
| | forward | backward | forward | backward | forward | backward | forward | backward |
| $p_{amp_avg}(\times 10^{-30} Cm^2)$ | 1.12 | 1.22 | 3.52 | 3.11 | 1.10 | 1.11 | 3.14 | 3.49 |
| $p_{amp_max}(\times 10^{-30} Cm^2)$ | 1.75 | 1.82 | 5.52 | 4.88 | 1.59 | 1.74 | 4.92 | 5.48 |
| $p_{xy_avg}(\times 10^{-30} Cm^2)$ | 1.11 | 1.21 | 3.47 | 3.04 | 1.09 | 1.11 | 3.11 | 3.45 |
| $p_{xy_avg}(\times 10^{-30} Cm^2)$ | 1.11 | 1.21 | 3.47 | 3.04 | 1.09 | 1.11 | 3.11 | 3.45 |
| $p_{z_avg}(\times 10^{-31} Cm^2)$ | 1.18 | 1.73 | 5.84 | 6.37 | 0.448 | 1.16 | 3.71 | 5.79 |
| $p_{z_max}(\times 10^{-31} Cm^2)$ | 1.86 | 2.71 | 9.17 | 9.98 | 0.705 | 1.82 | 5.82 | 9.09 |
| $p_{co_avg}(\times 10^{-33} Cm^2)$ | -0.301 | 1.09 | 0.476 | 1.72 | -2.91 | -2.56 | -3.49 | -5.69 |
| $p_{co_max}(\times 10^{-30} Cm^2)$ | 0.123 | 0.493 | 0.282 | 0.624 | 1.06 | 0.94 | 1.26 | 2.44 |
| $p_{cross_avg}(\times 10^{-32} Cm^2)$ | -0.473 | 0.476 | -1.25 | 1.31 | 0.321 | -0.399 | 1.31 | -1.10 |
| $p_{cross_max}(\times 10^{-30} Cm^2)$ | 1.73 | 1.77 | 5.44 | 4.73 | 1.26 | 1.46 | 4.72 | 4.83 |
| $\theta_{avg}$ (°) | 6.09 | 7.59 | 9.59 | 11.72 | 3.22 | 6.05 | 6.81 | 9.59 |
| $\theta_{max}$ (°) | 11.11 | 8.60 | 18.56 | 13.11 | 8.85 | 14.85 | 9.73 | 19.06 |

**Table S3.** Magnetic dipole characteristics. The average and maximal magnetic dipole amplitude ($p_{amp_avg\&max}$), in-plane projected ($p_{xy_avg\&max}$), along propagation direction ($p_{z_avg\&max}$), co-polarized ($p_{co_avg\&max}$) and cross-polarized ($p_{cross_avg\&max}$) components, and inclination ($\Theta_{avg\&max}$) measured from the xy-plane of the magnetic dipole.

The $p_{xy}$ components, as well as the inclinations of the $\boldsymbol{p}_m$ corresponding to the A and B sub-lattices are in intermediate ($\delta\phi$~30°) (almost the same) phase in forward propagation direction, while in backward propagation direction the $p_{xy}$ components, as well as the inclination are in the same (intermediate ($\delta\phi$~30°)) phase in Pattern-I (Pattern-II) (Fig. 7k,m and Fig. 8k,m). The $p_z$ components in the A and B sub-lattices of Pattern-I are also in intermediate ($\delta\phi$~30°) and almost in-phase in forward and backward propagation directions, similar to the $p_{xy}$ (Fig. 7l). In contrast, the $p_z$ components in the A and B sub-lattices in Pattern-II exhibit different phase tendencies, namely the A and B sub-lattices show slightly larger ($\delta\phi$~60°) and intermediate ($\delta\phi$~30°) phase difference in forward and backward propagation directions, respectively (Fig. 8l). The phase shift is intermediate ($\delta\phi$~180°+/-30°) between the forward and backward propagation directions for both of the A and B sub-lattices, except the $p_{xy}$ in A sub-lattice of Pattern-I and Pattern-II, which are in opposite phases; and the $p_z$ in A sub-lattice of Pattern-II, which shows larger phase difference ($\delta\phi$~180°+/-30°) (Fig. 7k-m, Fig. 8k-m, Table S3).

## Correlation between displacement and optical response asymmetry on the multilayers

| | | Pattern-I | | Pattern-II | |
|---|---|---|---|---|---|
| | γ (°) | 95.39 | 120.06 | 95.23 | 65.62 |
| 2-sided, linear, $D_z$ asym, cx | $D_{top}$ (C) ($\times 10^{-17}$) | 6.56 | 6.10 | 6.70 | 7.60 |
| | $D_{bottom}$ (C) ($\times 10^{-17}$) | 6.29 | 6.21 | 6.56 | 6.76* |
| | ΔD (C) ($\times 10^{-17}$) | 0.28 | -0.11 | 0.14 | 0.84 |
| | δD (%) | 4.28 | -1.70 | 2.04 | 11.75 |
| 2-sided, linear, $D_z$ asym, cv | $D_{top}$ (C) ($\times 10^{-16}$) | 1.98 | 2.06 | 1.97 | 1.77 |
| | $D_{bottom}$ (C) ($\times 10^{-16}$) | 1.75 | 1.92 | 1.75 | 1.60 |
| | ΔD (C) ($\times 10^{-17}$) | 2.32 | 1.48 | 2.21 | 1.73 |
| | δD (%) | 12.47 | 7.43 | 11.92 | 10.29 |
| 2-sided, linear, outflow asym | $out_{top}$ ($\times 10^{-2}$) | 6.5 | 4.9 | 6.5 | 7.7 |
| | $out_{bottom}$ ($\times 10^{-1}$) | 1.14 | 0.83 | 1.13 | 1.37 |
| | Δout ($\times 10^{-2}$) | -5.0 | -3.5 | -4.8 | -6.0 |
| | δout (%) | -55.6 | -52.3 | -53.8 | -55.9 |
| 1-sided asymmetry | rot (°) | -24.67 | -64.41 | 29.60 | 64.05 |
| | $AT_{co}^{classical}$ ($\times 10^{-5}$) | 0.948 | 1.81 | 5.47 | 2.49 |
| | $AT_{cross}^{classical}$ ($\times 10^{-4}$) | 0.611 | 1.16 | 0.122 | 5.29 |

**Table S4.** Correlation between the normal component of the displacement current - $D_z$ and optical responses asymmetry. The $\gamma$ azimuthal orientation, normal component of the displacement current vector on the top ($D_{top}$) and bottom ($D_{bottom}$) surface of the concave (cv) and convex (cx) layers in case of two-sided linearly polarized plane-wave illumination. The displacement current difference ($\Delta D = D_{top} - D_{bottom}$) and asymmetry ratio ($\delta D = \frac{2*(D_{top}-D_{bottom})}{D_{top}+D_{bottom}}\times 100$). The outgoing intensity on the top ($out_{top}$) and bottom ($out_{bottom}$) regions, their difference ($\Delta out = out_{top} - out_{bottom}$) and asymmetry ratio ($\delta out = \frac{2*(out_{top}-out_{bottom})}{out_{top}+out_{bottom}}\times 100$). The polarization rotation (*rot*), classical co-polarized ($AT_{co}^{classical}$) and cross-polarized ($AT_{cross}^{classical}$) asymmetric transmission (*AT*).

In Pattern-I at the GFR condition no well-defined correlation is observed between the normal component of the displacement vector and the rotation, the classical asymmetrical co-polarized transmission, either on the convex or on the concave layer. The asymmetry of the normal component of the displacement vector correlates solely with the |outflow| observed under two-sided illumination on both layers ($\delta D_{z_convex\&concave}$~outflow), which is larger in forward configuration, though the outflow is reversal. However, when the maximum values of the classical asymmetrical co-polarised transmission $AT_{co}^{classical}$ are inspected, the ratio is reversed for the two propagation directions for $AT_{co}^{classical}$ as well as for $\delta D_{z_convex}$ of Pattern-I. As a consequence, on the concave layer the classical asymmetrical co-polarised transmission shows correlation with the asymmetry of the normal component displacement vector ($\delta D_{z_concave} \sim AT_{co}^{classical}$) in the forward configuration, hence Pattern-I is similar to Pattern-II (Table S4).

In Pattern-II the rotation, the $AT_{cross}^{classical}$ /outflow under single /two sided illumination, correlates with the asymmetry of the normal component of the displacement vector on the convex layer ($\delta D_{z_convex} \sim \beta - \alpha$, $AT_{cross}^{classical}$/outflow), namely all of them are larger in backward configuration, however the outflow is reversal. In comparison, on the concave layer the classical asymmetrical co-polarised transmission for single-sided illumination shows correlation with the asymmetry of the normal component displacement vector ($\delta D_{z_concave} \sim AT_{co}^{classical}$), namely it is larger in the forward configuration (Table S4).

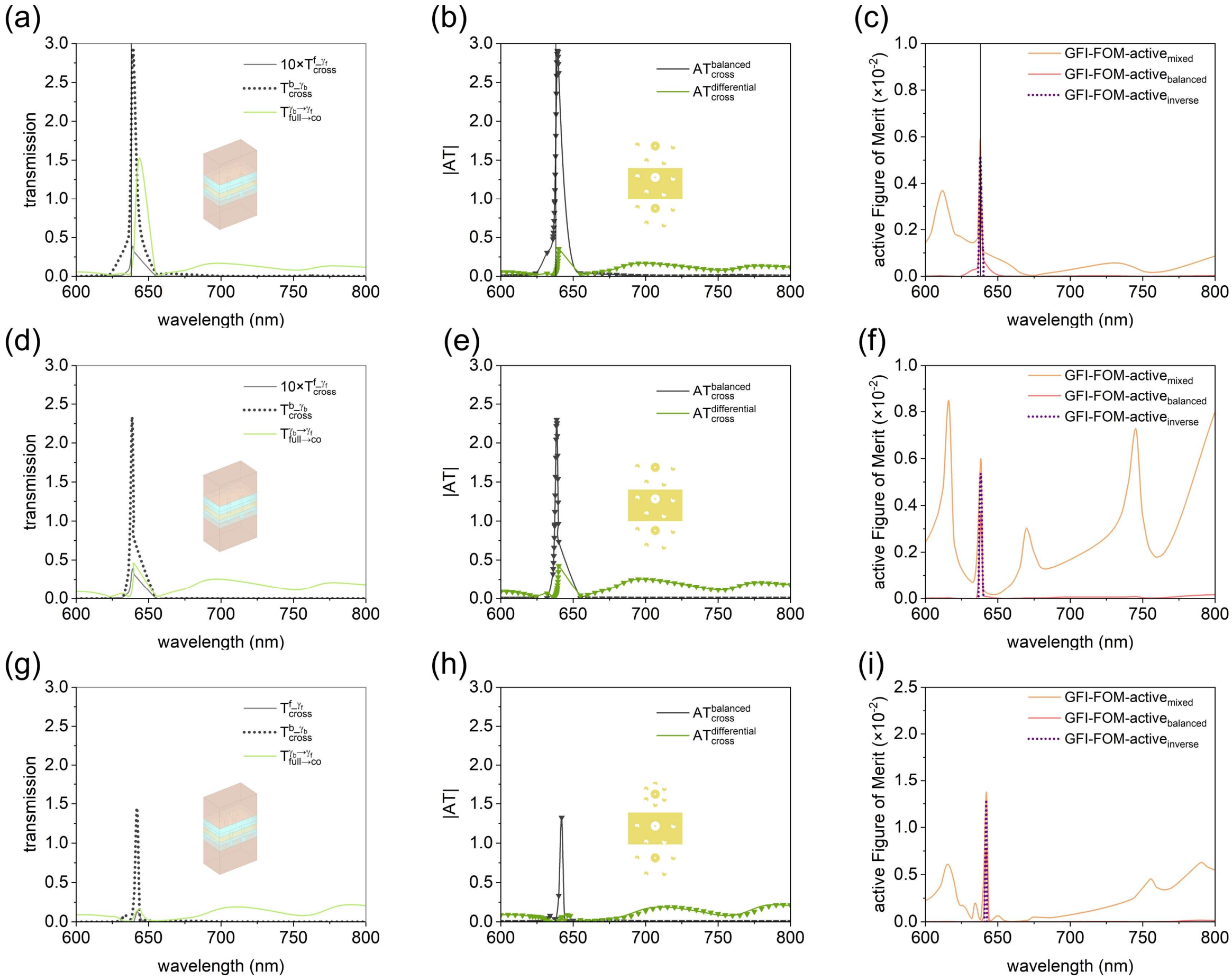

**Figure S15.** Optical responses of Pattern-II-active, transmission side. (a, d, g) Cross-polarized transmission in forward and backward propagation direction as well as the backward transmission projected onto the forward co-polarized component, (b, e, h) balanced and differential cross-polarized asymmetric transmission (triangular symbols indicate regions, where the backward transmission is larger), (c, f, i) *GFI-FOM-active*$_{mixed}$, *GFI-FOM-active*$_{balanced}$ and *GFI-FOM-active*$_{inverse}$. (a-c) Wavelength independent, (d-f) wavelength dependent active Pattern-II and (g-i) wavelength dependent active dimerized Pattern-II configuration.

| | Pattern-II-active | Pattern-II-dimerized |
|---|---|---|
| $T_{co}^{f_\gamma f}$ ($\times 10^{-2}$) | 4.71 | 0.88 |
| $T_{cross}^{f_\gamma f}$ ($\times 10^{-2}$) | 2.7 | 1.2 |
| $T_{full}^{f_\gamma f}$ ($\times 10^{-2}$) | 7.40 | 2.1 |
| $T_{co}^{b_\gamma b}$ | 1.4 | 1.01 |
| $T_{cross}^{b_\gamma b}$ | 1.84 | 0.40 |
| $T_{full}^{b_\gamma b}$ | 3.24 | 1.41 |
| $T_{full\to co}^{\gamma b\to\gamma f}$ ($\times 10^{-3}$) | 65.6 | 3.45 |
| $T_{full\to cross}^{\gamma b\to\gamma f}$ | 3.18 | 1.41 |
| $T_{full\to cross}^{\gamma b\to\gamma f}/T_{co}^{f_\gamma f}$ ($\times 10^{1}$) | 6.747 | 16.048 |
| $T_{co}^{b_\gamma b}/T_{co}^{f_\gamma f}$ ($\times 10^{1}$) | 2.973 | 11.488 |
| $T_{full}^{b_\gamma b}/T_{full}^{f_\gamma f}$ ($\times 10^{1}$) | 4.378 | 6.669 |
| $AT_{co}^{balanced}$ | 1.35 | 1.00 |
| $AT_{cross}^{balanced}$ | 1.82 | 0.29 |
| $AT_{co}^{differential}$ | 3.13 | 1.40 |
| $AT_{cross}^{differential}$ ($\times 10^{-3}$) | 38.6 | 8.97 |
| $rot_f$ (°) | -35.55 | -54.91 |
| $rot_b$ (°) | -49.02 | -32.32 |
| Σ (°) | -84.57 | -87.23 |
| $\tau_f$ (°) | -16.53 | -29.87 |
| $\tau_b$ (°) | 6.16 | 0.59 |
| GFR-FOM | 0.21 | 0.266 |
| GFI-FOM$_{\|backward\|}$ ($\times 10^{-3}$) | 5.97 | 8.13 |
| GFI-FOM-active$_{cross}$ ($\times 10^{-3}$) | 5.70 | 7.90 |
| GFI-FOM-active$_{mixed}$ ($\times 10^{-3}$) | 5.82 | 7.91 |
| GFI-FOM-active$_{balanced}$ ($\times 10^{-3}$) | 3.35 | 2.74 |
| GFI-FOM-active$_{inverse}$ ($\times 10^{-3}$) | 5.18 | 4.05 |
| Q-factor ($T_{co}^{b_\gamma b}$) | 291.91 | 448.66 |
| Q-factor ($T_{full}^{b_\gamma b}$) | 281.73 | 423.14 |
| Q-factor ($T_{full\to cross}^{\gamma b\to\gamma f}$) | 294.45 | 432.35 |

**Table S5.** Active multilayer optical response parameters: transmission, polarization rotation and FOMs at extrema. The co-polarized ($T_{co}$), cross-polarized ($T_{cross}$) and total ($T_{full}$) transmissions where the superscript indicates forward (f) and backward (b) propagation direction for azimuthal orientation corresponding to forward ($\gamma_f$) and backward ($\gamma_b$) configurations. The transmission in backward propagation direction projected onto the co-polarized ($T_{full\to co}^{\gamma b\to\gamma f}$) and cross-polarized ($T_{full\to cross}^{\gamma b\to\gamma f}$) component of forward direction. The $T_{full\to cross}^{\gamma b\to\gamma f}/T_{co}^{f_\gamma f}$, $T_{co}^{b_\gamma b}/T_{co}^{f_\gamma f}$ and $T_{tot}^{b_\gamma b}/T_{tot}^{f_\gamma f}$ transmission ratios indicating isolator properties. The balanced co-polarized ($AT_{co}^{balanced} = T_{co}^{f_\gamma_f} - T_{co}^{b_\gamma_b}$) and cross-polarized ($AT_{cross}^{balanced} = T_{cross}^{f_\gamma_f} - T_{cross}^{b_\gamma_b}$), the differential co-polarized ($AT_{co}^{differential} = T_{co}^{f_\gamma_f} - T_{full\to cross}^{\gamma_b\to\gamma_f}$) and cross-polarized ($AT_{cross}^{differential} = T_{cross}^{f_\gamma_f} - T_{full\to co}^{\gamma_b\to\gamma_f}$), and total ($AT_{full}^{differential} = T_{full}^{f_\gamma_f} - T_{full\to cross}^{\gamma_b\to\gamma_f}$) asymmetrical transmissions. The polarization rotations and ellipticity in forward ($rot_f$ and $\tau_f$) and backward ($rot_b$ and $\tau_b$) propagation directions, the cumulative rotation ($\Sigma$). The used FOMs qualifying the Generalized Faraday Rotation (GFR) and Generalized Faraday Isolator (GFI) properties of the active multilayers and the *Q*-factor of the emission peak for the quantities inside the brackets.